\documentclass{article}

\usepackage{amssymb}
\usepackage{amsmath}
\usepackage{latexsym}
\usepackage[dvips]{graphicx}

\def\abstract#1{\vskip 7mm 
        \begin{center}{\large Abstract}\par \smallskip
                \begin{minipage}[c]{15.5cm}
                       #1
                \end{minipage}
        \end{center}
}
\def\title#1{\begin{center}{\Large\bf #1}\end{center}}
\def\author#1{\vskip 5mm \begin{center}{#1}\end{center}}
\def\address#1{\begin{center}{\it #1}\end{center}}
\def\ead#1{\centerline{e-mail : #1}}

\numberwithin{equation}{section}

\newcommand{\eqb}{\begin{equation}}
\newcommand{\eqe}{\end{equation}}
\newcommand{\eqbnon}{\begin{equation*}}
\newcommand{\eqenon}{\end{equation*}}

\newcommand{\eqab}{\begin{eqnarray}}
\newcommand{\eqae}{\end{eqnarray}}
\newcommand{\eqabnon}{\begin{eqnarray*}}
\newcommand{\eqaenon}{\end{eqnarray*}}
\newcommand{\seqb}{\begin{subequations}}
\newcommand{\seqe}{\end{subequations}}

\newcommand{\defeq}{:=}
\newcommand{\defeqr}{=:}

\newcommand{\sigmap}{\sigma^{\prime}}
\newcommand{\gin}[1]{g_{in}(\vec{#1})}
\newcommand{\gout}[1]{g_{out}(\vec{#1})}
\newcommand{\gb}[1]{g_g(\vec{#1})}
\newcommand{\gh}[1]{g_h(\vec{#1})}
\newcommand{\jn}{J_{NE}}
\newcommand{\je}{J_{empty}}
\newcommand{\tn}{t_{NE}}
\newcommand{\te}{t_{empty}}

\newcommand{\anglm}{\theta_m}
\newcommand{\rt}{\tilde{t}}
\newcommand{\rtau}{\tilde{\tau}}

\begin{document}

\title{Black Hole Evaporation as a Nonequilibrium Process
\footnote{This is a slightly modified version of contribution (Chap.8) to an edited book; M.N.Christiansen and T.K.Rasmussen, {\it Classical and Quantum Gravity Research}, Nova Science Publishers, 2008}}

\author{Hiromi Saida}
\address{Department of Physics, Daido Institute of Technology, Minami-ku, Nagoya 457-8530, Japan
\footnote{Institution's name is changed to ``Daido University'' from April 2009.}}
\ead{saida@daido-it.ac.jp}

\abstract{
When a black hole evaporates, there arises a net energy flow from the black hole into its outside environment due to the Hawking radiation and the energy accretion onto black hole. 
Exactly speaking, any thermal equilibrium state has no energy flow, and therefore the black hole evaporation is a nonequilibrium process. 
To study details of evaporation process, nonequilibrium effects of the net energy flow should be taken into account. 
The nonequilibrium nature of black hole evaporation is a challenging topic which includes not only black hole physics but also nonequilibrium physics. 
In this article we simplify the situation so that the Hawking radiation consists of non-self-interacting massless matter fields and also the energy accretion onto the black hole consists of the same fields. 
Then we find that the nonequilibrium nature of black hole evaporation is described by a nonequilibrium state of that field, and we formulate nonequilibrium thermodynamics of non-self-interacting massless fields. 
By applying it to black hole evaporation, followings are shown: 
(1)~Nonequilibrium effects of the energy flow tends to accelerate the black hole evaporation, and, consequently, a specific nonequilibrium phenomenon of semi-classical black hole evaporation is suggested. 
Furthermore a suggestion about the end state of quantum size black hole evaporation is proposed in the context of information loss paradox. 
(2)~Negative heat capacity of black hole is the physical essence of the generalized second law of black hole thermodynamics, and self-entropy production inside the matter around black hole is not necessary to ensure the generalized second law. 
Furthermore a lower bound for total entropy at the end of black hole evaporation is given. 
A relation of the lower bound with the so-called covariant entropy bound conjecture is interesting but left as an open issue.
}

 \tableofcontents

\section{Introduction}
\label{sec-intro}

Black hole evaporation is one of interesting phenomena in black hole physics~\cite{ref-hr}. 
A direct treatment of time evolution of the evaporation process suffers from mathematical and conceptual difficulties; 
the mathematical one will be seen in the dynamical Einstein equation in which the source of gravity may be a quantum expectation value of stress-energy tensor of Hawking radiation, and the conceptual one will be seen in the definition of dynamical black hole horizon. 
Therefore an approach based on the black hole thermodynamics is useful~\cite{ref-bht,ref-gsl}.

Exactly speaking, dynamical evolution of any system is a nonequilibrium process. 
If and only if thermodynamic state of the system under consideration passes near equilibrium states during its evolution, its dynamics can be treated by an approximate method, the so-called {\it quasi-static process}. 
In this approximation, it is assumed that the thermodynamic state of the system evolves on a path lying in the state space which consists of only thermal equilibrium states, and the time evolution is described by a succession of different equilibrium states. 
However once the system comes far from equilibrium, the quasi-static approximation breaks down. 
In that case a nonequilibrium thermodynamic approach is necessary. 
For dissipative systems, the heat flow inside the system can quantify the degree of nonequilibrium nature \cite{ref-eit,ref-sst,ref-anti.sst,ref-anti.info}.

For the black hole evaporation, when its horizon scale is larger than Planck size, it is relevant to describe the black hole itself by equilibrium solutions of Einstein equation, Schwarzschild, Reissner-Nortstr\"{o}m and Kerr black holes, because the evaporation proceeds extremely slowly and those equilibrium solutions are stable under gravitational perturbations~\cite{ref-chandra}. 
The slow evolution is understandable by the Hawking temperature~\cite{ref-hr} which is regarded as an equilibrium temperature of black hole,
\eqb
 T_g \defeq \dfrac{m_{pl}^2}{8 \pi M} \, ,
\label{eq-intro.temperature}
\eqe
where $M$ is the black hole mass, $m_{pl}$ is the Planck mass and the units $c = \hbar = k_B = 1$ and $G = 1/m_{pl}^2$ are used. 
Obviously a classical size black hole ($M \gg m_{pl}$) has a very low temperature. 
This means a very weak energy emission rate by the Hawking radiation which is proportional to $(2 G M)^2\,T_g^4$ due to the Stefan-Boltzmann law. 
Therefore the quasi-static approximation works well for the black hole itself during its evaporation process. 
However the outside environment around black hole may not be described by the quasi-static approximation because of the energy flow due to the Hawking radiation. 
The Hawking radiation causes an energy flow in the outside environment, and that energy flow drives the outside environment out of equilibrium. 
As indicated by Eq.\eqref{eq-intro.temperature}, the black hole temperature and the energy emission rate by black hole increase as $M$ decreases along the evaporation. 
The stronger the energy emission, the more distant from equilibrium the outside environment. 
Therefore the nonequilibrium nature of the outside environment becomes stronger as the black hole evaporation proceeds. 
At the same time, the quasi-static approximation is applicable to the black hole itself since equilibrium black hole solutions are stable under gravitational perturbation. 
Hence, in studying detail of evaporation process, while the black hole itself is described by quasi-static approximation, but the nonequilibrium effects of the energy flow in the outside environment should be taken into account.

In the above paragraph, the energy accretion onto black hole is ignored. 
However if the temperature of outside environment is non-zero and lower enough than the black hole temperature, then the black hole evaporates under the effect of energy exchange due to the Hawking radiation and the energy accretion. 
In this case the same consideration explained above holds and we recognize the importance of the net energy flow from black hole to outside environment. 
Dynamical behaviors of black hole evaporation will be described well by taking nonequilibrium nature of the net energy flow into account.

In Sec.\ref{sec-model}, we introduce a simple model of black hole evaporation to examine the net energy flow in the outside environment, where the matter fields of Hawking radiation and energy accretion are represented by non-self-interacting massless fields for simplicity. 
Sec.\ref{sec-sst} is devoted to construction of nonequilibrium thermodynamics of that field. 
Then it is applied to the black hole evaporation. 
Sec.\ref{sec-evapo} reveals that the nonequilibrium effect tends to accelerate the evaporation process and, consequently, gives a suggestion about the end state of quantum size black hole evaporation in the context of the information loss paradox. 
Sec.\ref{sec-gsl} reveals that the generalized second law is guaranteed not by self-interactions of matter fields around black hole which cause self-production of entropy inside the matters, but by the self-gravitational effect of black hole appearing as its negative heat capacity in Eq.\eqref{eq-model.capacity}. 
Readers can read Secs.\ref{sec-evapo} and~\ref{sec-gsl} separately, and may skip over Sec.\ref{sec-evapo} to see discussions on generalized second law in Sec.\ref{sec-gsl}. 
Finally Sec.\ref{sec-conc} concludes this article with comments for future direction of this study.

Throughout this article except for Eq.\eqref{eq-intro.temperature}, Planck units $c = \hbar = G = k_B = 1$ are used. 
Then the Stefan-Boltzmann constant becomes $\sigma \defeq \pi^2/60$ which is appropriate for photon gas. 
When one consider non-self-interacting massless matter fields, as indicated in Sec.\ref{sec-sst}, it is necessary to replace $\sigma$ by its generalization,
\eqb
 \sigmap \defeq \dfrac{N\, \pi^2}{120} = \dfrac{N}{2}\, \sigma \, ,
\label{eq-intro.sigma}
\eqe
where $N \defeq n_b + ( 7/8 )\, n_f$. 
Here $n_b$ is the number of inner states of massless bosonic fields and $n_f$ is that of massless fermionic fields. 
($n_b = 2$ for photons.) 
Furthermore, at least when the black hole temperature is lower than 1 TeV (upper limit by present accelerator experiments), it is appropriate to estimate the order of $N$ by the standard particles (inner states of quarks, leptons and gauge particles of four fundamental interactions),
\eqb
 N \simeq 100 \, .
\label{eq-intro.N}
\eqe
This denotes $\sigmap \simeq 10$. 
Throughout this article, we simply assume that $N$ is independent of black hole temperature and this estimate~\eqref{eq-intro.N} holds always for semi-classical black hole evaporation ($T_g < 10^{16}$ TeV).

\section{Thermodynamic model of black hole evaporation}
\label{sec-model}

According to the black hole thermodynamics \cite{ref-hr,ref-bht,ref-gsl}, a stationary black hole is regarded as an object in thermal equilibrium, a black body. 
For simplicity, let us consider a Schwarzschild black hole. 
Its equations of states as a black body are
\eqb
 E_g = \frac{1}{8 \pi T_g} = \frac{R_g}{2} \quad , \quad
 S_g = \frac{1}{16 \pi T_g^2} = \pi R_g^2 \, ,
\label{eq-model.eos}
\eqe
where $E_g$, $R_g$, $T_g$, $S_g$ correspond respectively to mass energy, horizon radius, Hawking temperature and Bekenstein-Hawking entropy. 
Obviously the radius $R_g$ decreases when this body loses its energy $E_g$. 
The black hole evaporation is represented by the energy loss of this black body.

The heat capacity $C_g$ of this body is negative,
\eqb
 C_g \defeq \frac{dE_g}{dT_g} = - \frac{1}{8 \pi T_g^2} = - 2 \pi R_g^2 < 0 \, .
\label{eq-model.capacity}
\eqe
The negative heat capacity is a peculiar property of self-gravitating systems~\cite{ref-sgs}. 
Therefore the energy $E_g$ includes self-gravitational effects of a black hole on its own thermodynamic state. 
Furthermore it has already been revealed that, using the Euclidean path-integral method for a black hole spacetime and matter fields on it, an equilibrium entropy of whole gravitational field on a black hole spacetime is given by the Bekenstein-Hawking entropy~\cite{ref-entropy}. 
This means $S_g$ in Eq.\eqref{eq-model.eos} is the equilibrium entropy of whole gravitational field on black hole spacetime, and the gravitational entropy vanishes if there is no black hole horizon. 
Hence we find that energetic and entropic properties of a black hole are encoded in the equations of states~\eqref{eq-model.eos}. 
Hereafter we call this black body {\it the black hole}.

As mentioned Sec.\ref{sec-intro}, the nonequilibrium nature of black hole evaporation arises in the matter fields around black hole due to the net energy flow by Hawking radiation and energy accretion onto the black hole. 
When we consider arbitrary dissipative matter fields as the Hawking radiation and energy accretion, we immediately face a very difficult problem how to construct a nonequilibrium thermodynamics for arbitrary dissipative matters. 
This is one of the most difficult subjects in physics~\cite{ref-eit,ref-sst,ref-anti.sst,ref-anti.info}. 
To avoid such a difficult problem and for simplicity, let us consider non-self-interacting massless fields to represent the Hawking radiation and energy accretion. 
For example, photon, graviton, neutrino (if it is massless) and free Klein-Gordon field ($\Box \Phi = 0$) are candidates of such matter fields, and they possess the generalized Stefan-Boltzmann constant $\sigmap$ given in Eq.\eqref{eq-intro.sigma}. 
Hereafter we call these fields {\it the radiation fields}.

\begin{figure}[t]
 \begin{center}
 \includegraphics[height=30mm]{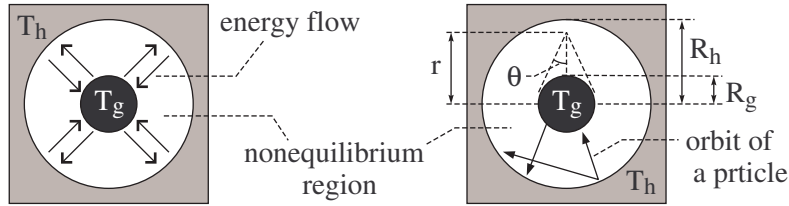}
 \end{center}
\caption{NE model. Left panel shows energy flow between black hole and heat bath. Right one shows some variables and particle orbits of radiation fields. The radiation fields sandwiched by black bodies are in two-temperature steady state.}
\label{fig-1}
\end{figure}

As mentioned above, nonequilibrium phenomenon is one of the most difficult subjects in physics. 
It is impossible at present to treat the nonequilibrium nature of black hole evaporation in a full general relativistic framework. 
Hence we resort to a simplified model to examine the nonequilibrium effects of net energy flow in the outside environment around black hole~\cite{ref-evapo.sst}:
\begin{description}
\item[Nonequilibrium Evaporation (NE) model:] Put a spherical black body of temperature $T_g$ in a heat bath of temperature $T_h (< T_g)$, where the equations of states of the spherical black body is given by Eq.\eqref{eq-model.eos} and we call the black body {\it the black hole}. 
Let the heat bath (the outer black body of temperature $T_h$) be made of ordinary materials of positive heat capacity. 
Then hollow a spherical region out of the heat bath around black hole as seen in Fig.\ref{fig-1}. 
The hollow region is a shell-like region which is concentric with the black hole and separates the black hole from the heat bath. 
This region is filled with matter fields emitted by black hole and heat bath. 
Those matter fields are the non-self-interacting massless fields possessing generalized Stefan-Boltzmann constant $\sigmap$ given in Eq.\eqref{eq-intro.sigma}, and we call these fields {\it the radiation fields}. 
This model consists of three parts, black hole, heat bath and radiation fields. 
Furthermore we consider the case that the whole system is isolated and the total energy of the three parts is conserved.
\end{description}
The isolated condition reflects the whole universe including an evaporating black hole. 
The temperature difference ($T_g > T_h$) causes a net energy flow from the black hole to the heat bath. 
This energy flow drives a relaxation process of the whole system because of the isolated condition. 
For ordinary systems of positive heat capacity, relaxation process reaches an equilibrium state at the end of the process. 
But the relaxation in NE model may not reach a total equilibrium state of the whole system, because the temperature difference $T_g - T_h$ may increase along the decrease of $E_g$ due to the negative heat capacity $C_g$ given in Eq.\eqref{eq-model.capacity}. 
This causes the decrease of $R_g$ due to the equations of states~\eqref{eq-model.eos}. 
Therefore the relaxation process in NE model corresponds to the black hole evaporation. 

Here let us comment about terminology. 
There may be an objection that the term ``relaxation'' is not suitable for the case of increasing temperature difference. 
But in this article, please understand it means the time evolution arising in isolated inhomogeneous systems.

If a full general relativistic treatment is possible, the Hawking radiation experiences the curvature scattering to form a spacetime region filled with interacting matters and some fraction of Hawking radiation is radiated back to the black hole from that region. 
The heat bath in the NE model is understood as a simple representation of not only matters like accretion disk but also such region formed by curvature scattering.

Here, as an objection to the NE model, one may remember the so-called {\it Tolman factor} which appears in the ``equilibrium'' temperature of radiation fields around black hole: 
Exactly speaking, equilibrium of any matter fields around black hole is ``local'' equilibrium. 
It has already been known that, when the radiation fields around black hole are in local equilibrium with the black hole, the local equilibrium temperature $T_{eq}(r)$ of radiation fields (not of black hole) at a spacetime point of areal radius $r$ from the center of black hole is $T_{eq}(r) = T_g/\sqrt{1-R_g/r}$, where $T_g$ is the Hawking temperature and $R_g$ is the horizon radius. 
This $T_{eq}(r)$ is obtained in a full general relativistic framework of equilibrium thermodynamics, and the factor $1/\sqrt{1-R_g/r}$ is the Tolman factor~\cite{ref-tolman}. 
One may think that, because $T_{eq}~\to~\infty$ as $r~\to~R_g$, it is unreasonable to assign $T_g$ to black hole as its temperature. 
But let us emphasize that $T_{eq}(r)$ is not black hole temperature but the equilibrium temperature of radiation fields. 
We can understand $T_{eq}(r)$ as follows: 
In order to retain equilibrium of radiation fields against external gravitational force by black hole, a temperature of radiation fields higher than the asymptotic value $T_g$ is required, since the higher temperature denotes the higher pressure against external gravitational force. 
The Tolman factor describes the effect of external gravity on the equilibrium radiation fields, and  becomes unity if the external gravity vanishes. 
The local equilibrium temperature of radiation fields may count an ``intrinsic'' temperature of black hole and an additional gravitational effect in Tolman factor. 
It may be reasonable to regard the asymptotic value $T_g$ as an intrinsic black hole temperature. 
Hence we assign $T_g$ to the black hole in NE model. 
But it is ture that the NE model is not a full general relativistic model, and ignoring, for example, gravitational redshift and curvature scattering on the ``nonequilibrium'' radiation fields propagating in the hollow region. 
Although the NE model may be too simple, let us try to investigate nonequilibrium effects of the net energy flow from black hole to its outside environment in the framework of NE model.

Furthermore, we put the quasi-equilibrium assumption to utilize the ``equilibrium'' equations of states~\eqref{eq-model.eos} for black hole, and the fast propagation assumption to treat the nonequilibrium nature as simple as possible:
\begin{description}
\item[Quasi-equilibrium assumption:] Time evolution in the NE model is not so fast that the evolutions of black hole and heat bath are approximated well by the quasi-static process individually, while the quasi-static approximation is not valid for the radiation fields due to the net energy flow from black hole to heat bath. 
Then, it is reasonable to use Eq.\eqref{eq-model.eos} as the equations of states for black hole. 
Furthermore, since Schwarzschild black hole is not a quantum one, following relation is required,
\eqb
 R_g \gtrsim 1 \, .
\label{eq-model.Rg}
\eqe
\item[Fast propagation assumption:] The volume of hollow region is not so large that particles of radiation fields travel very quickly across the hollow region. 
Then the retarded effect on radiation fields during propagating in the hollow region is ignored.
\end{description}
There are two points which we should note here. 
The first point is about the thermodynamic states of black hole and heat bath under the quasi-equilibrium assumption. 
This denotes the temperatures $T_g$ and $T_h$ are given by equilibrium temperatures at each moment of evaporation process. 
Therefore we can regard $T_g$ and $T_h$ as constants within a time scale that one particle of radiation fields travels in the hollow region until absorbed by black hole or heat bath. 
This is consistent with the fast propagation assumption.

The second point is about the thermodynamic state of radiation fields. 
In the hollow region, the radiation fields of different temperatures $T_g$ and $T_h$ are simply superposed, since the radiation fields are of non-self-interacting (collisionless particles gas). 
This means the radiation fields are in a two-temperature nonequilibrium state. 
Furthermore, because $T_g$ and $T_h$ are constant while one particle of radiation fields travel across the hollow region, it is reasonable to consider that the radiation fields have a stationary energy flow from black hole to heat bath within that time scale. 
Hence, at each moment of time evolution of NE model, the thermodynamic state of radiation fields is well approximated to a macroscopically stationary nonequilibrium state, which we call {\it the steady state} hereafter. 
Consequently, time evolution of radiation fields is described by {\it the quasi-steady process} in which the thermodynamic state of radiation fields evolves on a path lying in the state space which consists of steady states, and the time evolution is described by a succession of different steady states. 
Therefore we need a thermodynamic formalism of two-temperature steady states for radiation fields. 
The steady state thermodynamics for radiation fields has already been formulated in~\cite{ref-sst.rad}, which is summarized in next section.

\section{Steady state thermodynamics for radiation fields}
\label{sec-sst}

Before proceeding to the black hole evaporation, two-temperature steady state thermodynamics for radiation fields~\cite{ref-sst.rad} is explained in this section. 
Subsec.\ref{sec-sst.min} introduces the minimum tools required to apply to the NE model. 
Readers may skip over Subsec.\ref{sec-sst.detail} to read remaining Secs.\ref{sec-evapo},~\ref{sec-gsl} and~\ref{sec-conc}. Furthermore, since Secs.\ref{sec-evapo} and~\ref{sec-gsl} are written separately, one can also skip over Sec.\ref{sec-evapo} to see the generalized second law in Sec.\ref{sec-gsl}.

Subsec.\ref{sec-sst.detail} exhibits a more detail of two-temperature steady state thermodynamics for radiation fields. 
Although a full understanding of it is not necessary for black hole evaporation, but it may be helpful to understand, for example, the free streaming in the universe like cosmic microwave background and/or the radiative energy transfer inside a star and among stellar objects. 
Keeping future expectation of such applications in mind, Subsec.\ref{sec-sst.detail} is placed here.

\subsection{Minimum tools for applying to NE model}
\label{sec-sst.min}

\begin{figure}[t]
 \begin{center}
 \includegraphics[height=25mm]{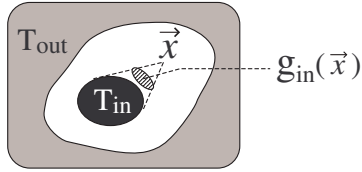}
 \end{center}
\caption{SST model. Radiation fields between outer and inner black bodies are in steady state. }
\label{fig-2}
\end{figure}

To concentrate on investigating two-temperature steady states of radiation fields, we consider a model which can be realized in laboratory experiments. 
According to the NE model, we introduce the following model named SST after the Steady State Thermodynamics. 
Then, after constructing the steady state thermodynamics for radiation fields, we will modify the SST model to the NE model in Secs.\ref{sec-evapo} and~\ref{sec-gsl}.
\begin{description}
\item[SST model:] Make a vacuum cavity in a large black body of temperature $T_{out}$ and put an another smaller black body of temperature $T_{in} \, ( \neq T_{out} )$ in the cavity as seen in Fig.\ref{fig-2}. 
For the case $T_{in} > T_{out}$, the radiation fields emitted by two black bodies causes a net energy flow from the inner black body to the outer one. 
When the outer black body is isolated from outside world and the heat capacities of two black bodies are positive definite, the whole system which consists of two black bodies and radiation fields between them relaxes to a total equilibrium state in which two black bodies and radiation fields have the same equilibrium temperature. 
\end{description}
It should be emphasized that $T_{in}$ and $T_{out}$ are ``equilibrium'' temperatures of black bodies, respectively. 
Outer body is in an equilibrium state, and inner one is also. But their equilibrium sates are different from each other since $T_{in} \neq T_{out}$. 
Then the radiation fields between them is in a nonequilibrium state. 
By keeping temperatures $T_{in}$ and $T_{out}$ constant, the nonequilibrium state of radiation fields becomes a macroscopically stationary nonequilibrium state, which we call a steady state.

The difference of SST model from the NE model is that heat capacities of two black bodies are positive definite and the system does not necessarily has symmetric geometry. 
Because of the positive heat capacity, the relaxation process in the SST model leads the whole system to a total equilibrium state, while the black hole evaporation increases the temperature difference between black hole and heat bath due to the negative heat capacity of black hole. 
But as for the NE model, the radiation fields in the SST model are also non-self-interacting massless fields. 
The SST model ignores gravitational interactions among two bodies and radiation fields. 
Furthermore we consider the case satisfying following two conditions according to the quasi-equilibrium and fast propagation assumptions; 
the first is that the evolution of each black body is of quasi-static during the relaxation process of the whole system, and the second condition is that the volume of cavity is so small that retarded effect on radiation fields is ignored. 
Then, as discussed Sec.\ref{sec-model}, the evolution of radiation fields is of quasi-steady. 
Due to the Stefan-Boltzmann law, each steady state composing the quasi-steady process of the radiation fields has a net energy flow
\eqb
 J_{sst} = \sigma \, (\, T_{in}^{\,4} - T_{out}^{\,4} \, ) \, A_{in} \, ,
\label{eq-sst.min.J}
\eqe
where $A_{in}$ is the surface area of inner black body. 
This $J_{sst}$ equals the net energy exchanged par a unit time between the two black bodies via the radiation fields.

A consistent thermodynamic framework for the steady states of radiation fields has already been constructed~\cite{ref-sst.rad}. 
The outline of the construction of steady state internal energy and steady state entropy is as follows.

For the first we consider a massless bosonic gas. The energy of the gas $E_b$ is given by
\eqb
 E_b \defeq \int \frac{dp^3}{(2\, \pi)^3} dx^3 \,
  g_{\vec{p},\vec{x}} \,\, \epsilon_{\vec{p},\vec{x}} \,\, d_{\vec{p},\vec{x}} \, ,
\label{eq-sst.min.energy.ll}
\eqe
where $\vec{p}$ is a momentum of particle, $\vec{x}$ is a spatial point, $\epsilon_{\vec{p},\vec{x}}$ is an energy of particle of $\vec{p}$ at $\vec{x}$, and $g_{\vec{p},\vec{x}}$ and $d_{\vec{p},\vec{x}}$ are respectively the number of states and the average number of particles at a point $(\vec{p},\vec{x})$ in the phase space of the gas. 
For the entropy, we refer to the Landau-Lifshitz type definition of nonequilibrium entropy $S_b$ for bosonic gas (see \S55 in chapter 5 of~\cite{ref-ll}),
\eqb
 S_b \defeq
  \int \frac{dp^3}{(2\, \pi)^3} dx^3 \, g_{\vec{p},\vec{x}}
  \left[\, \left(\, 1 + d_{\vec{p},\vec{x}} \,\right) \,
             \ln\left(\, 1 + d_{\vec{p},\vec{x}} \,\right)
         - d_{\vec{p},\vec{x}} \, \ln d_{\vec{p},\vec{x}}
  \,\right] \, .
\label{eq-sst.min.entropy.ll}
\eqe
It has also been shown in \S55 of~\cite{ref-ll} that the maximization of $S_b$ for an isolated system ($\delta S_b = 0$) gives the equilibrium Bose distribution. 
This is frequently referred in many works on nonequilibrium systems as the {\it H-theorem}. 
However in \S55 of~\cite{ref-ll}, concrete forms of $g_{\vec{p},\vec{x}}$ and $d_{\vec{p},\vec{x}}$ are not specified, since an arbitrary system is considered.

In the SST model, the radiation fields are sandwiched by two black bodies, and its particle is massless and ``collisionless''. Then we can determine $\epsilon_{\vec{p},\vec{x}}$, $g_{\vec{p},\vec{x}}$ and $d_{\vec{p},\vec{x}}$ to be
\eqb
 \epsilon_{\vec{p},\vec{x}} = \omega \quad , \quad
 g_{\vec{p},\vec{x}} = n_b \quad , \quad
 d_{\vec{p},\vec{x}} = \frac{1}{\exp[\omega / T(\vec{p},\vec{x})] - 1} \, ,
\label{eq-sst.min.distribution}
\eqe
where the frequency of a particle $\omega = \left| \vec{p} \right|$, $n_b$ is the number of inner states of bosonic gas which is assumed to be constant as mentioned in Eq.\eqref{eq-intro.N}, and $T(\vec{p},\vec{x})$ is given by
\eqb
 T(\vec{p},\vec{x}) \defeq
 \begin{cases}
  T_{in}  & \text{for} \,\,\, \vec{p} = \vec{p}_{in} \,\,\, \text{at $\vec{x}$} \\
  T_{out} & \text{for} \,\,\, \vec{p} = \vec{p}_{out} \,\,\, \text{at $\vec{x}$}
 \end{cases} \, ,
\eqe
where $\vec{p}_{in}$ is momentum of a particle emitted by the inner black body and $\vec{p}_{out}$ emitted by the outer one. 
The $\vec{x}$-dependence of $T(\vec{p},\vec{x})$ arises, because the directions in which particles of $\vec{p}_{in}$ and $\vec{p}_{out}$ can come to a point $\vec{x}$ vary from point to point.

From the above, we obtain the steady state energy and entropy,
\seqb
\label{eq-sst.min.E.S.boson}
\eqab
 E_b = \int dx^3 e_b(\vec{x}) \quad &,& \quad
 e_b(\vec{x}) \defeq 4 \, \sigma_b \,
  \left(\, \gin{x}\, T_{in}^{\,4} + \gout{x}\, T_{out}^{\,4} \,\right) \\
 S_b = \int dx^3 s_b(\vec{x}) \quad &,& \quad
 s_b(\vec{x}) \defeq \frac{16 \, \sigma_b}{3} \,
  \left(\, \gin{x}\, T_{in}^{\,3} + \gout{x}\, T_{out}^{\,3} \,\right) \, ,
\eqae
\seqe
where integrals in Eq.\eqref{eq-sst.min.integrals} are used, $\sigma_b \defeq n_b \pi^2/120$, $\gin{x}$ is the solid angle (divided by $4 \pi$) covered by directions of $\vec{p}_{in}$ at $\vec{x}$ as shown in Fig.\ref{fig-2}, and $\gout{x}$ is similarly defined with $\vec{p}_{out}$. 
By definition, we have $\gin{x} + \gout{x} \equiv 1$. 
On the other hand, if the radiation fields are in equilibrium of temperature $T$, the ordinary equilibrium thermodynamics determine equilibrium energy density $e$ and entropy density $s$ to be $e = 4 \sigma_b \, T^4$ and $s = ( 16 \sigma_b/3 ) \, T^3$. 
We find that the steady state energy and entropy are given by a simple linear combination of the values which are calculated as if the radiation fields are in equilibrium of temperature $T_{in}$ or $T_{out}$. 
This is consistent with the collisionless nature of radiation fields.

Next we consider a massless fermionic gas, the formula of energy \eqref{eq-sst.min.energy.ll} holds for fermions as well. But the formula of entropy \eqref{eq-sst.min.entropy.ll} is replaced by~\cite{ref-ll}
\eqb
 S_f \defeq
 -\int \frac{dp^3}{(2\, \pi)^3} dx^3 \, g_{\vec{p},\vec{x}}
  \left[\, \left(\, 1 - d_{\vec{p},\vec{x}} \,\right) \,
             \ln\left(\, 1 - d_{\vec{p},\vec{x}} \,\right)
         + d_{\vec{p},\vec{x}} \, \ln d_{\vec{p},\vec{x}}
  \,\right] \, .
\label{eq-sst.min.entropy.ll.fermion}
\eqe
Then, following the same procedure as for the massless bosonic gas with replacing the distribution function by $d_{\vec{p},\vec{x}} = [\,\exp[\omega / T(\vec{p},\vec{x})] + 1\,]^{-1}$, we can obtain the steady state energy and entropy of a massless fermionic gas,
\eqb
 E_f = \frac{7}{8} \, \frac{n_f}{n_b}\, E_b \quad , \quad
 S_f = \frac{7}{8} \, \frac{n_f}{n_b}\, S_b \, ,
\label{eq-sst.fermion}
\eqe
where integrals in Eq.\eqref{eq-sst.min.integrals} are used, and $n_f$ is the number of inner states of fermionic gas which is assumed to be constant as mentioned in Eq.\eqref{eq-intro.N}. 
Hence, when the black bodies in SST model emit $n_f$ massless fermionic modes and $n_b$ massless bosonic modes, the steady state energy $E_{sst}$ and entropy $S_{sst}$ of radiation fields are given by Eqs.\eqref{eq-sst.min.E.S.boson} with replacing $\sigma_b$ by $\sigmap$ given in Eq.\eqref{eq-intro.sigma},
\seqb
\label{eq-sst.min.energy.entropy}
\eqab
 E_{sst} = \int dx^3 e_{sst}(\vec{x}) \quad &,& \quad
 e_{sst}(\vec{x}) \defeq 4 \, \sigmap \,
  \left(\, \gin{x}\, T_{in}^{\,4} + \gout{x}\, T_{out}^{\,4} \,\right)
\label{eq-sst.min.energy} \\
 S_{sst} = \int dx^3 s_{sst}(\vec{x}) \quad &,& \quad
 s_{sst}(\vec{x}) \defeq \frac{16 \, \sigmap}{3} \,
  \left(\, \gin{x}\, T_{in}^{\,3} + \gout{x}\, T_{out}^{\,3} \,\right) \, ,
\label{eq-sst.min.entropy}
\eqae
\seqe
At equilibrium limit $T_{in} = T_{out}$, these densities $e_{sst}$ and $s_{sst}$ become equilibrium ones $e = 4 \sigmap T^4$ and $s = (16 \sigma/3)T^3$ respectively, where an identical equation $\gin{x} + \gout{x} \equiv 1$ is used.

By defining the other variables like free energy with somewhat careful discussions, it has already been checked that the 0th, 1st, 2nd and 3rd laws of ordinary equilibrium thermodynamics are extended to include two-temperature steady states of radiation fields~\cite{ref-sst.rad}. 
This means that the steady state thermodynamics for radiation fields has already been constructed in a consistent way. 
Especially on the steady state entropy, the total entropy of the whole system which consists of two black bodies and radiation fields increases monotonously during the relaxation process of the whole system,
\eqb
 dS_{in} + dS_{out} + dS_{sst} \ge 0 \, ,
\label{eq-sst.min.2nd}
\eqe
where the equality holds for total equilibrium states, $S_{in}$ and $S_{out}$ are entropies of two black bodies which are given by ordinary equilibrium thermodynamics, and a simple sum of total entropy $S_{in} + S_{out} + S_{sst}$ is assumed for not only equilibrium states but also steady states. 
This indicates that the second law holds for steady states of radiation fields.

Here for the convenience to follow Eqs.\eqref{eq-sst.min.E.S.boson} and \eqref{eq-sst.fermion}, we list useful formulae; 
\begin{align}
 &\int^{\infty}_0 dx \, \frac{x^3}{e^x - 1} = \frac{\pi^4}{15}
 &\int^{\infty}_0 dx & \frac{x^3}{e^x + 1} = \frac{7 \pi^4}{120}
\nonumber \\
 &\int^{\infty}_0 dx \, \frac{x^2}{1 - e^{-x}} \, \ln( 1 - e^{-x} ) = - \frac{\pi^4}{36}
 &\int^{\infty}_0 dx & \frac{x^2}{1 + e^{-x}} \, \ln( 1 + e^{-x} ) = \frac{11 \pi^4}{180} - F
\label{eq-sst.min.integrals}  \\
 &\int^{\infty}_0 dx \, \frac{x^2}{e^x - 1} \, \ln( e^x - 1 ) = \frac{11 \pi^4}{180}
 &\int^{\infty}_0 dx & \frac{x^2}{e^x + 1} \, \ln( e^x + 1 ) = \frac{\pi^4}{60} + F
\nonumber
\end{align}
where $F = [ (\ln 2)^2 - \pi^2 ]\,(\ln 2)^2/6 + 4 \, \phi(4,1/2) + (7 \ln 2/2)\, \zeta(3)$. Here $\zeta(z)$ is zeta function and $\phi(z,s)$ is modified zeta function (Apell's function).

\subsection{A more detail}
\label{sec-sst.detail}

This subsection exhibits a more detail of two-temperature steady state thermodynamics for radiation fields. 
A peculiar point of radiation fields is the collisionless nature of composite particles. 
Radiation fields are non-dissipative matter. 
A representative of them is a photon gas. 
Before a construction of steady state thermodynamics for radiation fields, there were some existing works on nonequilibrium radiation fields.

A traditional treatment of radiative energy transfer, for example that in a star~\cite{ref-sgs}, has been applied to a mixture of radiation fields (photon gas) with a matter like a dense gas or other continuum medium. 
In such a traditional treatment, successive absorption and emission of photons by components of medium matter makes it possible to consider radiation fields as if in local equilibrium states whose temperatures equal those of local equilibrium states of medium matter. 
However, when the radiation fields are in ``vacuum'' space, the idea of local equilibrium is never applicable to radiation fields because they are non-dissipative (see \S63 in~\cite{ref-ll}).

Here let us look at ordinary dissipative systems very briefly. 
For ordinary dissipative systems, the so-called {\it extended irreversible thermodynamics} treats successfully nonequilibrium states~\cite{ref-eit}. 
However this is applicable not to any highly nonequilibrium state but to a state whose entropy flux is well approximated up to second order in the expansion by the heat flux of the nonequilibrium state. 
On the other hand, a steady state thermodynamics has already been suggested for dissipative systems like heat conduction, shear flow, electrical conduction and so on~\cite{ref-sst}. 
Also a heat flux appears as a consistent state variable in those steady state thermodynamics. 
However the range of its application is limited~\cite{ref-anti.sst,ref-anti.info}. 
Although present nonequilibrium thermodynamics for dissipative systems have some restriction on their applicability, the point is that a heat flux plays the role of consistent state variable which quantifies a degree of nonequilibrium nature of ordinary dissipative systems. 
When one deals with nonequilibrium dissipative systems, it is usual to place an interest on the heat flux.

Notable works on nonequilibrium radiation fields in ``vacuum'' are given by Essex~\cite{ref-noneq.rad}. 
Here it may be helpful to point out that a ``heat'' arises by dissipation. 
No heat flux exists in nonequilibrium radiation fields, but an energy flux in radiation fields may correspond to a heat flux in dissipative systems. 
It was natural that Essex considered an energy flux in nonequilibrium radiation fields (photon gas) in vacuum. 
Essex has shown that, contrary to the success in extended irreversible thermodynamics, the entropy flux for nonequilibrium radiation fields is NOT expressed by the expansion by energy flux of those radiation fields. 
Even if the same method of extended irreversible thermodynamics is applied to nonequilibrium radiation fields, the energy flux becomes inconsistent with the nonequilibrium free energy. 
This inconsistency will be looked over in Eq.\eqref{eq-sst.detail.vanish} later in this section. 
This means the energy flux does not work as a consistent state variable for nonequilibrium radiation fields in vacuum.

Apart from Essex's works, there were other works on nonequilibrium radiation fields (photon gas) in vacuum emitted by nonequilibrium ordinary matters~\cite{ref-noneq.rad.info}. 
They used the {\it information theory}. 
The basis of information theory is the assumption that nonequilibrium entropy is given by $\ln d_{ne}$, where $d_{ne}$ is a nonequilibrium distribution function defined case by case according to the system under consideration (see for example~\cite{ref-eit} in which the information theory is also explained). 
Applying the information theory to the total system composed by nonequilibrium radiation fields and its source matter, complicated distribution functions for general nonequilibrium radiation fields and source matter have been suggested in~\cite{ref-noneq.rad.info}. 
However, after those works were reported, it is revealed in~\cite{ref-anti.info} that, at least for a matter whose components are colliding and interacting with each other, the distribution function for a steady state of that matter derived by the information theory does NOT qualitatively agree with that derived by a steady state Boltzmann equation. 
Furthermore it is also concluded in~\cite{ref-anti.info} that the nonequilibrium temperature of that matter determined by the information theory has no physical meaning. 
The information theory does not always work well. 
Therefore, because the suggested distribution function of nonequilibrium radiation fields depends on nonequilibrium temperature of source matter derived by the information theory, the reliability of the distribution function may not be given in~\cite{ref-noneq.rad.info}. 
There is no confirmed form of distribution function of nonequilibrium radiation fields emitted by nonequilibrium matter. 
Hence, to avoid the difficult problem on nonequilibrium state of source matter, we simply assume in the SST model that the source bodies are in equilibrium states.

From the above, we recognize the following three facts: 
(1) The traditional treatment of radiative transfer is applicable only to a mixture of radiation fields with a matter which is dense enough to ignore the vacuum region among components of the matter. 
(2) Energy flux is not a consistent state variable for nonequilibrium radiation fields in vacuum, and therefore a special approach different from that to dissipative systems is required to understand nonequilibrium radiation fields. 
(3) A consistent thermodynamic formulation for a system including nonequilibrium radiation fields in vacuum has not been accomplished, and therefore a consistent nonequilibrium order parameter for radiation fields has not been obtained so far.

At least for two-temperature steady states of radiation fields in vacuum, a thermodynamic formalism is accomplished and a consistent nonequilibrium order parameter for steady states is obtained in~\cite{ref-sst.rad} in the framework of SST model. 
Exactly speaking, the radiation fields in SST model is in {\it local steady states}, because the distribution function in Eq.\eqref{eq-sst.min.distribution} and its fermion version have $\vec{x}$-dependence. 
The radiation fields in a sufficiently small region are in a local steady state, but that local steady state may be different from a local steady state in the other small region. 
Therefore state variables for SST model should be defined as a function of spatial point $\vec{x}$. 
The extensive variable is to be understood as a density.

Let us exhibit consistent steady state variables for radiation fields. 
See~\cite{ref-sst.rad} for detail discussions to justify the following definitions of state variables.

\subsubsection*{Steady state internal energy density $e_{sst}(\vec{x})$ and entropy density $s_{sst}(\vec{x})$}

These are already defined in Eq.\eqref{eq-sst.min.energy.entropy}.

\subsubsection*{Steady state Pressure tensor $P_{sst}^{i j}$ {\rm (in 3-dim. space)}}

One may naively expect that the pressure of ``steady'' state is a global quantity, since a pressure gradient in an ordinary dissipative system accelerates components of that system to cause a dynamical evolution. 
However this is not true of radiation fields whose particle is ``collisionless''. 
As seen below, $P_{sst}^{i j}$ becomes a function of $\vec{x}$ because of $\vec{x}$-dependence of distribution function $d_{\vec{p},\vec{x}}$.

In general, pressure is defined by the momentum flux, the amount of momentum carried by composite particles par unit area and unit time. 
For equilibrium states, momentum flux is homogeneous and isotropic, then equilibrium pressure becomes a scalar quantity. 
However for nonequilibrium states, momentum flux is not homogeneous and/or isotropic, then the pressure should be defined as a tensor,
\eqb
 P_{sst}^{i j}(\vec{x}) \defeq N\,\int dp^3 \, \dfrac{1}{p}\, p^i \, p^j \, d_{\vec{p},\vec{x}} \, ,
\eqe
where $p$ and $p^i$ are, respectively, spatial magnitude and components of momentum of a particle of radiation fields. 
Trace of $P_{sst}^{i j}$ becomes
\eqb
 \frac{1}{3} \Sigma_i P_{sst}^{i i}(\vec{x})
 = \frac{4 \sigmap}{3} \, \left( \gin{x}\,T_{in}^4 + \gout{x} \,T_{out}^4 \right) \, .
\eqe
At equilibrium limit $T_{in} = T_{out}$, this trace becomes equilibrium pressure, $(4 \sigmap/3)T^4$, where an identical equation $\gin{x} + \gout{x} \equiv 1$ is used.

\subsubsection*{Steady state free energy density $f_{sst}$}

By a requirement that the differential of free energy by volume gives the minus of pressure as for ordinary equilibrium thermodynamics, we can obtain
\eqb
 f_{sst}(\vec{x}) \defeq -\frac{1}{3} \Sigma_i P_{sst}^{i i}
 = -\frac{4 \sigmap}{3} \, \left( \gin{x}\,T_{in}^4 + \gout{x} \,T_{out}^4 \right) 
 = -\frac{1}{3} e_{sst}(\vec{x}) \, .
\eqe
At equilibrium limit $T_{in} = T_{out}$, this becomes equilibrium free energy, $-(4 \sigmap/3)T^4$.

\subsubsection*{Steady state chemical potential}

Chemical potential in general can be interpreted as a work needed to add one particle to the system under consideration. 
Because particles of radiation fields are collisionless, no work is needed to add a new one into radiation fields. 
This is the case for either equilibrium or steady states. 
Indeed, the chemical potential of radiation fields (photon gas) in equilibrium is zero. 
Therefore the steady state chemical potential of radiation fields is zero as well.

\subsubsection*{Steady state temperature $T_{sst}(\vec{x})$}

By a requirement that the differential of $f_{sst}$ by $T_{sst}$ gives the minus of entropy density ($\partial f_{sst}/\partial T_{sst} = -s_{sst}$) as for equilibrium ordinary thermodynamics, we can obtain
\eqb
 T_{rad}(\vec{x}) \defeq \gin{x}\,T_{in} + \gout{x}\,T_{out} \, .
\eqe
At equilibrium limit $T_{in} = T_{out}$, this becomes equilibrium temperature.

\subsubsection*{Intensive steady state order parameter $\tau$}

Energy flux $\vec{j}(\vec{x})$ is defined by
\eqb
 \vec{j}(\vec{x}) \defeq j \, \vec{n} \, ,
\eqe
where $\vec{n}$ is a unit vector in the direction of total momentum of particles at $\vec{x}$, and
\eqb
 j \defeq N\int dp^3 \omega\, d_{\vec{p},\vec{x}}\, \cos\phi \, ,
\eqe
where $\omega = p$ is the energy (frequency) of a particle of radiation fields, and $\phi$ is the angle between $\vec{n}$ and $\vec{p}$. 
If $\vec{j}(\vec{x})$ is adopted as an intensive steady state order parameter, its conjugate state variable should also be defined well. 
Such a conjugate variable would be defined by the differential of free energy by $j(\vec{x})$. 
However it has shown in~\cite{ref-sst.rad} such a conjugate variable vanishes,
\eqb
 \frac{\partial f_{sst}}{\partial j} \equiv 0 \, .
\label{eq-sst.detail.vanish}
\eqe
This means $\vec{j}(\vec{x})$ is not a consistent state variable, since its thermodynamic conjugate variable does not exist. \\
Hence, instead of energy flux, we adopt the temperature difference as an intensive steady state order parameter,
\eqb
 \tau \defeq T_{in} - T_{out} \, .
\eqe
This is obviously intensive variable and satisfies a natural requirement $\tau = 0$ at equilibrium limit $T_{in} = T_{out}$. This $\tau$ is consistent with the first law and the concavity of free energy as looked over in Eqs.\eqref{eq-sst.detail.1st} and~\eqref{eq-sst.detail.consistency}.

\subsubsection*{Extensive steady state order parameter density $\psi(\vec{x})$}

After ordinary equilibrium thermodynamics, we define an extensive steady state order parameter as a thermodynamic conjugate variable to $\tau$ using the differential of free energy density. 
Hence we define as
\eqb
 \psi(\vec{x}) \defeq -\frac{\partial f_{sst}}{\partial \tau}
 = \frac{16 \sigmap}{3}\, \gin{x} \, \gout{x} \,\left( T_{in}^3 - T_{out}^3 \right) \, .
\eqe
This satisfies a natural requirement $\psi = 0$ at equilibrium limit $T_{in} = T_{out}$. \\
It may be useful to rewrite $\psi(\vec{x})$ as
\eqb
 \psi(\vec{x}) = \gin{x} \, \gout{x} \,\left( s_{eq}(T_{in}) - s_{eq}(T_{out}) \right) \, ,
\eqe
where $s_{eq}(T) = (16 \sigmap/3) T^3$ is the equilibrium entropy density of radiation fields of temperature $T$. 
It seems very natural and reasonable that a difference of entropies quantifies a degree of nonequilibrium nature of steady states.

\subsubsection*{From zeroth to third laws and concavity of free energy}

Zeroth law, the existence of steady states, is the existence of systems which realize steady state radiation fields as shown in Fig.\ref{fig-2}. \\
First law can be checked from the above definitions of state variables. 
We can obtain the following equation,
\eqb
 de_{sst}(\vec{x})\Bigr|_{\vec{x} = \text{fixed}}
 = T_{sst}(\vec{x})\, ds_{sst}(\vec{x})
   + \tau\, d\psi(\vec{x}) \,\Bigr|_{\vec{x}=\text{fixed}} \, ,
\label{eq-sst.detail.1st}
\eqe
where we fixed $\vec{x}$ since a local steady state is considered, and consequently a ``work term'' including pressure does not explicitly appear in this relation. 
A work term will appear if the above equation is integrated over the hollow region in SST model. 
Eq.\eqref{eq-sst.detail.1st} denotes the first law. \\
Second law is satisfied as mentioned in Eq.\eqref{eq-sst.min.2nd}. \\
Third law is satisfied by definition of $T_{sst}(\vec{x})$, if the third law of ordinary equilibrium thermodynamics holds for inner and outer black bodies. \\
Furthermore, as for ordinary equilibrium thermodynamics, we can find the free energy density is concave with intensive state variables,
\eqb
 \frac{\partial^2 f_{sst}}{\partial T_{sst}^2} \le 0 \quad , \quad
 \frac{\partial^2 f_{sst}}{\partial \tau^2} \le 0 \, .
\label{eq-sst.detail.consistency}
\eqe

From the above, we have obtained a consistent two-temperature steady state thermodynamics for radiation fields.

\section{Black hole evaporation with energy accretion}
\label{sec-evapo}

Now we apply the steady state thermodynamics for radiation fields to the NE model. 
Contents of this section are based on~\cite{ref-evapo.sst}. 
This section is not necessary to read next Sec.\ref{sec-gsl}. 
Readers interested in the generalized second law may skip over this section.

\subsection{From SST to NE model}
\label{sec-evapo.model}

The NE model is obtained from the SST model by setting the system spherically symmetric and assigning Eq.\eqref{eq-model.eos} to the inner black body as its equations of states. 
Then the inner black body is regarded as a black hole, and the net energy flow from black hole to heat bath via radiation fields causes the black hole evaporation.

Before proceeding to the NE model, let us review here about existing works. 
In the framework of ordinary equilibrium thermodynamics, equilibrium states of black hole in a heat bath have already been investigated. 
It has already been revealed that an equilibrium state of the total system composed of a black hole and a heat bath is unstable for a sufficiently small black hole and stable for a sufficiently large black hole~\cite{ref-trans.1,ref-trans.2}. 
If the instability occurs for a small black hole and the system starts to evolve towards the other stable state, there are two possibilities of its evolution: 
The first is that, due to the statistical (and/or quantum) fluctuation, the temperature of heat bath exceeds that of black hole and a net energy flow into black hole arises. 
Then the black hole swallows a part of heat bath and settles down to a stable equilibrium state of a larger black hole in heat bath. 
The second possible evolution is that, due to the statistical (and/or quantum) fluctuation, the temperature of heat bath becomes lower than that of black hole and a net energy flow from black hole arises. 
Then the black hole evaporates and settles down to some other stable state. 
However we do not know the detail of end state of the second possibility, since the final fate of black hole evaporation is an unresolved issue at present.

When one distinguishes the phase of the equilibrium system by a criterion whether a black hole can exist stably in an equilibrium with a heat bath or not, the phase transition of the system occurs in varying the black hole radius. 
This phenomenon is known as {\it the black hole phase transition}~\cite{ref-trans.1,ref-trans.2}. 
So far cosmological constant is not considered. 
But the black hole phase transition has also been found for asymptotically anti-de~Sitter black holes, which is known as {\it Hawking-Page phase transition}~\cite{ref-trans.ads}. 
However, we do not consider cosmological constant throughout this article.

The black hole phase transition is one of interesting issues in black hole thermodynamics. 
However this section concentrates on a black hole evaporation in a heat bath after an instability of equilibrium occurs. 
We investigate a detail of black hole evaporation in the framework of NE model and try to extract some insight into the final fate of black hole evaporation. 
Readers interested also in the black hole phase transition may see~\cite{ref-evapo.sst} in which its equilibrium and nonequilibrium versions are also discussed.

\subsection{Energy transport in the NE model}
\label{sec-evapo.trans}

We discuss energetics of NE model. 
Total energy of the whole system is
\eqb
 E_{tot} \defeq E_g + E_h + E_{rad} \, ,
\eqe
where $E_g$ is the energy of black hole given in Eq.\eqref{eq-model.eos}, $E_h$ is the energy of heat bath defined by ordinary thermodynamics, and $E_{rad}$ is the steady state energy of radiation fields given in Eq.\eqref{eq-sst.min.energy},
\eqb
 E_{rad} = 4 \sigmap \left( G_g \, T_g^4 + G_h \, T_h^4 \right) \, ,
\eqe
where
\eqb
 G_g \defeq \int_{V_{rad}} dx^3\, \gb{x} \quad , \quad
 G_h \defeq \int_{V_{rad}} dx^3\, \gh{x} \, ,
\eqe
where $\gb{x}$ is the solid angle (divided by $4 \pi$) covered by directions of particles which are emitted by black hole and come to a point $\vec{x}$ (see figs.\ref{fig-1} and~\ref{fig-2}), and $\gh{x}$ is defined similarly by particles emitted by heat bath. 
By definition $\gb{x} + \gh{x} \equiv 1$ holds, and consequently $V_{rad} \defeq G_g + G_h$ gives the volume of hollow region. 
Furthermore, since the black hole is concentric with the hollow region, we obtain $\gb{x} = \left(\, 1 - \cos\theta \,\right)/2$ and $\gh{x} = \left(\, 1 + \cos\theta \,\right)/2$, where $\theta$ is the zenith angle which covers the black hole at a point of radial distance $r$ (see right panel in Fig.\ref{fig-1}). 
Then $G_g$ and $G_h$ are expressed as
\seqb
\label{eq-evapo.trans.G}
\eqab
 G_g &=& \frac{2\,\pi}{3}
       \left[\, R_h^3 - R_g^3 - \left(\, R_h^2 - R_g^2 \, \right)^{3/2} \, \right] \\
 G_h &=& \frac{2\,\pi}{3}
       \left[\, R_h^3 - R_g^3 + \left(\, R_h^2 - R_g^2 \, \right)^{3/2} \, \right] \, ,
\eqae
\seqe
where $R_h$ is the outermost radius of hollow region (see right panel in Fig.\ref{fig-1}).

To understand the energy flow in NE model, we divide the whole system into two sub-systems X and Y as follows: 
Sub-system X is composed of the black hole and the ``out-going'' radiation fields emitted by black hole, and sub-system Y is composed of the heat bath and the ``in-going'' radiation fields emitted by heat bath (see left panel in Fig.\ref{fig-1}). 
The sub-system X is a combined system of components of NE model which share the temperature $T_g$, and Y is that which share the temperature $T_h$. 
Then the total energy is expressed as
\eqb
 E_{tot} = E_X + E_Y \, ,
\label{eq-evapo.trans.Etot}
\eqe
where $E_X$ and $E_Y$ are respectively the energies of sub-systems X and Y, 
\seqb
\label{eq-evapo.trans.Exy}
\eqab
  E_X = E_g + E_{rad}^{(g)} \quad &,& \quad E_{rad}^{(g)} \defeq 4 \sigmap \, G_g \, T_g^4 \\
  E_Y = E_h + E_{rad}^{(h)} \quad &,& \quad E_{rad}^{(h)} \defeq 4 \sigmap \, G_h \, T_h^4 \, ,
\eqae
\seqe
where $E_{rad}^{(g)}$ and $E_{rad}^{(h)}$ are respectively the energies of out-going and in-going radiation fields. 
It is easily found that $E_X$ has no $T_h$-dependence, while $E_Y$ has $T_g$- and $T_h$-dependence. 
The energy flow in NE model can be understood as an energy transport between sub-systems X and Y. 
Because this energy transport is carried by the out-going and in-going radiation fields, the Stefan-Boltzmann law works well to give an explicit expression of energy transport,
\seqb
\label{eq-evapo.trans.transport.1}
\eqab
 \frac{dE_X}{dt} &=& - \sigmap \left( T_g^4 - T_h^4 \right) A_g
\label{eq-evapo.trans.transport.1.a} \\
 \frac{dE_Y}{dt} &=& \sigmap \left( T_g^4 - T_h^4 \right) A_g \, ,
\label{eq-evapo.trans.transport.1.b}
\eqae
\seqe
where $A_g = 4 \pi R_g^2$ is the surface area of black hole, and $t$ is a time coordinate which corresponds to a proper time of a rest observer at asymptotically flat region if we can extend the NE model to a full general relativistic model. 
Because some particles emitted by heat bath are not absorbed by the black hole but return to the heat bath (see right panel in Fig.\ref{fig-1}), the effective surface area through which Y exchanges energy with X is equal to the surface area of black hole. 
Therefore $A_g$ appears in Eq.\eqref{eq-evapo.trans.transport.1.b}. 
Furthermore Eqs.\eqref{eq-evapo.trans.transport.1} are formulated to be consistent with the isolated setting of the NE model, $E_{tot} \equiv constant$.

It is useful to rewrite the energy transport \eqref{eq-evapo.trans.transport.1} to a more convenient form for later discussions. 
By Eqs.\eqref{eq-model.capacity}, \eqref{eq-evapo.trans.G} and \eqref{eq-evapo.trans.Exy}, we find Eq.\eqref{eq-evapo.trans.transport.1} becomes
\eqb
 C_X\,\frac{dT_g}{dt} = - J \quad , \quad
 C_X\,C_Y^{(h)}\,\frac{dT_h}{dt} = \left(\,C_X + C_Y^{(g)} \,\right)\,J \, ,
\label{eq-evapo.trans.transport.2}
\eqe
where
\eqb
 J \defeq \sigmap \left(\, T_g^4 - T_h^4 \,\right) A_g \, ,
\label{eq-evapo.trans.J}
\eqe
and
\seqb
\label{eq-evapo.trans.capacity}
\eqab
 C_X &\defeq&
  \dfrac{dE_X}{dT_g} = C_g + C_{rad}^{(g)} \\
 C_{rad}^{(g)} &\defeq&
    \dfrac{dE_{rad}^{(g)}}{dT_g}
  = 16\,\sigmap\,G_g\,T_g^3
  + \dfrac{\sigmap}{2\,\pi}\left(R_g - \sqrt{R_h^2 - R_g^2} \right)T_g \\
 C_Y^{(g)} &\defeq&
    \dfrac{\partial E_Y}{\partial T_g}
  = \dfrac{\sigmap}{2\,\pi}\left(R_g + \sqrt{R_h^2 - R_g^2} \right) \dfrac{T_h^4}{T_g^3} \\
 C_Y^{(h)} &\defeq&
    \dfrac{\partial E_Y}{\partial T_h}
  = C_h + 16\,\sigmap\,G_h\,T_h^3 \\
 C_h &\defeq& \dfrac{dE_h}{dT_h} \, ( > 0) \, ,
\eqae
\seqe
where $C_g = - 2 \pi R_g^2$ is given in Eq.\eqref{eq-model.capacity} and it is assumed for simplicity that $R_h \equiv constant$ and $E_h$ depends on $T_h$ but not on $T_g$. 
$C_h$ is the heat capacity of heat bath, and we assume $C_h \equiv constant > 0$ for simplicity. 
$C_Y^{(g)}$ is the heat capacity of sub-system Y under the change of $T_g$ with fixing $T_h$, and $C_Y^{(h)}$ is that under the change of $T_h$ with fixing $T_g$. 
$C_{rad}^{(g)}$ is the heat capacity of out-going radiation fields, and $C_X$ is the heat capacity of sub-system X. 
In analyzing the nonlinear differential equations \eqref{eq-evapo.trans.transport.2}, behaviors of various heat capacities \eqref{eq-evapo.trans.capacity} are used. 
Some useful properties of these heat capacities are explained in next Subsec.\ref{sec-evapo.capa}.

From the above, we find that an inequality $C_X + C_Y^{(g)} < 0$ has to hold in order to guarantee the validity of NE model. 
To understand this requirement, consider the case $T_g > T_h$ for the first. 
Due to the temperature difference, energy flows from black hole to heat bath via radiation fields, $dE_g < 0$ and $dE_h > 0$. 
Then $dT_g > 0$ and $dT_h > 0$ hold due to $C_g < 0$ and $C_h > 0$. 
Recall that the whole system is isolated, $E_{tot} \equiv constant$, which means $( C_X + C_Y^{(g)} ) \,dT_g + C_Y^{(h)}\,dT_h = 0$. 
Therefore, because of $C_Y^{(h)} > 0$ by definition, it is concluded that the inequality $C_X + C_Y^{(g)} < 0$ must hold. 
And an inequality $C_X < 0$ follows immediately due to $C_Y^{(g)} > 0$ by definition. 
The similar discussion holds for the case $T_g < T_h$, and gives the same inequality. 
Hence the following inequality must hold in the framework of NE model,
\eqb
 C_X + C_Y^{(g)} < 0 \quad ( \, \Rightarrow \,C_X < 0 \, )  \, .
\label{eq-evapo.trans.validity.1}
\eqe
This inequality is the condition which guarantees the validity of NE model. 
A more detailed property of the combined heat capacity $C_X + C_Y^{(g)}$ is explained in next Subsec.\ref{subsubsec-CXCYg}, which shows that inequality \eqref{eq-evapo.trans.validity.1} can hold for a sufficiently small $T_h$. 
Therefore we assume $T_h$ is small enough so that the inequality \eqref{eq-evapo.trans.validity.1} holds.

Concerning the validity of NE model, what the quasi-equilibrium assumption implies is important. 
This assumption requires the time evolution is not so fast. 
Therefore, when a black hole evaporates, the shrinkage speed of black hole surface is less than unity,
\eqb
 v := \left| \frac{dR_g}{dt} \right| < 1 \, .
\label{eq-evapo.trans.validity.2}
\eqe
This inequality is also the condition which guarantees the validity of NE model. 
Hence, in the framework of NE model, our analysis should be restricted within the situations satisfying conditions \eqref{eq-evapo.trans.validity.1} and \eqref{eq-evapo.trans.validity.2}.

It is helpful for later discussions to consider what a violation of validity conditions \eqref{eq-evapo.trans.validity.1} and \eqref{eq-evapo.trans.validity.2} denotes. 
Firstly consider if condition \eqref{eq-evapo.trans.validity.1} is not satisfied. 
Then the system, especially the radiation fields, can never be described with steady state thermodynamics. 
The radiation fields are neither equilibrium nor steady (stationary nonequilibrium). 
This means that the radiation fields should be in a highly nonequilibrium dynamical state. 
Furthermore the quasi-equilibrium assumption is violated, because it is this assumption that lead us to utilize the steady state thermodynamics. 
Therefore highly nonequilibrium radiation fields make a black hole dynamical, and the black hole can not be treated by equilibrium solutions of Einstein equation. 
Next consider if condition \eqref{eq-evapo.trans.validity.2} is not satisfied. 
Then the black hole evolves so fast that the quasi-equilibrium assumption is violated. 
The black hole can not be described by equilibrium solutions of Einstein equation, but described by some unknown dynamical solution. 
Therefore, because the source of radiation fields becomes dynamical, radiation fields evolve into a highly nonequilibrium dynamical state and the steady state thermodynamics is not applicable. 
Hence, when one of the conditions \eqref{eq-evapo.trans.validity.1} or \eqref{eq-evapo.trans.validity.2} is violated, the system evolves into a highly nonequilibrium dynamical state which can not be treated in the framework of NE model.

\subsection{Properties of some heat capacities}
\label{sec-evapo.capa}

This subsection summarizes properties of various heat capacities \eqref{eq-evapo.trans.capacity} which we will use in remaining subsections.

\subsubsection{$C_X(R_g)$ as a function of $R_g$}
\label{subsubsec-CX.Rg}

Here we show a behavior of heat capacity $C_X$ of sub-system X as a function of black hole radius $R_g$. 
By Eqs.\eqref{eq-model.eos} and \eqref{eq-evapo.trans.G}, $C_X(R_g)$ is rewritten into the following form,
\eqb
 C_X(x) = - 2 \pi R_h^2\, x^2 + \frac{\sigmap}{8 \pi^2}\, f(x) \, ,
\label{eq-CX.CX}
\eqe
where $x \defeq R_g/R_h$ and
\eqb
 f(x) \defeq
    \frac{4}{3}\,\frac{1}{x^3}\, \left[\, 1 - x^3 - \left( 1 - x^2 \right)^{3/2} \,\right]
  + \frac{1}{x}\,\left(\, x - \sqrt{1 - x^2} \,\right) \, .
\label{eq-CX.f}
\eqe
By definition, $0 < R_g < R_h$, i.e., $0 < x < 1$, and we find
\eqb
 \begin{cases}
  C_X \, \to \, \infty                           & \text{as $x \,\to\,0$} \\
  C_X = - 2 \pi R_h^2 + \dfrac{\sigmap}{8 \pi^2} & \text{at $x = 1$}
 \end{cases} \, .
\label{eq-CX.limit}
\eqe
By Eq.\eqref{eq-intro.N} and Eq.\eqref{eq-model.Rg} which denotes $R_h > 1$, we find $C_X(x=1) < 0$ holds for the NE model.

\begin{figure}[t]
 \centerline{\includegraphics[height=35mm]{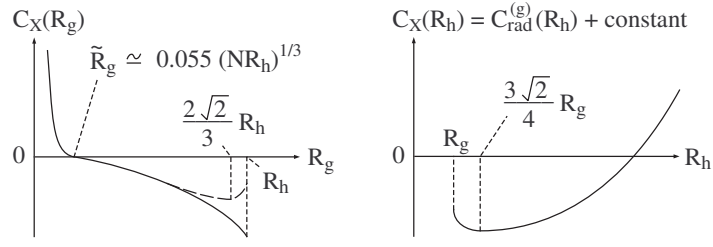}}
\caption{Left panel is for $C_X(R_g)$ as a function of $R_g$, where $2 \sqrt{2}/3 \simeq 0.943$. Right is for $C_X(R_h)$ as a function of $R_h$, where $3 \sqrt{2}/4 \simeq 1.06$. $C_X(R_g = R_h) < 0$ holds due to $\sigmap \simeq 10$ and $1 \lesssim R_g \le R_h$.}
\label{fig-3}
\end{figure}

The differential of $C_X(x)$ is
\eqb
 \frac{d C_X(x)}{dx} = - 4 \pi R_h^2\, x + \frac{\sigmap}{8 \pi^2}\, \frac{d f(x)}{dx} \, ,
\eqe
where
\eqb
 \dfrac{d f(x)}{dx} = \dfrac{4 - 3\, x^2 - 4\,\sqrt{1-x^2}}{x^4 \,\sqrt{1-x^2}}
 \quad \Rightarrow \quad
 \begin{cases}
  \dfrac{d f(x)}{dx} < 0 & \text{for $x < \dfrac{2 \sqrt{2}}{3}$} \\
  \dfrac{d f(x)}{dx} > 0 & \text{for $x > \dfrac{2 \sqrt{2}}{3}$}
 \end{cases} \, ,
\label{eq-CX.df}
\eqe
where $2\sqrt{2}/3 \simeq 0.943$. 
A schematic graph of $C_X(R_g)$ is shown in Fig.\ref{fig-3} (left panel), where $\tilde{R}_g$ is the solution of $C_X = 0$, 
\eqb
 \tilde{R}_g \defeq R_h\, \tilde{x} \quad , \quad
 \tilde{x} \defeq \left\{ x\, |\, C_X(x) = 0 \right\} \, .
\eqe
Because of $C_X(x=1) < 0$, equation $C_X = 0$ has only one solution.

Finally, we estimate the value of $\tilde{x}$. 
Since $x < 1$ by definition, we apply the Taylor expansion, $(1-x^2)^{\alpha} = 1 - \alpha\, x^2 + O(x^4)$, to Eq.\eqref{eq-CX.CX} and obtain
\eqb
 C_X(x) = - 2 \pi R_h^2 \,
           \left[\, x^2 - \varepsilon \, \frac{1}{x}\,
                          \left( 1 - \frac{x}{3} + O(x^2) \right)  \,\right] \, ,
\eqe
where $\varepsilon = \sigmap/16 \pi^3 R_h^2 = N/1920 \pi R_h^2$. 
By Eq.\eqref{eq-intro.N} and Eq.\eqref{eq-model.Rg} which denotes $R_h > 1$, we find $\varepsilon < 1$. 
Then $C_X(x)$ becomes
\eqb
 C_X(x) = - 2 \pi R_h^2 \,\frac{1}{x}\,
          \left(\, x^3 - \varepsilon + \varepsilon\, O(x) \,\right) \, .
\eqe
Hence, by taking leading terms of $x$ and $\varepsilon$, we find an approximate expression for $\tilde{x}$ as
\eqb
 \tilde{x} \simeq \varepsilon^{1/3} \, .
\eqe
This gives an approximate value of $\tilde{R}_g$,
\eqb
 \tilde{R}_g \simeq \left( \frac{N R_h}{1920 \pi} \right)^{1/3}
             \simeq 0.055 \times \left( N R_h \right)^{1/3} \, .
\label{eq-CX.tRg}
\eqe

\subsubsection{$C_X(R_h)$ as a function of $R_h$}
\label{subsubsec-CX.Rh}

Here we show a behavior of heat capacity $C_X$ of sub-system X as a function of outermost radius $R_h$ of the hollow region. 
We make use of the calculations done in previous subsection. 
According to Eq.\eqref{eq-CX.CX}, $C_X(R_h)$ is expressed as
\eqb
 C_X(R_g) = - 2 \pi R_g^2 + C_{rad}^{(g)}(R_h) \quad , \quad
 C_{rad}^{(g)}(R_h) = \frac{\sigmap}{8 \pi^2} \, f(x) \, ,
\eqe
where $x \defeq R_g/R_h$, $0 < x < 1$ by definition, and $f(x)$ is given in Eq.\eqref{eq-CX.f}. 
This denotes $C_X(R_h)$ behaves as $C_{rad}^{(g)}(R_h) + constant$. 
The differential becomes
\eqb
 \frac{d C_X(R_h)}{dR_h} = - \frac{\sigmap}{8 \pi} \, x^2 \, \frac{d f(x)}{dx} \, .
\eqe
Using Eq.\eqref{eq-CX.df}, we find
\eqb
 \begin{cases}
  \dfrac{dC_X}{dR_h} \to - \infty               & \text{as $R_h \to R_g + 0$ ($x \to 1+0$)} \\[3mm]
  \dfrac{dC_X}{dR_h} \to \dfrac{\sigmap}{8 \pi^2} & \text{as $R_h \to \infty$ ($x \to 0$)}
 \end{cases} \, .
\eqe
Hence referring to limit values \eqref{eq-CX.limit}, a schematic graph of $C_X(R_h)$ is obtained as shown in Fig.\ref{fig-3} (right panel). 
$C_X(R_h)$ is monotone increasing for $R_h > (3 \sqrt{2}/4) R_g \simeq 1.06 R_g$.

\subsubsection{Proof of the inequality $C_g/C_X > 1$}
\label{subsubsec-Cg/CX}

Here we prove inequality $C_g/C_X > 1$ under the condition $C_X < 0$ (see condition \eqref{eq-evapo.trans.validity.1}). 
We make use of the calculations done in Subsec.\ref{subsubsec-CX.Rg}. 
By Eq.\eqref{eq-CX.CX} and definition of $C_X$ given in Eq.\eqref{eq-evapo.trans.capacity}, $C_{rad}^{(g)}$ is expressed as $C_{rad}^{(g)}(x) = (\sigmap/8 \pi^2) \, f(x)$, where $x \defeq R_g/R_h$, $0 < x < 1$ by definition, and $f(x)$ is given in Eq.\eqref{eq-CX.f}. 
Then Eq.\eqref{eq-CX.df} indicates $C_{rad}^{(g)}(x) \ge C_{rad}^{(g)}(2\sqrt{2}/3)$, and Eq.\eqref{eq-CX.f} gives $f(2\sqrt{2}/3) \simeq 0.845 > 0$. 
Therefore we find $C_{rad}^{(g)} > 0$ for $0 < x <1$, which is consistent with a naive expectation that an ordinary matter like radiation fields has a positive heat capacity.

On the other hand a required condition $C_X \,(= C_g + C_{rad}^{(g)}) < 0$ indicates $0 < C_{rad}^{(g)} < \left| C_g \right|$. 
This gives $\left| C_X \right| = \left| C_g \right| - C_{rad}^{(g)} < \left| C_g \right|$. 
Hence we find the inequality, $C_g/C_X > 1$.

\subsubsection{$C_X + C_Y^{(g)}$ as a function of $R_g$}
\label{subsubsec-CXCYg}

\begin{figure}[t]
 \begin{center}
  \includegraphics[height=45mm]{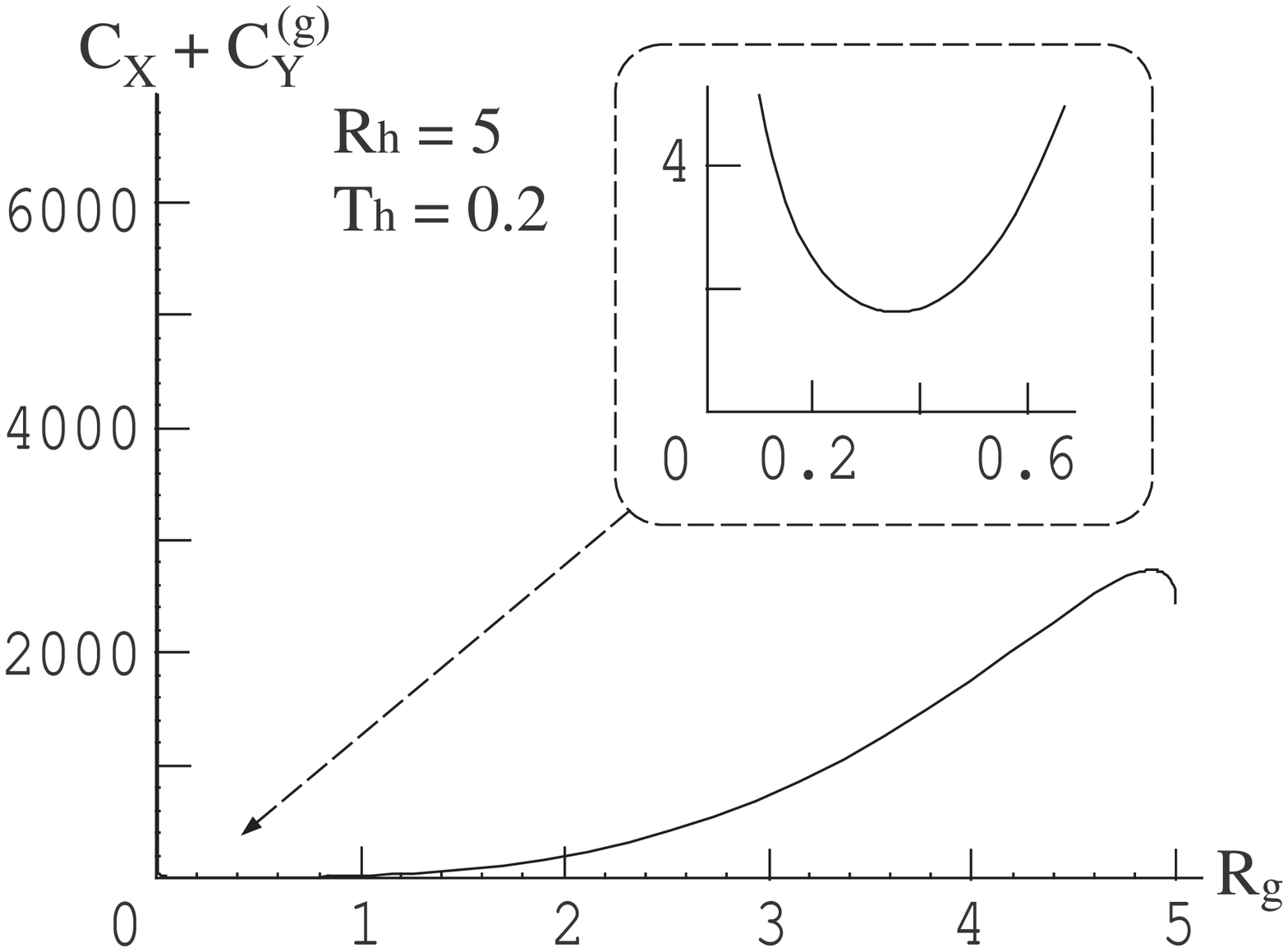} \qquad
  \includegraphics[height=45mm]{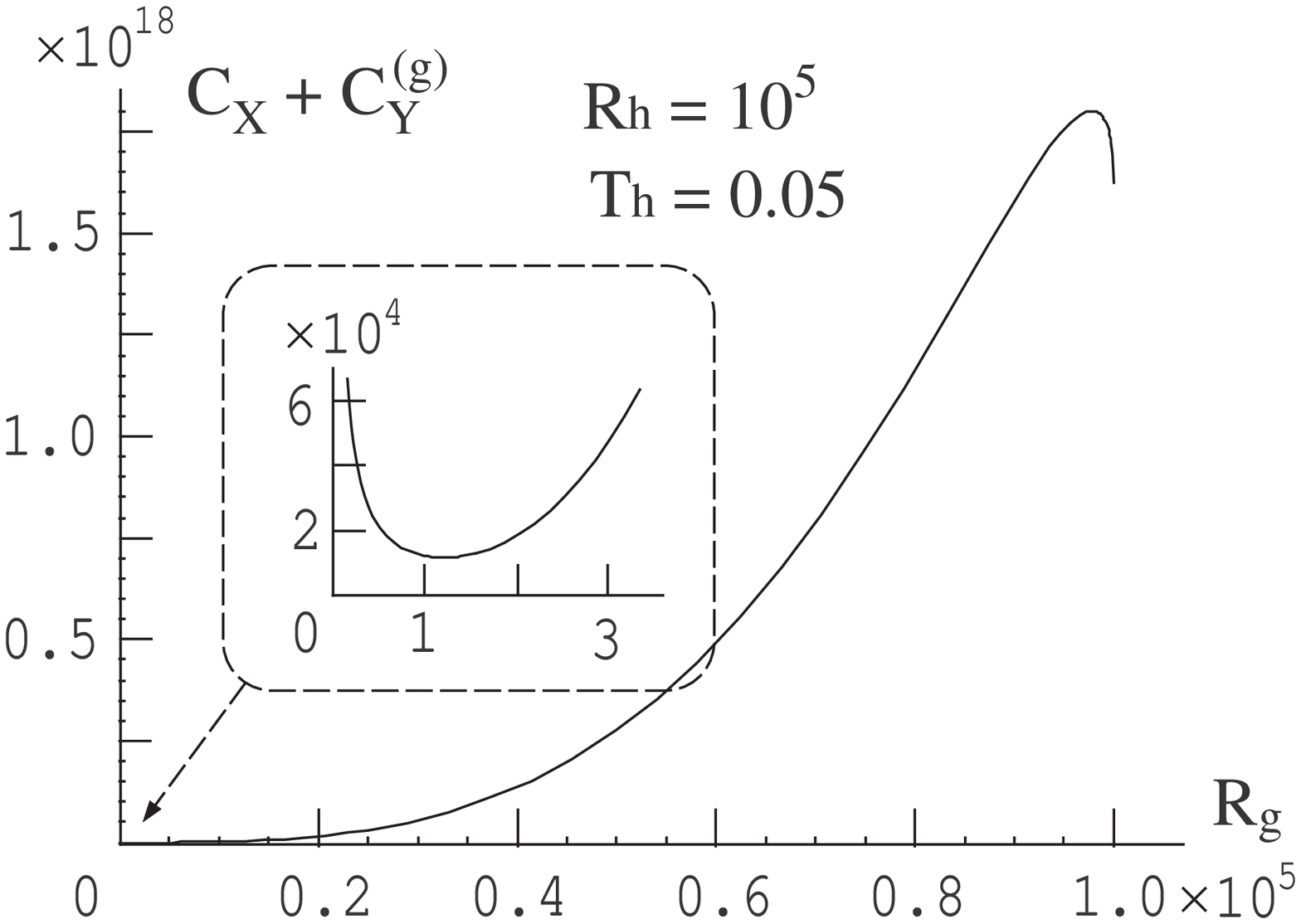} \\
  \includegraphics[height=45mm]{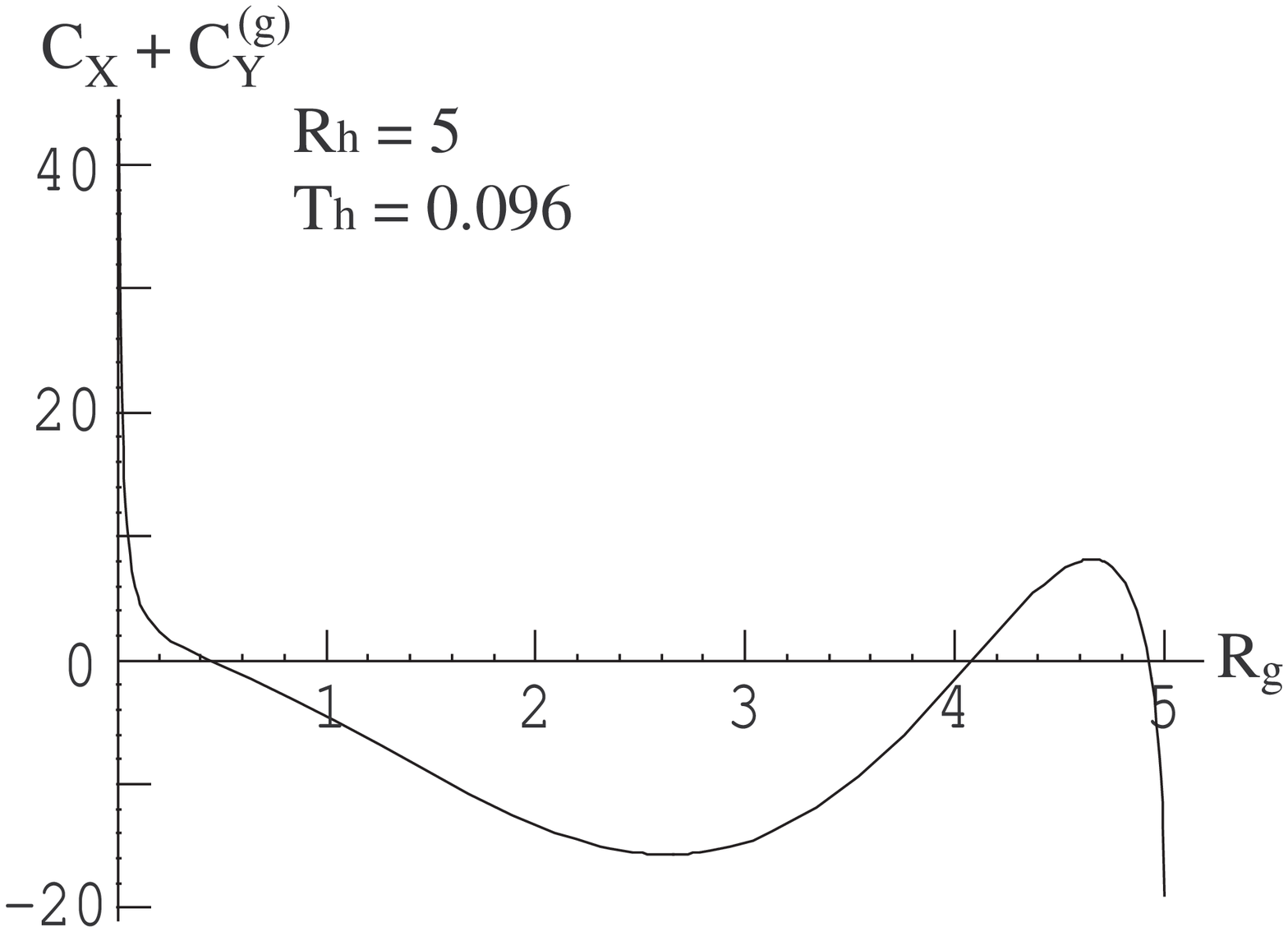} \qquad
  \includegraphics[height=45mm]{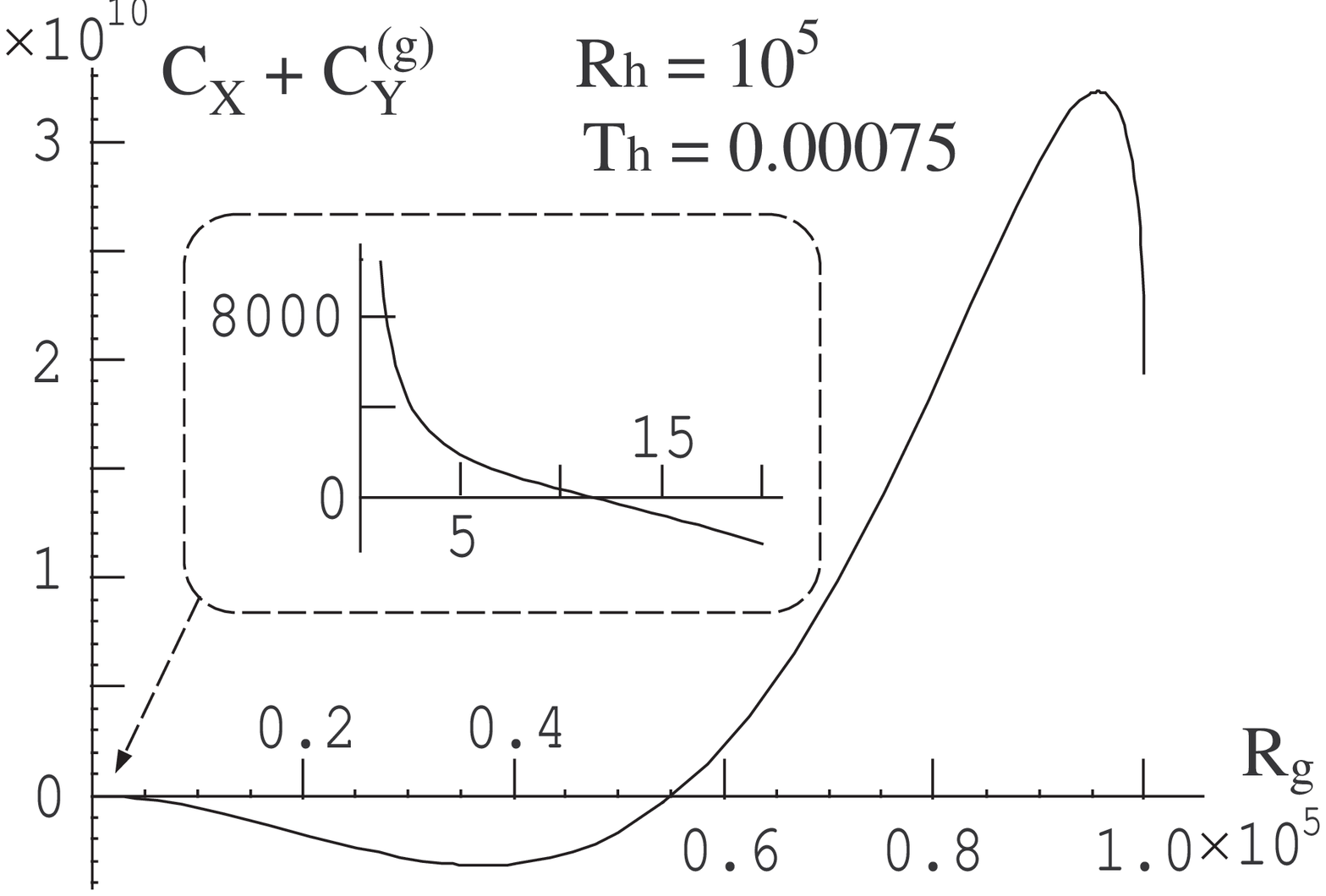} \\
  \includegraphics[height=45mm]{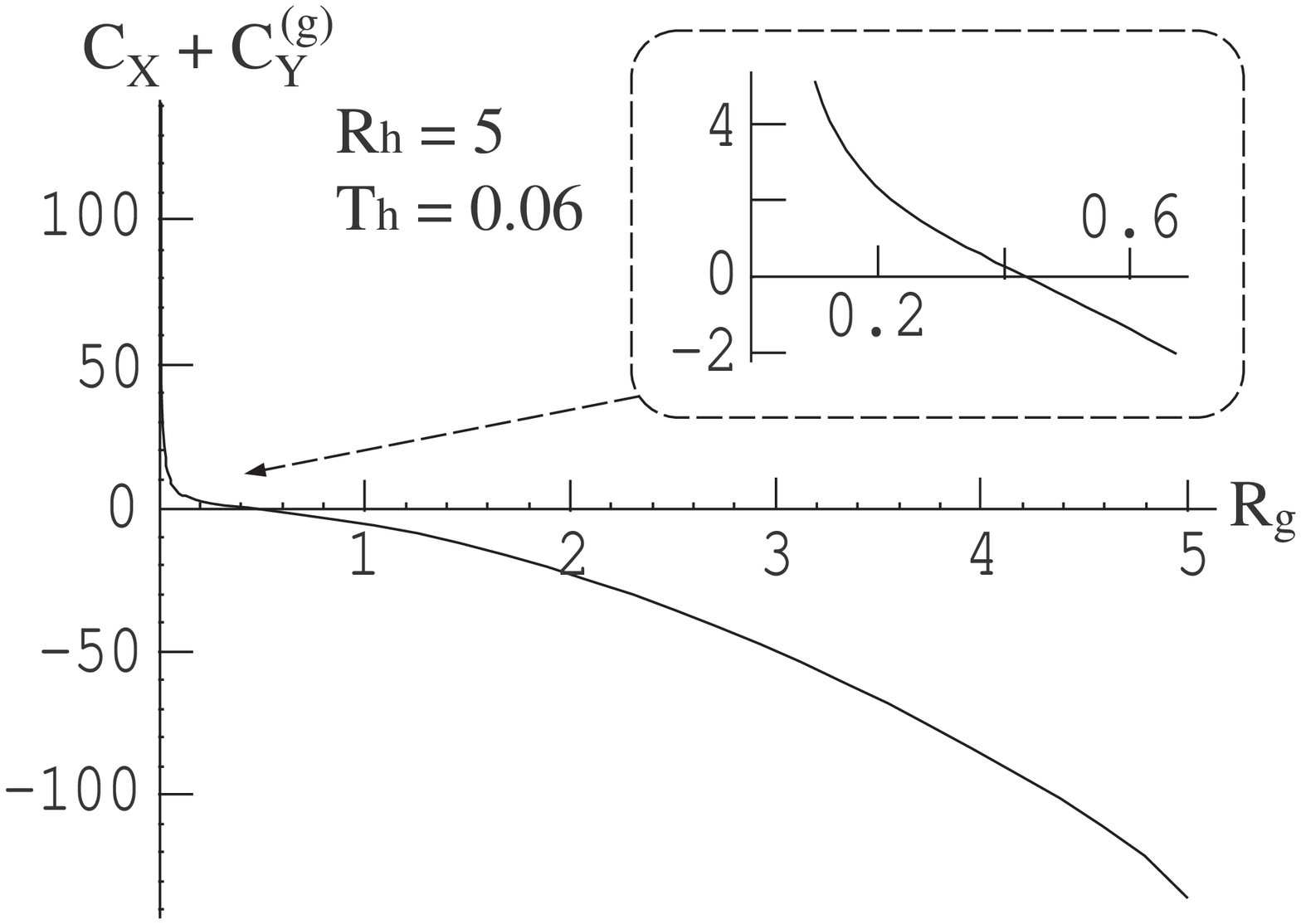} \qquad
  \includegraphics[height=45mm]{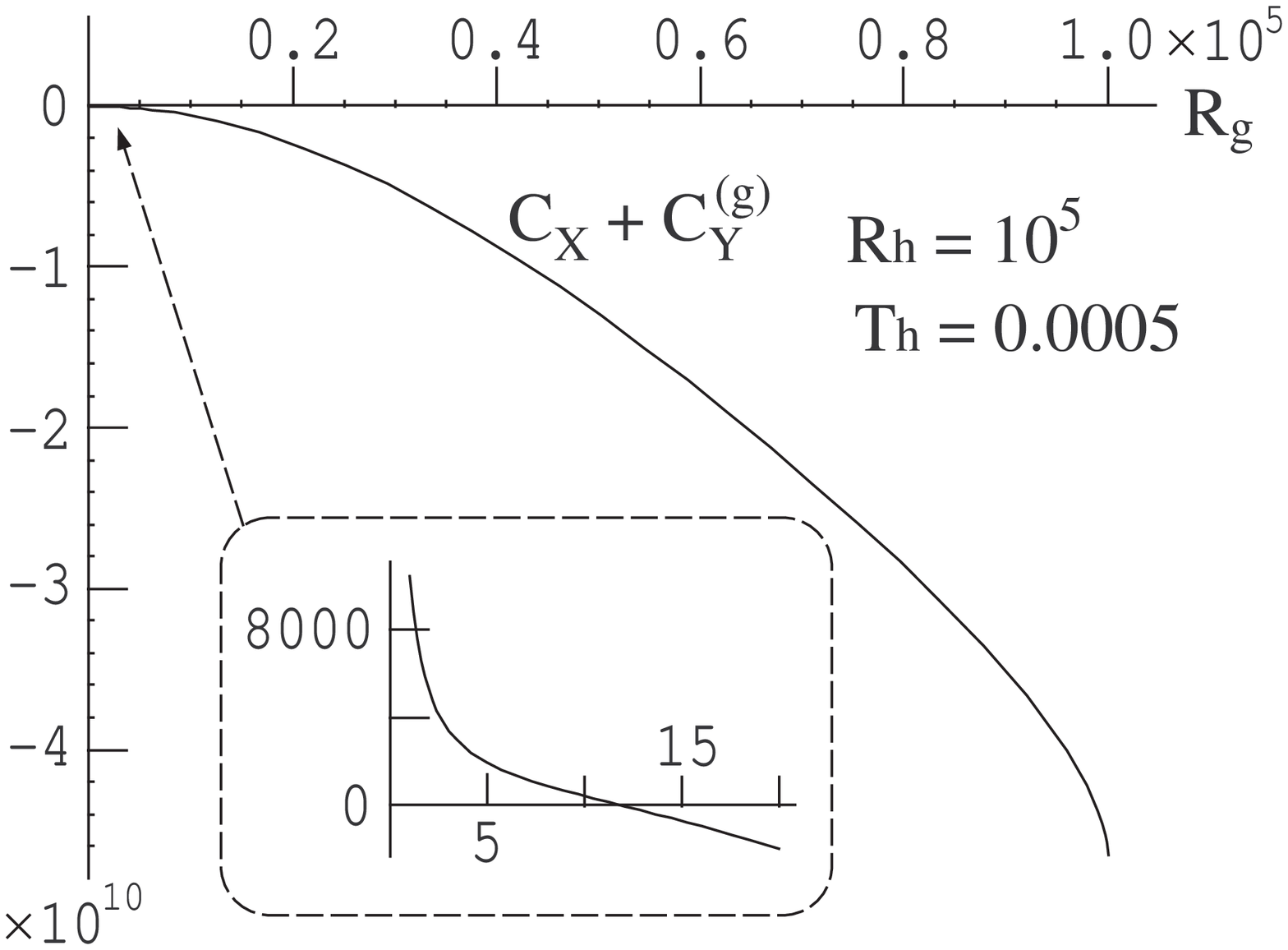}
 \end{center}
\caption{Numerical plots of $C_X + C_Y^{(g)}$ as a function of $R_g$ with $N = 100$. Left three panels are for $R_h = 5$ and $T_h = 0.2$, $0.096$ and $0.06$ downwards. Right three panels are for $R_h = 10^5$ and $T_h = 0.05$, $0.00075$ and $0.0005$ downwards. As far as the author checked, the same behavior is observed for every value of $R_h > 1$. $C_X + C_Y^{(g)} < 0$ holds for a sufficiently small $T_h$.}
\label{fig-4}
\end{figure}

Here we show a behavior of combined heat capacity $C_X + C_Y^{(g)}$ as a function of black hole radius $R_g$. 
We make use of the calculations done in Subsec.\ref{subsubsec-CX.Rg}. 
By Eq.\eqref{eq-CX.CX} and definitions of $C_X$ and $C_Y^{(g)}$ given in Eq.\eqref{eq-evapo.trans.capacity}, $C_X + C_Y^{(g)}$ is expressed as
\eqb
 C_X(R_g) + C_Y^{(g)}(R_g) =
 - 2 \pi R_h^2\, x^2 + \frac{\sigmap}{8 \pi^2}\,
 \left[\, f(x) + q\, x^3 \,\left(\, x + \sqrt{1-x^2} \,\right) \,\right] \, ,
\eqe
where $q \defeq \left( 4 \pi R_h \, T_h \right)^4$, $x \defeq R_g/R_h$, $0 < x < 1$ by definition, and $f(x)$ is given in Eq.\eqref{eq-CX.f}. 
Then, using limit values \eqref{eq-CX.limit}, we find
\eqb
 \begin{cases}
  C_X + C_Y^{(g)} \, \to \, \infty & \text{as $x \,\to\,0$} \\
  C_X + C_Y^{(g)} =
    - 2 \pi R_h^2 + \dfrac{\sigmap}{8 \pi^2} + \dfrac{\sigmap}{8 \pi^2} \, q & \text{at $x = 1$}
 \end{cases} \, .
\eqe
The first two terms in $C_X + C_Y^{(g)}$ at $x=1$ is negative in the framework of NE model as discussed in Eq.\eqref{eq-CX.limit}. 
Therefore, if $T_h$ is sufficiently small, then $q$ can be small enough so that $C_X + C_Y^{(g)}$ at $x=1$ becomes negative.

The differential of $C_X + C_Y^{(g)}$ is complicated and not suitable for analytical utility. 
Instead of analytical discussion, we show some numerical examples of $C_X + C_Y^{(g)}$ in Fig.\ref{fig-4}. 
It is recognized with this figure that, for a sufficiently small $T_h$, the condition \eqref{eq-evapo.trans.validity.1} is satisfied for a certain range of $R_g$. 
Although Fig.\ref{fig-4} shows only some examples, the same behavior is observed for every value of $R_h > 1$ as far as the author checked. 
Therefore the validity condition \eqref{eq-evapo.trans.validity.1}, $C_X + C_Y^{(g)} < 0$, holds for a sufficiently small $T_h$. 
During a semi-classical and quasi-equilibrium stage of black hole evaporation ($R_g(t) > 1$), it is reasonable to require $T_g < 1$ due to Eq.\eqref{eq-model.eos}, and also $T_h < T_g < 1$ for black hole evaporation. 
We assume throughout this article that $T_h$ is small enough so that the equation $C_X(R_g) + C_Y^{(g)}(R_g) = 0$ has one, two or three solutions of $R_g$.

Finally we discuss what happens along a black hole evaporation for the case that the equation $C_X(R_g) + C_Y^{(g)}(R_g) = 0$ has two or three solutions. 
For the first consider the case of three solutions, and denotes these solutions by $R_{g0}^{(s)}$, $R_{g0}^{(m)}$ and $R_{g0}^{(l)}$ in increasing order, $R_{g0}^{(s)} < R_{g0}^{(m)} < R_{g0}^{(l)}$. 
If the evaporation starts with initial black hole radius $R_g(0)$ in the range $R_{g0}^{(l)} < R_g(0) < R_h$ (see for example left-center panel in Fig.\ref{fig-4}), then the evaporation process is treated in the framework of NE model until $R_g$ decreases to $R_{g0}^{(l)}$. 
Then, when $R_g$ reaches $R_{g0}^{(l)}$, the NE model becomes inapplicable to the evaporation process because the validity condition \eqref{eq-evapo.trans.validity.1} is violated in the range $R_{g}^{(m)} < R_g < R_{g0}^{(l)}$. 
However we can expect $R_g$ decreases to $R_{g0}^{(m)}$ even if NE model is not applicable. 
Then the NE model becomes applicable again after $R_g$ decreases less than $R_{g0}^{(m)}$. 
The NE model is applied to the evaporation process in the range $R_{g0}^{(s)} < R_g < R_{g0}^{(m)}$.

Next consider the case that the equation $C_X(R_g) + C_Y^{(g)}(R_g) = 0$ has two solutions. 
When the black hole evaporation starts with initial radius larger than the larger solution of $C_X(R_g) + C_Y^{(g)}(R_g) = 0$ (see for example right-center panel in Fig.\ref{fig-4}), then the similar discussion given in previous paragraph holds. 
The NE model is not applicable to the evaporation process until $R_g$ becomes smaller than the larger solution of $C_X(R_g) + C_Y^{(g)}(R_g) = 0$. 
However after $R_g$ becomes less than the larger solution, the NE model becomes applicable until $R_g$ decreases to the smaller solution of $C_X(R_g) + C_Y^{(g)}(R_g) = 0$.

From the above we find that, for both of the cases that equation $C_X(R_g) + C_Y^{(g)}(R_g) =~0$ has two and three solutions, it is the lowest solution of that equation at which the evaporation process goes out of the framework of NE model and proceeds to a highly nonequilibrium dynamical stage (see last paragraph in Subsec.\ref{sec-evapo.trans}).

\subsection{Nonequilibrium effects of energy flow}
\label{sec-evapo.ne}

\subsubsection{General aspect of the NE model}
\label{subsubsec-general}

We discuss the black hole evaporation in the framework of NE model. 
To analyze energy transport equations \eqref{eq-evapo.trans.transport.2} from energetic viewpoint, we consider the energy emission rate $\jn$ by black hole,
\eqb
 \jn \defeq - \frac{dE_g}{dt}
     = \sigmap \, \frac{C_g}{C_X} \, \left(T_g^4 - T_h^4\right) \, A_g \, ,
\label{eq-evapo.ne.rate.ne}
\eqe
where Eqs.\eqref{eq-model.capacity} and \eqref{eq-evapo.trans.transport.2} are used. 
The stronger $\jn$, the more rapidly the mass energy of black hole $E_g$ decreases along its evaporation process. 
The stronger emission rate $\jn$ denotes the acceleration of black hole evaporation.

As mentioned in Eq.\eqref{eq-evapo.trans.capacity}, we assume $R_h \equiv constant$ for simplicity. 
$R_h$ is the parameter which controls the size of nonequilibrium region around black hole. 
To understand the nonequilibrium nature of black hole evaporation, it is useful to compare two situations which differ only by the value of $R_h$ with sharing the same values of the other parameters of NE model, $R_g$, $T_h$, $C_h$ and $N$. 
To do this comparison, we note the following three points; 
firstly $C_g < 0$ by definition, secondly $C_X < 0$ given in the condition \eqref{eq-evapo.trans.validity.1}, and finally that $\left| C_X \right|$ is monotone decreasing function of $R_h$ for $R_h \ge \left(3\sqrt{2}/4\right) R_g \simeq 1.06 R_g$ under the condition $C_X < 0$ (see Subsec.\ref{subsubsec-CX.Rh}). 
The first and second points denote $C_g/C_X > 0$, then the third point concludes that $C_g/C_X$ is monotone increasing function of $R_h$ for $R_h \ge \left(3\sqrt{2}/4\right) R_g$ under the condition $C_X < 0$. 
Hence it is recognized that, for the case $R_h > \left(3\sqrt{2}/4\right) R_g$, the larger we set the nonequilibrium region, the stronger the emission rate $\jn$ and the faster the black hole evaporation proceeds. 
Numerical examples are shown later in Subsec.\ref{subsubsec-ne.example}.

The above discussion is a comparison of NE model of a certain value of $R_h$ with that of a different value of $R_h$. 
In the following subsections, we compare the NE model with the other models of black hole evaporation, the equilibrium model used in~\cite{ref-trans.1} and the black hole evaporation in an empty space (a situation without heat bath originally considered by Hawking in~\cite{ref-hr}).

\subsubsection{Comparison with the equilibrium model used in~\cite{ref-trans.1}}
\label{subsubsec-ne.eq}

\begin{figure}[t]
 \begin{center}
  \includegraphics[height=25mm]{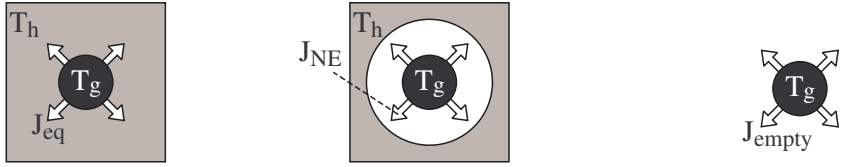}
 \end{center}
\caption{Energy emission rate by black hole $-dE_g/dt$. Left panel is $J_{eq}$ for equilibrium model extrapolated to a black hole evaporation, $T_h < T_g$. Center is $\jn$ for NE model. Right is $\je$ for a black hole evaporation in an empty space with ignoring grey body factor. White arrow in each panel expresses energy flow from black hole.}
\label{fig-5}
\end{figure}

In the original work~\cite{ref-trans.1} suggesting the black hole phase transition, only the equilibrium of the system which consists of a black hole and a heat bath is considered. 
This equilibrium model is obtained from the NE model by setting $R_h = R_g$ (no hollow region) and $T_h = T_g$ (equilibrium). 
Obviously the equilibrium model does not include nonequilibrium nature of black hole evaporation, since the radiation fields disappear. 
Exactly speaking, the evaporation process is not described by the ``equilibrium model''. 
However even in the framework of equilibrium model, we can find the black hole evaporation occurs for a sufficiently small black holes as mentioned in Subsec.\ref{sec-evapo.model}. 
By extrapolating the equilibrium model to the evaporation process, we may set $T_h < T_g$ with keeping the condition $R_g = R_h$ (see left panel in Fig.\ref{fig-5}). 
Then the energy emission rate $J_{eq}$ by black hole in equilibrium model is given by setting $R_h = R_g$ in $\jn$,
\eqb
 J_{eq} \defeq \left.\jn\right|_{R_h = R_g} = \sigmap \left(T_g^4 - T_h^4\right) A_g = J \, ,
\eqe
where $J$ is given in Eq.\eqref{eq-evapo.trans.J}. 
We find $\jn = \left( C_g/C_X \right) \, J_{eq}$. 
Here note that $C_g/C_X > 1$ is shown in Subsec.\ref{subsubsec-Cg/CX}. 
Therefore, when the values of $R_g$, $T_h$ and $N$ are the same for the NE and equilibrium models, then $\jn > J_{eq}$ holds. 
This implies that the black hole evaporation in NE model proceeds faster than that in equilibrium model. 
We can recognize that the nonequilibrium effect of energy exchange between black hole and heat bath accelerates the black hole evaporation.

\subsubsection{Comparison with the black hole evaporation in an empty space with ignoring grey body factor}
\label{subsubsec-ne.empty}

In the original work~\cite{ref-hr} of the Hawking radiation, Hawking considered mainly a simple situation as seen in right panel in fig.~\ref{fig-5}; 
a black hole in an empty space (a situation without heat bath) with ignoring curvature scattering of Hawking radiation. 
This describes a black hole evaporation in an empty space with ignoring the so-called {\it grey body factor}. 
There is no energy accretion onto black hole in this simple situation, and time evolution is given by the Stefan-Boltzmann law, $dE_g/dt = - \sigmap T_g^4 A_g$. 
Usually in most of the existing works on black hole physics, the time scale of black hole evaporation is estimated by assuming this simple situation.

It is interesting to compare the NE model with the black hole evaporation in an empty space with ignoring grey body factor. 
The energy emission rate $\je$ by black hole in an empty space with ignoring grey body factor is given by the Stefan-Boltzmann law as follows,
\eqb
 \je \defeq \sigmap \, T_g^4 \, A_g \, ,
\label{eq-evapo.ne.rate.empty}
\eqe
where it is assumed that matter fields of Hawking radiation is the non-self-interacting massless matter fields as for the NE model. 
Then we find
\eqb
 \jn = \frac{C_g}{C_X} \left( 1 - \frac{T_h^4}{T_g^4} \right) \je \, .
\label{eq-evapo.ne.jn.je}
\eqe
Recall that $T_g > T_h$ holds generally for a black hole evaporation, and $C_g/C_X > 1$ holds in the framework of NE model (see Subsec.\ref{subsubsec-Cg/CX}). 
Then the factor $\left(C_g/C_X\right) \left( 1 - T_h^4/T_g^4 \right)$ may be greater or less than unity. 
It is not definitely clear which of $\jn$ or $\je$ is larger than the other.

One may naively expect that the incoming energy flow from heat bath to black hole in NE model never enhance the energy emission rate by black hole, and that the relation $\jn > \je$ is impossible but $\jn < \je$ must hold necessarily. 
It is true if the black hole heat capacity $C_g$ is positive. 
However in the NE model, $C_g < 0$ as shown in Eq.\eqref{eq-model.capacity} and a naive sense based on ordinary systems of positive heat capacity is not always true. 
An inverse sense against the naive sense may be offered; the more amount of energy is extracted from black hole by heat bath, the more rapidly the black hole emits its mass energy. 
Furthermore the energy emitted by Hawking radiation is absorbed by the heat bath and affects the incoming energy flow from heat bath to black hole. 
The energetic interaction (energy exchange) between black hole and heat bath determines the energy emission rate $\jn$. 
When the negative heat capacity and energetic interaction are taken into account, a naively unexpected relation $\jn > \je$ is also possible as discussed in following paragraphs:

To analyze the energy emission rate by black hole $\jn$, it is useful to recall the decomposition of the whole system of NE model into sub-systems X and Y, as considered in Subsec.\ref{sec-evapo.trans}. 
On the other hand, the black hole evaporation in an empty space with ignoring grey body factor is regarded as the system obtained by removing the sub-system Y from the NE model. 
This means that, from energetic viewpoint, the black hole evaporation in an empty space with ignoring grey body factor can be thought of as a relaxation process of the ''isolated sub-system X'' keeping $E_X \equiv constant$. 
Therefore, for the black hole evaporation in an empty space with ignoring grey body factor, the energy emission rate by black hole $\je$ is the energy transport just inside the sub-system X (from black hole to out-going radiation fields), and no energy flows out of X. 
However the energy transport~\eqref{eq-evapo.trans.transport.1} in NE model is the energy exchange between sub-systems X and Y. 
This indicates that, in the NE model, the energy $E_X$ of X is extracted by Y due to the temperature difference $T_g > T_h$, and energy flows from X to Y. 
The energetic interaction (energy exchange) between X and Y makes the black hole evaporation in NE model quit different from the black hole evaporation in an empty space with ignoring grey body factor. 
This difference is recognized significantly by considering a limit of the energy transport \eqref{eq-evapo.trans.transport.1} as follows: 
One may expect that the energy emission by black hole in an empty space with ignoring grey body factor, $dE_g/dt = - \je$, should be obtained from Eq.\eqref{eq-evapo.trans.transport.1} by the limit operations, $T_h \to 0$, $E_h \to 0$ (remove the sub-system Y) and $R_h \to \infty$ (infinitely large volume of out-going radiation fields). 
However these operations transform Eq.\eqref{eq-evapo.trans.transport.1} into the set of equations, $dE_g/dt = - \je$ and $0 \equiv \je$. 
This gives an unphysical result $E_g \equiv constant\, (= \infty)$ which contradicts the ``evaporation'', $dE_g/dt < 0$. 
The black hole evaporation in an empty space with ignoring grey body factor can not be described as some limit situation of the NE model.

In addition to the naive expectation $\jn < \je$, the opposite relation $\jn > \je$ may be expected due to the negative heat capacity of black hole~\eqref{eq-model.capacity} and energetic interaction as follows: 
In the NE model, because the energy extraction ($dE_X < 0$) occurs along the black hole evaporation ($dT_g > 0$), the heat capacity of X is always negative $C_X = dE_X/dT_g = C_g + C_{rad}^{(g)} < 0$, where $C_{rad}^{(g)} > 0$ as indicated in Subsec.\ref{subsubsec-Cg/CX}. 
Furthermore the larger the volume of hollow region, the larger the heat capacity $C_{rad}^{(g)}$ and the smaller the absolute value $\left|C_X\right|$ because of $C_g < 0$. 
This implies that the more thick the hollow region, the more accelerated the increase of $T_g$ due to the relation $dT_g = \left|dE_X/C_X\right|$. 
Therefore, for sufficiently large $R_h$, the energy extraction from X by Y (the increase of $T_g$) can dominate over the in-coming energy flow onto black hole. 
This means the energy emission rate $\jn$ is enhanced by the energy extraction from X by Y, then $\jn > \je$ is implied.

The above discussion can also be supported by the following rough analysis: 
When a black hole evaporates in NE model, the temperature difference $\delta T \defeq T_g - T_h$ should grow infinitely, $\delta T \to \infty$, due to the negative heat capacity of black hole. 
Then, because of Eq.\eqref{eq-evapo.ne.jn.je} together with the facts $1 - \left(T_h/T_g\right)^4 \to 1$ (as $\delta T \to \infty$) and $C_g/C_X > 1$ (see Subsec.\ref{subsubsec-Cg/CX}), the larger the temperature difference $\delta T$, the larger the ratio $\jn/\je$. 
Hence for the black hole evaporation in NE model, it is expected that the relation $\jn~>~\je$ comes to be satisfied during the evaporation process even if the relation $\jn < \je$ holds at initial time. 
Furthermore, if the relation $\jn > \je$ holds for a sufficiently long time during the evaporation process, the evaporation time scale in NE model can be shorter than that in an empty space. 
Hence it is possible that the black hole evaporation in NE model proceeds faster than that in an empty space, where black holes of the same initial mass are considered in both cases. 
In next subsection, numerical examples support this discussion.

\subsubsection{Numerical example}
\label{subsubsec-ne.example}

\begin{figure}[t]
 \begin{center}
  \includegraphics[height=50mm]{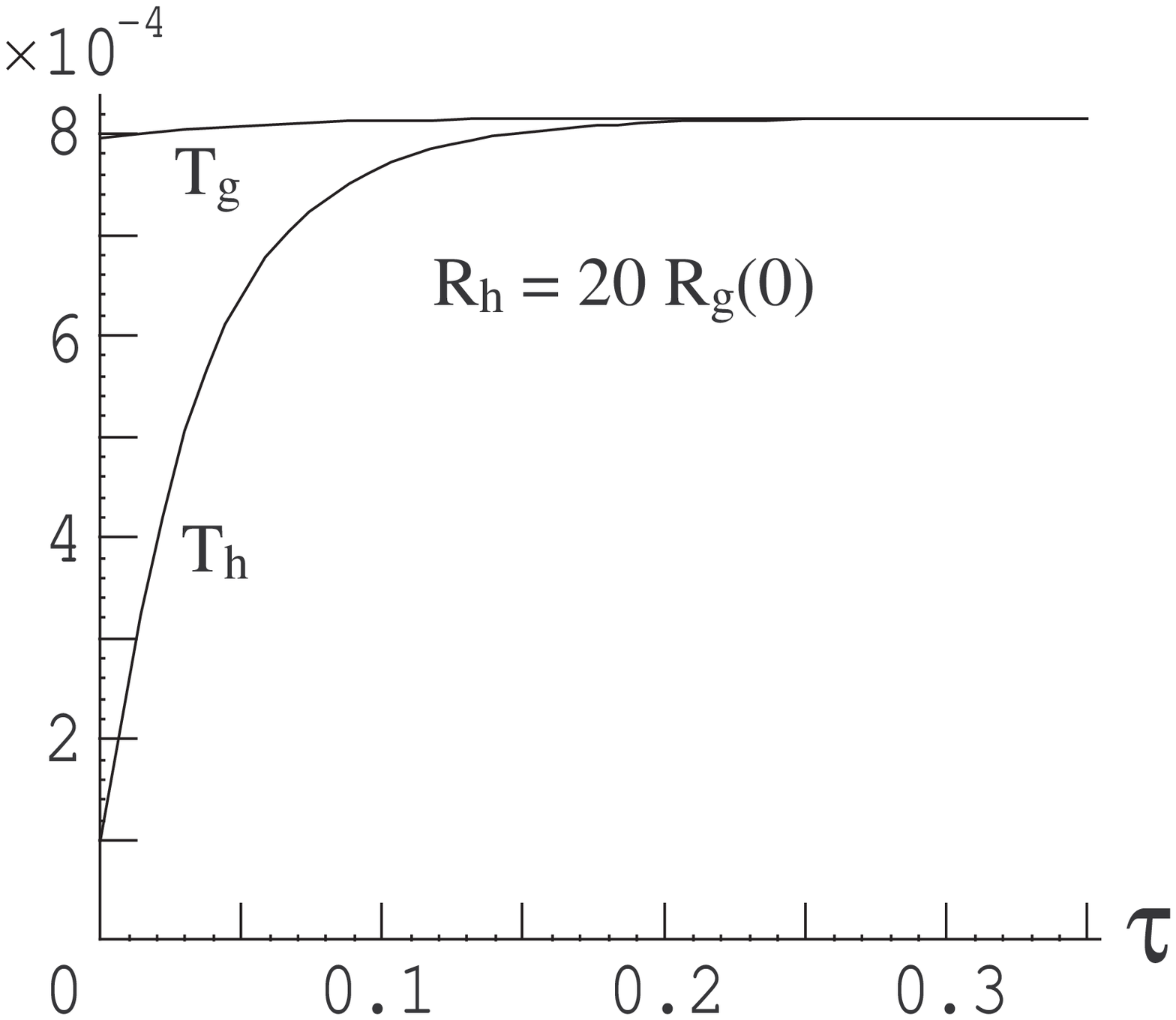} \qquad
  \includegraphics[height=50mm]{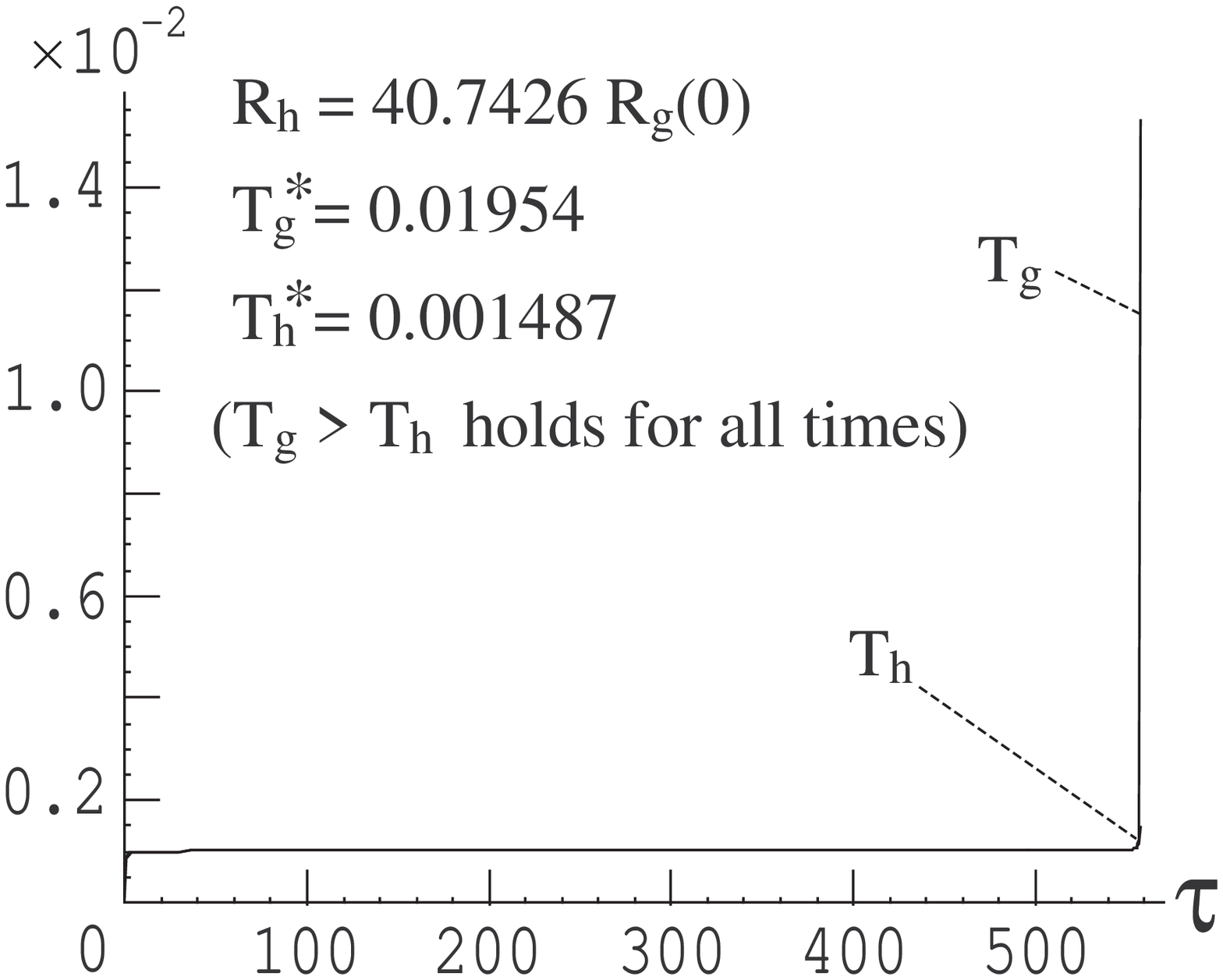} \\
  \includegraphics[height=50mm]{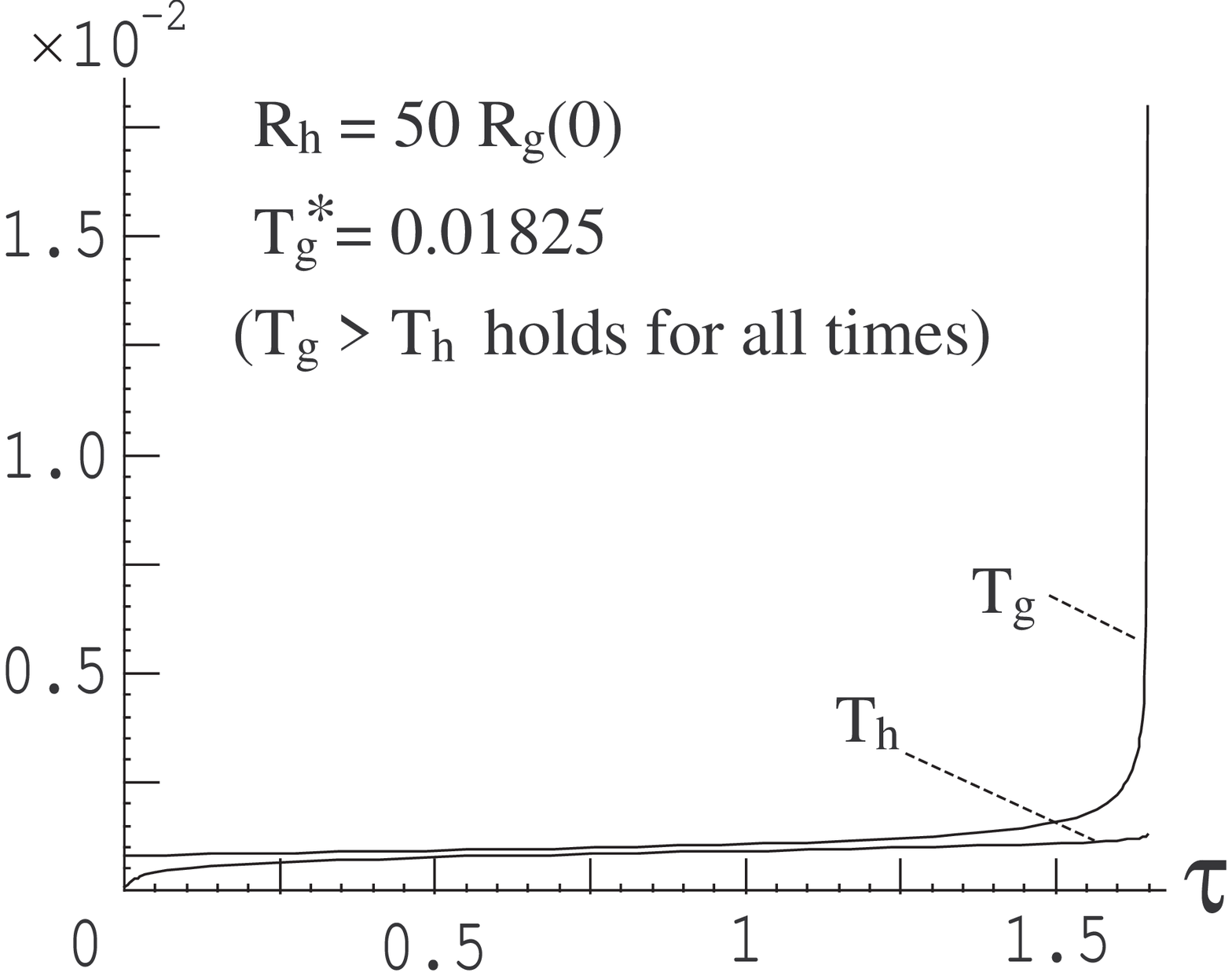} \qquad
  \includegraphics[height=50mm]{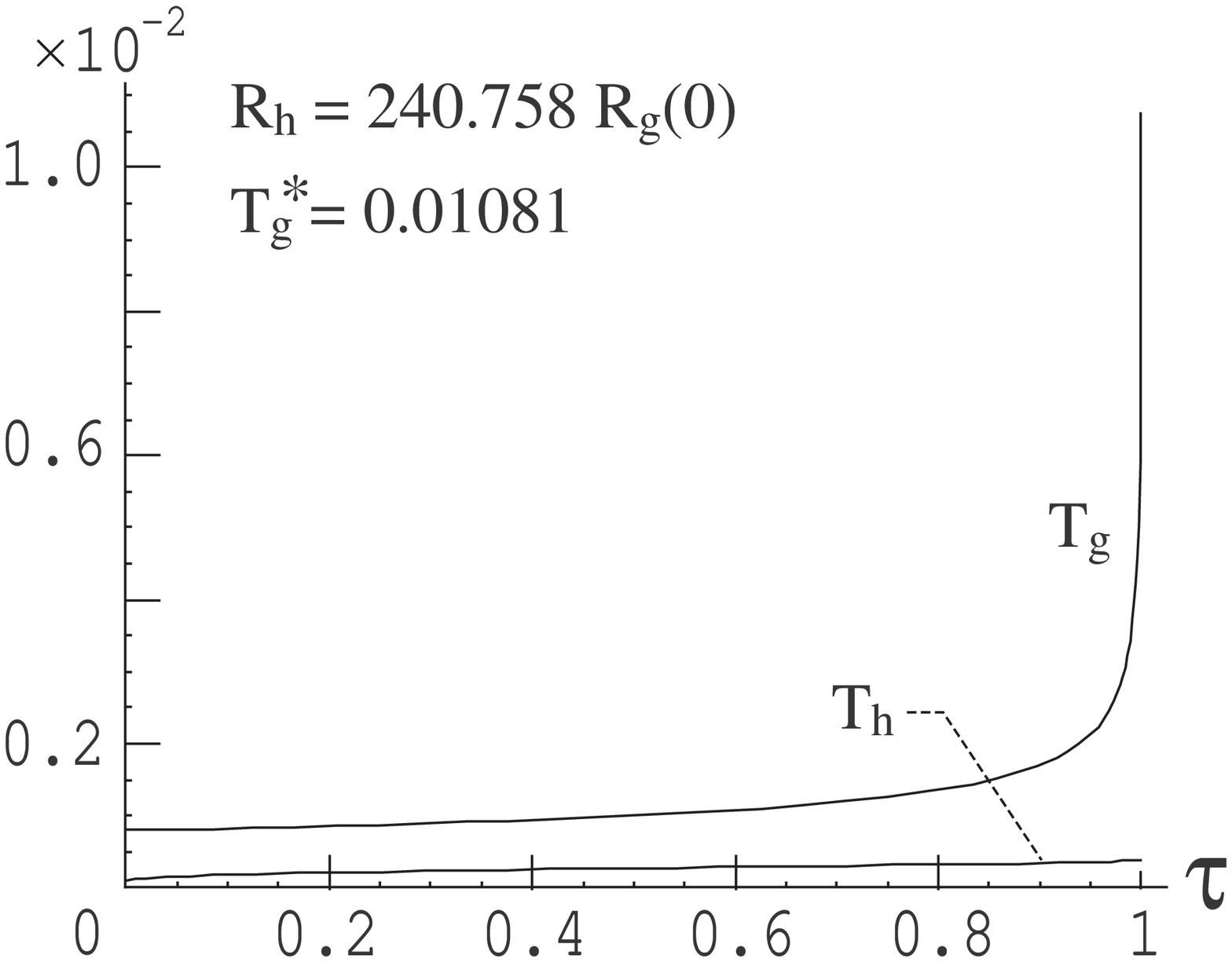} \\
  \includegraphics[height=50mm]{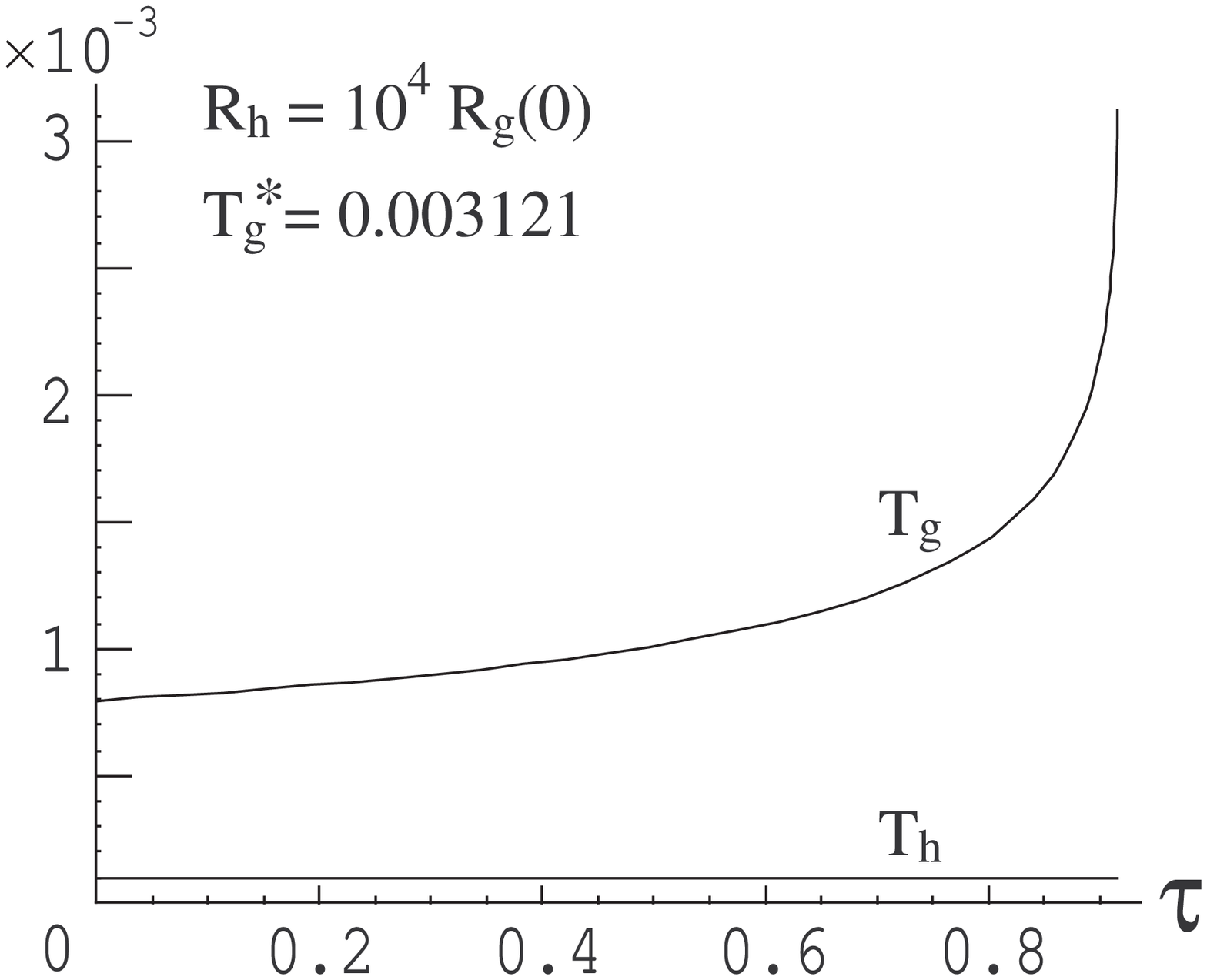} \qquad
  \includegraphics[height=50mm]{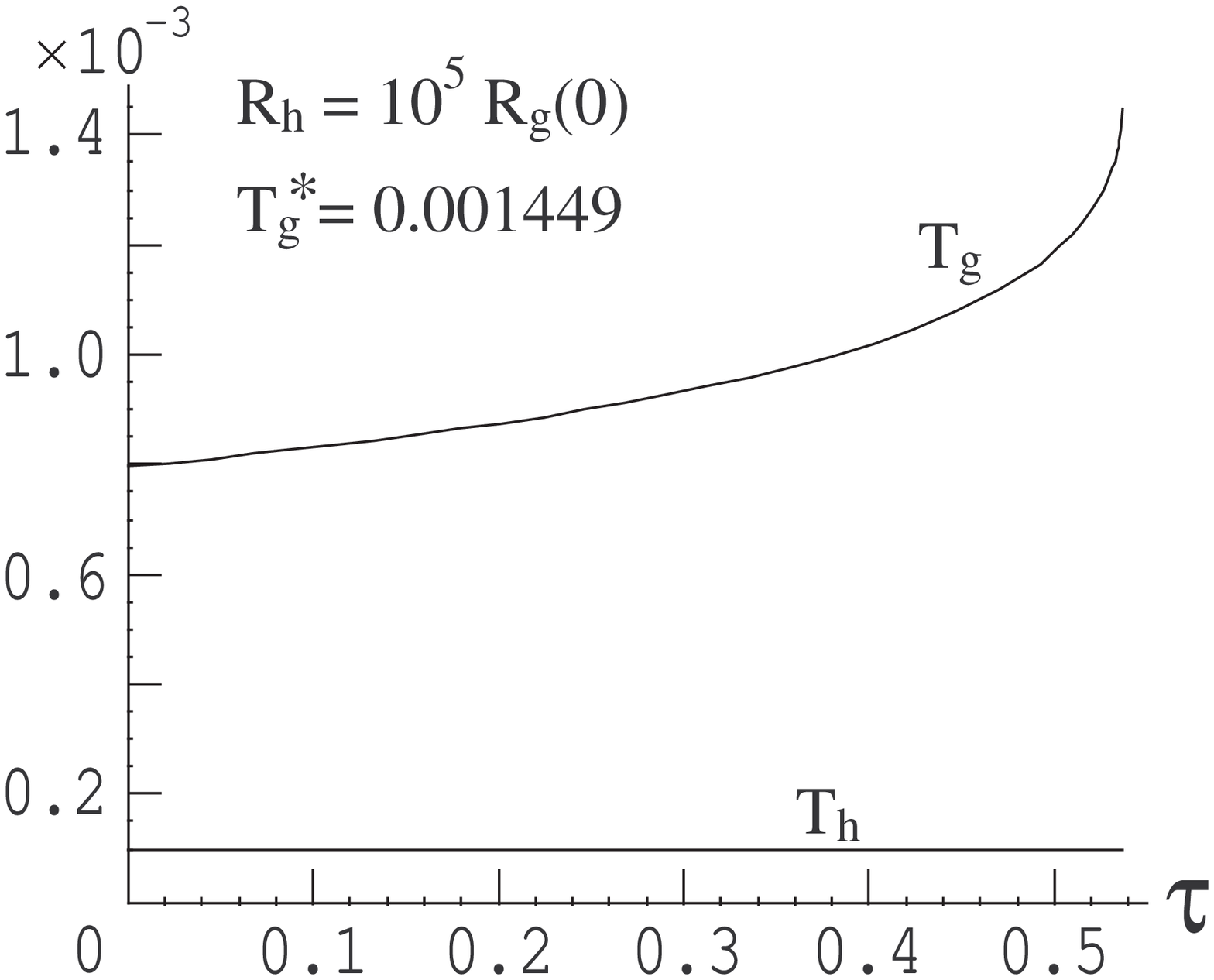}
 \end{center}
\caption{Numerical solutions of energy transport \eqref{eq-evapo.trans.transport.2} for $C_h = 10^3$, $N = 100$ and the initial conditions, $R_g(0) = 100$ and $T_h(0) = 0.0001$. Horizontal line denotes the normalized time $\tau \defeq t/\te$. The outermost radius of hollow region $R_h$ determines the size of nonequilibrium region.}
\label{fig-6}
\end{figure}

We show numerical solutions $T_g(t)$ and $T_h(t)$ of energy transport equations \eqref{eq-evapo.trans.transport.2}. 
The initial conditions are
\eqb
 R_g(0) = 100 \quad , \quad T_h(0) = 0.0001 \, .
\label{eq-evapo.ne.ic}
\eqe
$R_g(0) = 100$ gives $T_g(0) \simeq 0.00079$. 
The other parameters are set
\eqb
 C_h = 1000 \quad , \quad N = 100 \, ,
\label{eq-evapo.ne.Ch.N}
\eqe
where see Eq.\eqref{eq-intro.N} for $N$. 
Furthermore we have to specify the outermost radius $R_h$ of hollow region. 
As mentioned in Eq.\eqref{eq-evapo.ne.rate.ne}, by comparison of a numerical solution of energy transport \eqref{eq-evapo.trans.transport.2} of a certain value of $R_h$ with that of a different value of $R_h$, we can observe the nonequilibrium effect of energy exchange between black hole and heat bath. 
The numerical results are shown in Fig.\ref{fig-6}, and the value of $R_h$ is attached in each panel. 
Time coordinate $\tau$ in this figure is a time normalized as
\eqb
 \tau \defeq \dfrac{t}{\te} \, ,
\eqe
where $\te$ is the evaporation time (life time) of a black hole in an empty space with ignoring grey body factor. 
$\te$ is determined by the energy emission rate by black hole in an empty space,
\eqb
 \frac{dE_g}{dt} = -\je \, ,
\label{eq-evapo.ne.transport.empty}
\eqe
where no energy accretion exists due to the absence of heat bath and ignoring grey body factor, $E_g$ corresponds to the mass energy of black hole and $\je$ is given in Eq.\eqref{eq-evapo.ne.rate.empty}. 
This and Eq.\eqref{eq-model.eos} give
\eqb
 R_g(t) = R_g(0)\, \left( 1 - \frac{N\, t}{1280 \, \pi\, R_g(0)^3} \right)^{1/3} \, ,
\eqe
and we obtain
\eqb
\te \defeq \frac{1280 \, \pi}{N} \, R_g(0)^3 \simeq 4.02124 \times 10^7 \, ,
\eqe
where conditions \eqref{eq-evapo.ne.ic} and \eqref{eq-evapo.ne.Ch.N} are used in the second equality. 
This $\te$ is usually adopted as the time scale of black hole evaporation in many existing works on black hole physics.

\begin{figure}[t]
 \begin{center}
  \includegraphics[height=41mm]{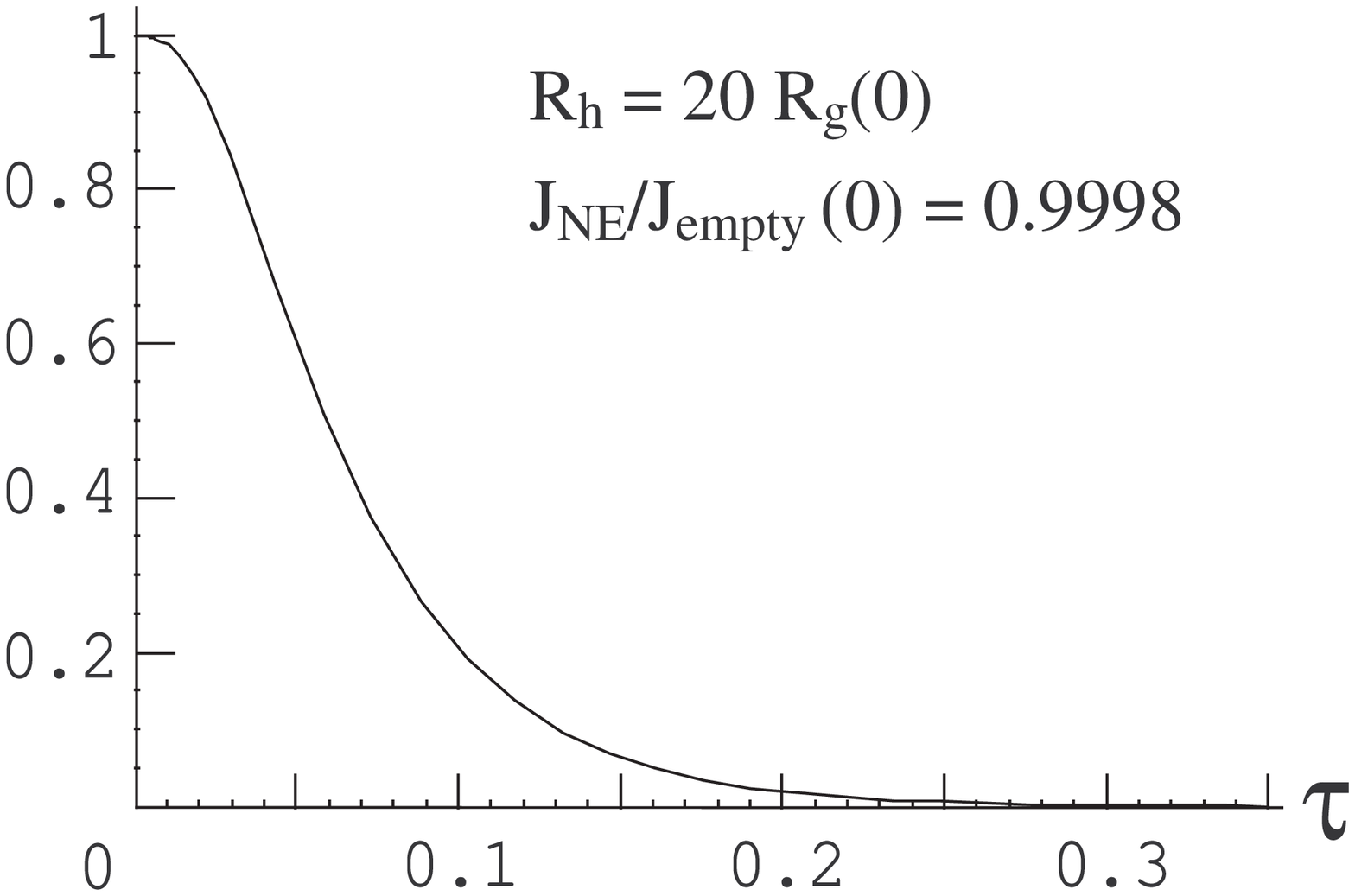} \qquad
  \includegraphics[height=41mm]{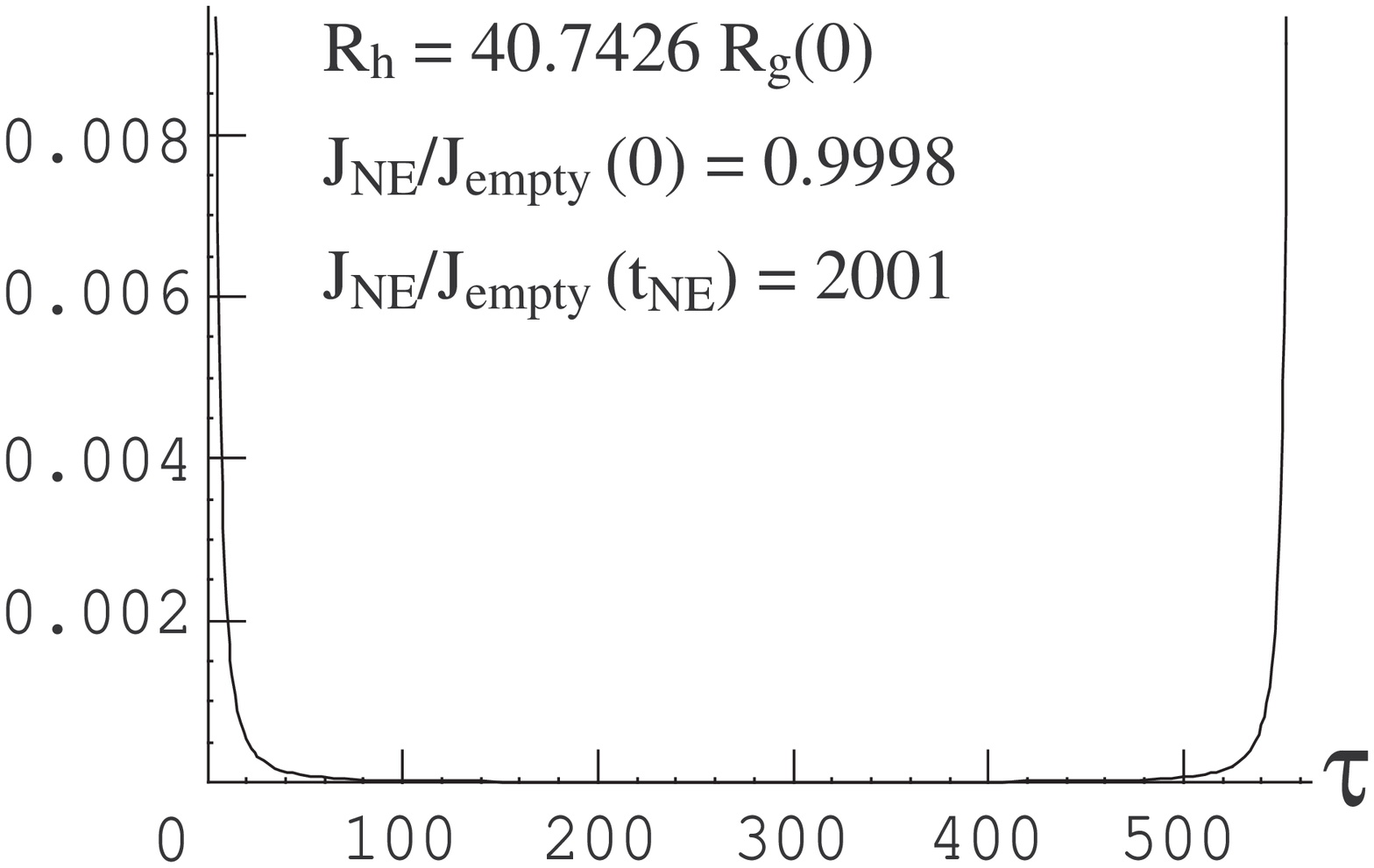} \\
  \includegraphics[height=41mm]{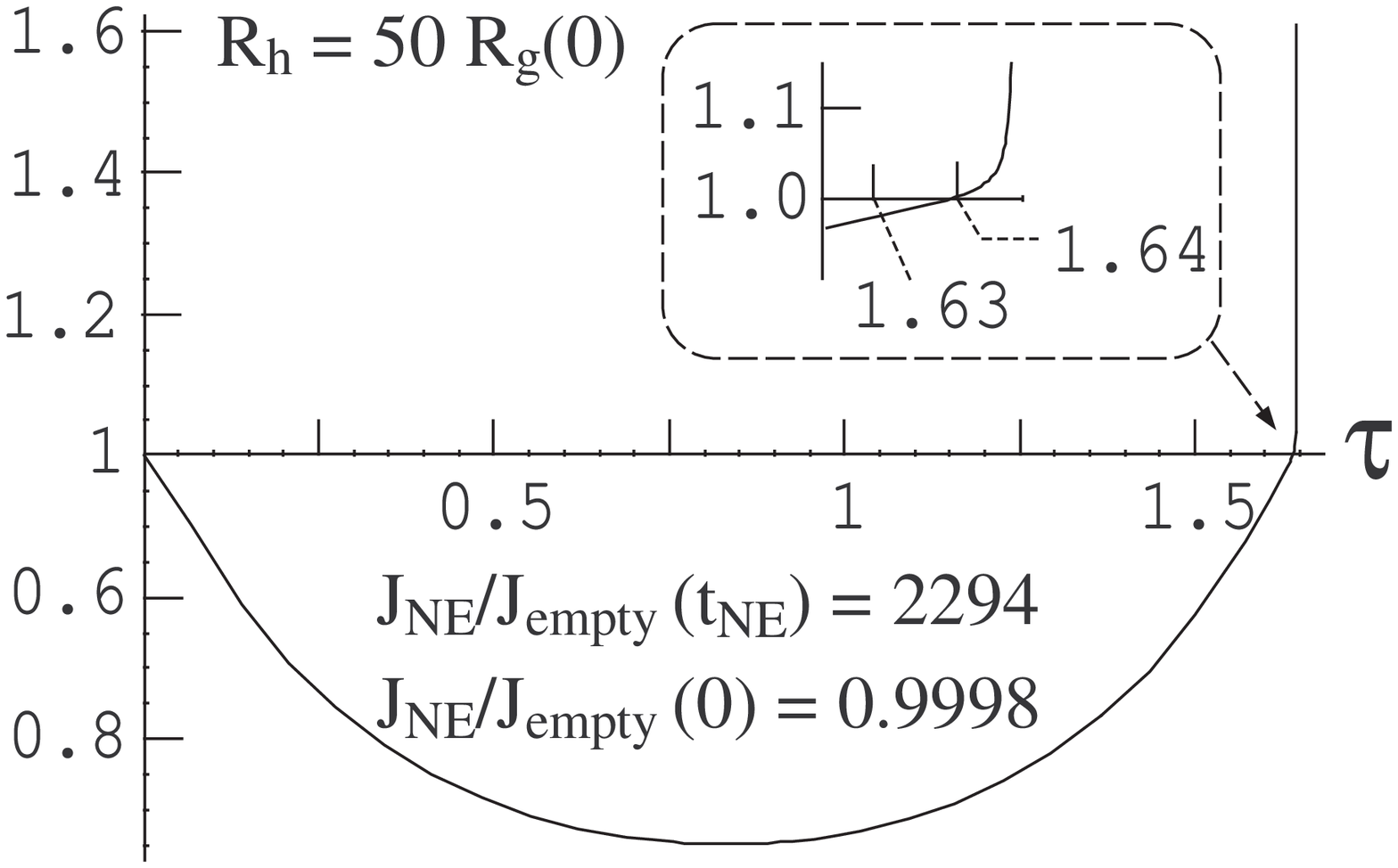} \qquad
  \includegraphics[height=41mm]{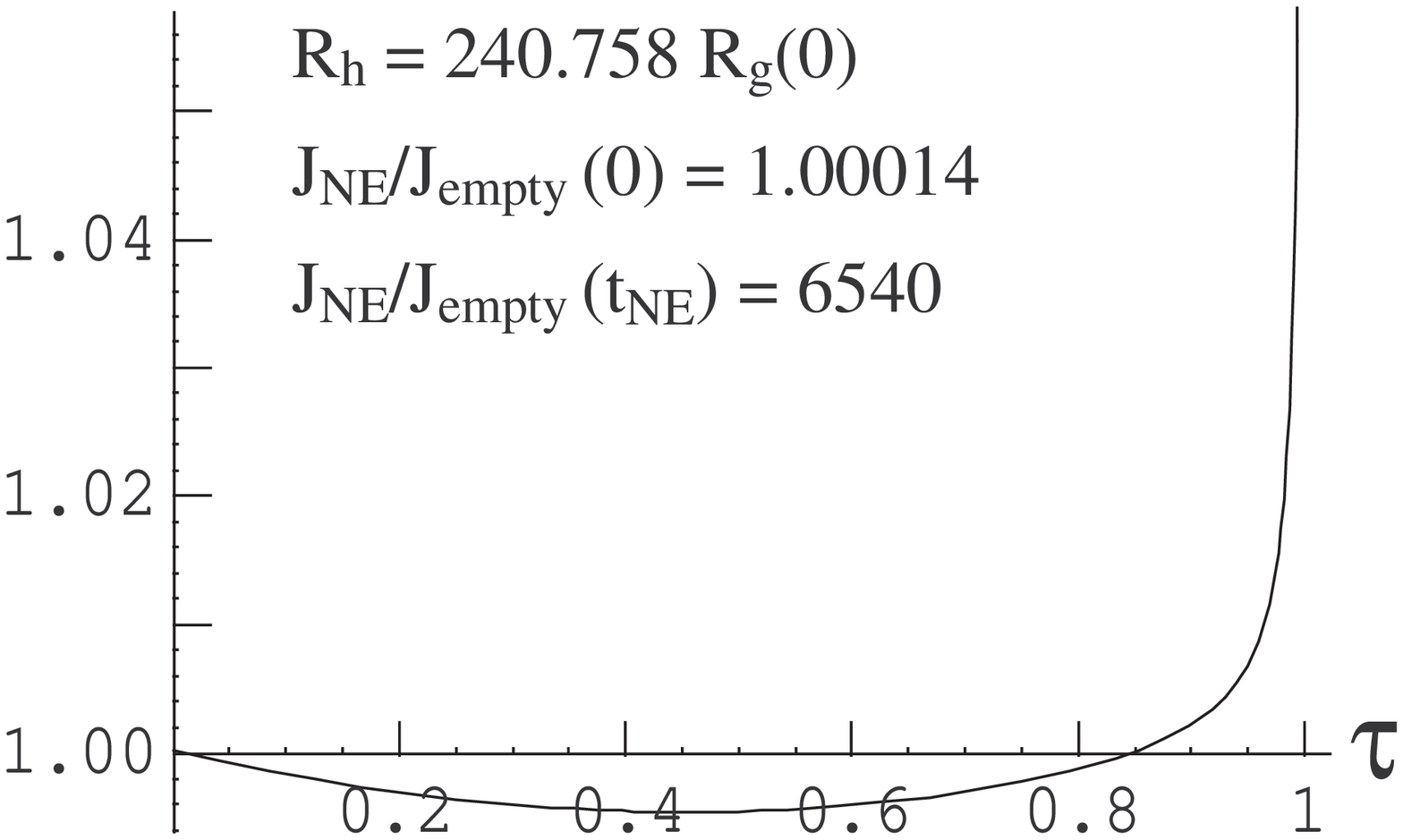} \\
  \includegraphics[height=41mm]{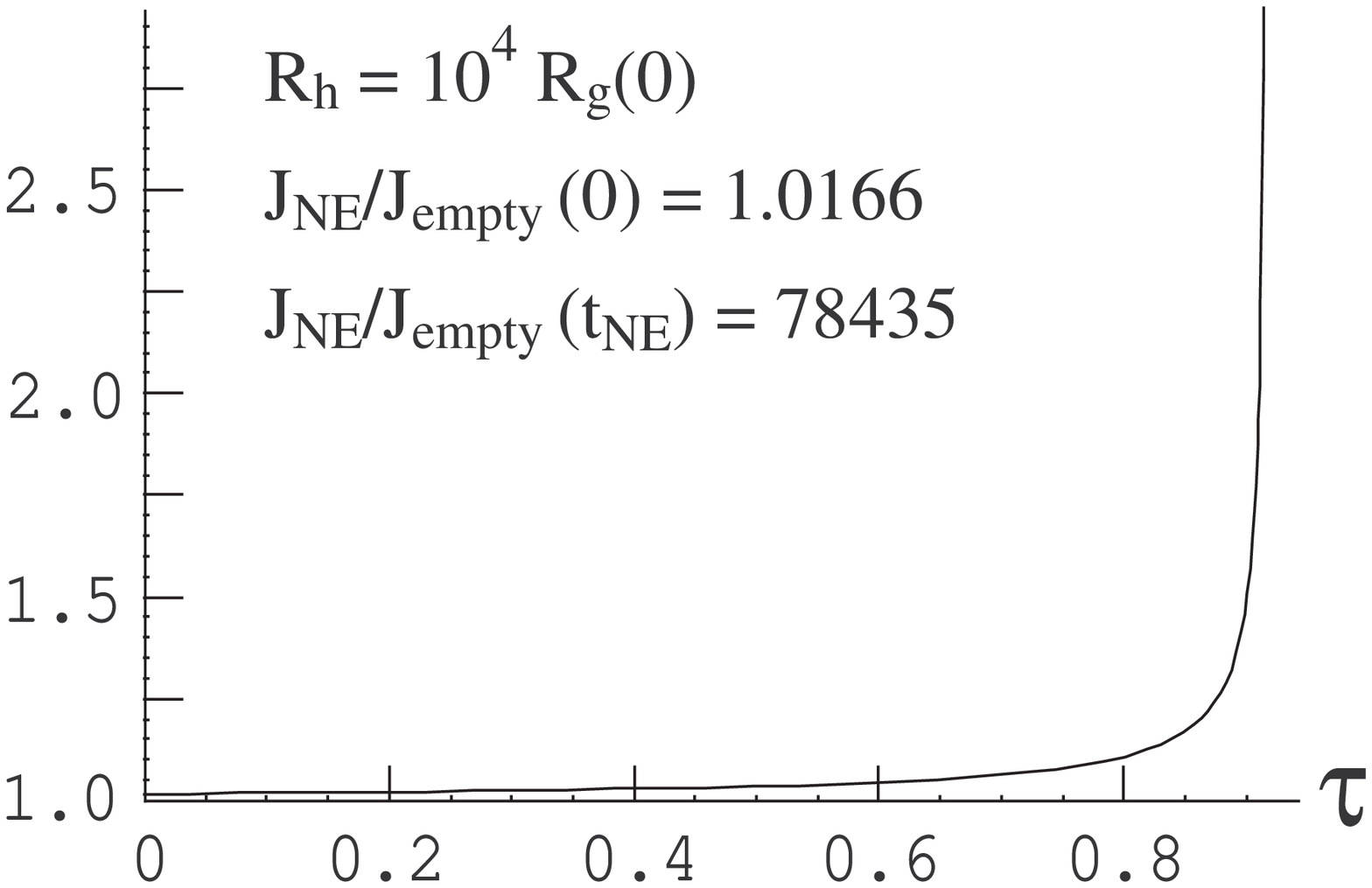} \qquad
  \includegraphics[height=41mm]{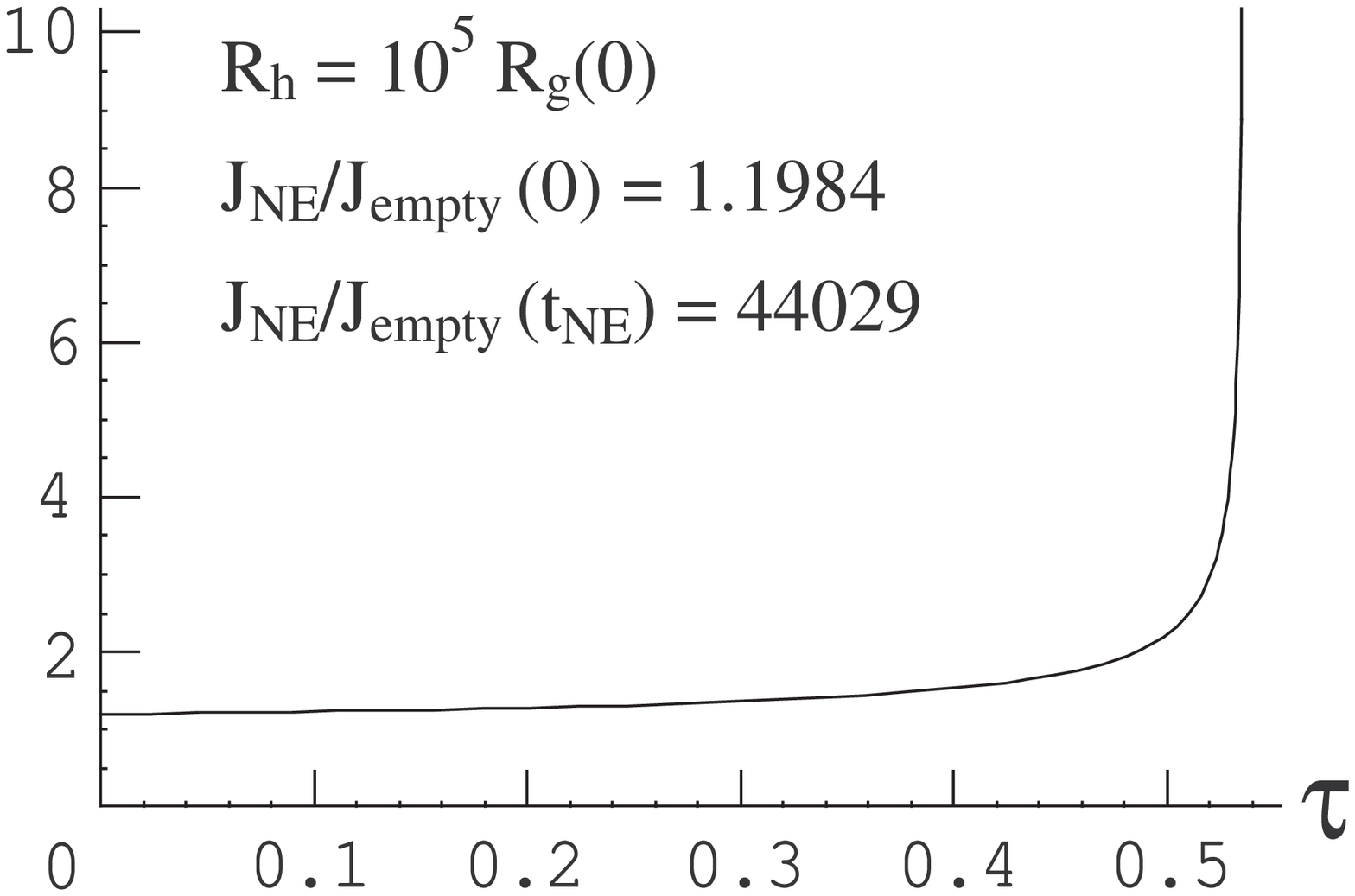}
 \end{center}
\caption{$\jn/\je$ for each case in Fig.\ref{fig-6}. The graph in left-center panel is continuous and smooth.}
\label{fig-7}
\end{figure}

Furthermore we consider the other time $\tn$ at which one of the validity conditions of NE model \eqref{eq-evapo.trans.validity.1} or \eqref{eq-evapo.trans.validity.2} breaks down,
\eqb
 \tn \defeq
 \min\left[ \, t_1 , t_2 \, \left| \,
           C_X(t_1) + C_Y^{(g)}(t_1) = 0 \,,\, v(t_2) = 1 \right. \right] \, ,
\label{eq-evapo.ne.tne}
\eqe
where $C_X + C_Y^{(g)}$ is regarded as a function of time $t$ through $R_g(t)$, and $v(t) \defeq \left|dR_g(t)/dt\right|$ is the shrinkage speed of black hole radius. 
Each panel in Fig.\ref{fig-6} shows time evolutions of $T_g(t)$ and $T_h(t)$ for $0 < t < \tn$. 
The black hole temperature at $\tn$ is denoted by $T_g^{\ast}$, 
\eqb
 T_g(\tn) \defeqr T_g^{\ast} = \frac{1}{4 \pi R_g^{\ast}}\, ,
\eqe
which is also attached in each panel in Fig.\ref{fig-6}. 
The time $\tn$ and the other quantities obtained from our numerical results are listed;
\eqb
\begin{array}{|c||*{6}{c|}} \hline
  R_h/R_g(0) & 40.7426  & 50      & 240.758 & 10^4     & 10^5   \\ \hline\hline
  \tn/\te    & 557.370  & 1.64705 & 1.00000 & 0.915560 & 0.536364 \\ \hline
  R_g^{\ast} & 4.07277  & 4.36041 & 7.36293 & 25.4986  & 54.9354  \\ \hline
  \jn/\je(\tn)
               & 2001
               & 2294
               & 6540
               & 78435
               & 44029 \\ \hline
  R_h/\tn    & 1.8 \times 10^{-7}
             & 7.5 \times 10^{-5}
             & 6.0 \times 10^{-4}
             & 2.7 \times 10^{-2}
             & 4.6 \times 10^{-1} \\ \hline
\end{array}
\label{eq-evapo.ne.list}
\eqe
For cases of $R_h/R_g(0) = 40.7462$, $50$, $240.758$ and $10^4$, the numerical plots stopped by $v(\tn) = 1$, but for a case of $R_h/R_g(0) = 10^5$, it stopped by $C_X(\tn) + C_Y^{(g)}(\tn)~=~0$. 
The second line in this list for $\tn/\te$ supports the discussion given in Subsec.\ref{subsubsec-general} that the larger the radius $R_h$, the faster the black hole evaporation proceeds and the shorter the time $\tn$. 
The third and fourth lines in list \eqref{eq-evapo.ne.list} for $R_g^{\ast}$ and $\jn/\je(\tn)$ give an important information in next subsection. 
The lowest line in list \eqref{eq-evapo.ne.list} for $R_h/\tn$ shows that our numerical results are consistent with the fast propagation assumption. 
To understand this, consider a typical time scale $t_{rad}$ in which one particle of radiation fields travels in the hollow region from black hole to heat bath. 
This $t_{rad}$ is given by $t_{rad} \defeq R_h$, since the radiation fields are massless. 
If $t_{rad} < \tn$, then the fast propagation assumption is reasonable. 
In fact, the ratio $t_{rad}/\tn \, ( = R_h/\tn )$ shown at the lowest line indicates the validity of fast propagation assumption.

Concerning a case $R_h = 20\,R_g(0)$, it is helpful to recognize that our conditions \eqref{eq-evapo.ne.ic} and \eqref{eq-evapo.ne.Ch.N} give $2\, \pi \times 100^2 > 1000$ which denotes $\left| C_g \right| > C_h$. 
This $\left| C_g \right| > C_h$ is the condition for stable equilibrium of black hole and heat bath in the framework of the equilibrium model (see~\cite{ref-trans.1} or~\cite{ref-evapo.sst}). 
Therefore, if the nonequilibrium region is ignored and the equilibrium model used in~\cite{ref-trans.1} is considered with the same setting parameters of Eqs.\eqref{eq-evapo.ne.ic} and \eqref{eq-evapo.ne.Ch.N}, then the black hole stabilizes with heat bath to settle down into a total equilibrium state $T_g - T_h \to 0$. 
Indeed, for the case $R_h = 20 \, R_g(0)$, the hollow region is not large enough and the black hole stabilizes with heat bath. 
Hence the occurrence of accelerated increase of temperature $T_g$ in Fig.\ref{fig-6} is obviously the nonequilibrium effect of energy exchange between black hole and heat bath.

Then as the outermost radius $R_h$ of hollow region is set larger and the nonequilibrium region becomes larger, the black hole evaporation comes to be observed as an accelerated increase of temperature difference $T_g - T_h$ for the case $R_h = 40.7426 \, R_g(0)$. 
$\tn$ is very longer than $\te$ in this case. 
Furthermore as $R_h$ is set larger, the time $\tn$ becomes shorter and we find $\tn \simeq \te$ for $R_h \simeq 240.758 \, R_g(0)$. 
At last $\tn$ becomes shorter to $\tn \simeq 0.536364 \, \te$ for $R_h = 10^5 \, R_g(0)$. 
If we set $R_h \gtrsim 10^6$, the combined heat capacity becomes positive $C_X + C_Y^{(g)} > 0$ which violates the validity condition \eqref{eq-evapo.trans.validity.1}. 
A black hole with a large nonequilibrium region of $R_h \gtrsim 10^6$ can not be treated in the framework of NE model, and, as discussed in the last paragraph in Subsec.\ref{sec-evapo.trans}, a black hole evaporation for such case is a highly nonequilibrium dynamical process in which the black hole can not be treated with equilibrium black hole solutions of Einstein equation. 
The larger the nonequilibrium region, the faster the black hole evaporation process evolves into a highly nonequilibrium dynamical stage.

Fig.\ref{fig-7} shows time evolutions of the ratio of energy emission rates $\jn/\je$ given by Eq.\eqref{eq-evapo.ne.jn.je}. 
This figure indicates that, when a black hole evaporates, the relation $\jn > \je$ comes to hold even if the converse relation $\jn < \je$ holds at initial time. 
Therefore the discussion given in previous Subsec.\ref{subsubsec-ne.empty} is supported. 
For the case $R_h = 20 R_g(0)$ in which a black hole stabilizes with heat bath, the energy emission rate $\jn$ disappears, $\jn/\je \to 0$, as easily expected by the behavior $T_g \to T_h \, \Rightarrow \, \jn \to 0$.

\subsection{Beyond the NE model I: abrupt catastrophic evaporation}
\label{sec-evapo.ace}

When the nonequilibrium region around black hole is not so large, the evaporating black hole is well approximated to equilibrium solutions of Einstein equation and the NE model is applicable to such quasi-equilibrium evaporation stage. 
However the condition \eqref{eq-evapo.trans.validity.1} or \eqref{eq-evapo.trans.validity.2} is violated at the time $\tn$ and the evaporation process becomes highly nonequilibrium dynamical stage (see last paragraph in Subsec.\ref{sec-evapo.trans}). 
Therefore, since a relation $R_g^{\ast} > 1$ is expected from the list \eqref{eq-evapo.ne.list}, we find the semi-classical evaporation stage ($R_g \gtrsim 1$) is divided into two stages, quasi-equilibrium one and highly nonequilibrium dynamical one. 
The former is described by the NE model, but the latter is beyond the range of NE model. 
This subsection extrapolates NE model to the latter stage and suggests a specific nonequilibrium phenomenon.

On the other hand, for the equilibrium model used in~\cite{ref-trans.1} and the black hole evaporation in an empty space with ignoring grey body factor, the highly nonequilibrium dynamical stage does not exist and the semi-classical stage is always described as the quasi-equilibrium stage. 
This is explained in next Subsec.\ref{subsubsec-sc}. 
Then the highly nonequilibrium dynamical stage and a suggestion about that stage in the framework of NE model are discussed in Subsec.\ref{subsubsec-ace}.

\subsubsection{On semi-classical evaporation stage, except for NE model}
\label{subsubsec-sc}

This subsection shows that the highly nonequilibrium dynamical stage does not occur for the equilibrium model used in~\cite{ref-trans.1} and for the black hole evaporation in an empty space with ignoring grey body factor. 

For the first consider the equilibrium model used in~\cite{ref-trans.1}. 
As explained in Subsec.\ref{subsubsec-ne.eq}, this model is obtained from NE model by setting $R_h = R_g$ and $T_g = T_h$. 
Exactly speaking, the evaporation process is not described by the ``equilibrium model''. 
However by extrapolating the equilibrium model to the evaporation process, we may set $T_h < T_g$ with keeping the condition $R_g = R_h$. 
Then the energy transport equations from black hole to heat bath are given by
\eqb
 \frac{d E_g}{dt} = - \sigmap \left(\, T_g^4 - T_h^4 \,\right) A_g \quad , \quad
 \frac{d E_h}{dt} = \sigmap \left(\, T_g^4 - T_h^4 \,\right) A_g \, ,
\eqe
where $A_g = 4 \pi R_g^2$ is the surface area of black hole. 
The equation of $dE_g/dt$ together with Eq.\eqref{eq-model.eos} give
\eqb
 v \defeq \left| \frac{d R_g}{dt} \right| = 2 \left| \frac{d E_g}{dt} \right|
   = 2 \sigmap \left(\, T_g^4 - T_h^4 \,\right) A_g \, .
\eqe
Since equations of states \eqref{eq-model.eos} are used, the quasi-equilibrium assumption is also necessary here and $v < 1$ is required. 
The inequality $v < 1$ corresponds to the validity condition \eqref{eq-evapo.trans.validity.2} of NE model. 
Obviously $v = 1$ occurs for non-zero radius $R_g > 0$. 
On the other hand, from Eq.\eqref{eq-model.capacity}, vanishing heat capacity of black hole $C_g = 0$ corresponds to zero radius $R_g = 0$. 
This means that the equilibrium model has no validity condition which corresponds to the condition \eqref{eq-evapo.trans.validity.1} of NE model. 
Hence we can recognize that, if $v = 1$ occurs for a black hole of semi-classical size $R_g > 1$, it is concluded that the highly nonequilibrium dynamical stage occurs at semi-classical level. 
But if $v = 1$ does not occur for a semi-classical black hole, then the highly nonequilibrium dynamical stage does not occur at semi-classical level in the framework of equilibrium model.

The validity condition $v < 1$ is rewritten as
\eqb
 \sigmap \left(\, T_g^4 - T_h^4 \,\right) < 2 \pi \, T_g^2 \, ,
\eqe
and this gives
\eqb
 R_g^2 > \frac{1}{8 \pi^2 \beta}
         \left(\, 1 + \sqrt{1 + \left( 2 T_h^2/\beta \right)^2} \,\right)^{-1} \, ,
\label{eq-sc.eq}
\eqe
where $\beta \defeq 2 \pi/\sigmap = 240/\pi N$ and $R_g = 1/4 \pi T_g$ is used. 
Here Eq.\eqref{eq-intro.N} gives $\beta \simeq 1$. 
Recall that $T_g > T_h$ holds generally for any evaporation process and $T_g < 1$ holds due to the quasi-equilibrium assumption. 
Then we find $T_h < 1$ and $(2 T_h^2/\beta)^2 < 1$. 
Therefore we can approximate inequality \eqref{eq-sc.eq} to $R_g^2 > 1/(16 \pi^2 \beta) \simeq 10^{-2}$, where $\beta \simeq 1$ is used. This gives
\eqb
 R_g \gtrsim 0.1 \, .
\label{eq-sc.fast.1}
\eqe
This denotes that the fast evaporation $v = 1$ occurs at $R_g \simeq 0.1$. 
Hence, as discussed in previous paragraph, the highly nonequilibrium dynamical stage does not occur in the framework of equilibrium model used in~\cite{ref-trans.1}.

Next consider a black hole evaporation in an empty space with ignoring grey body factor. 
The Stefan-Boltzmann law gives Eq.\eqref{eq-evapo.ne.transport.empty}. 
This together with equations of states \eqref{eq-model.eos} give
\eqb
 v \defeq \left| \frac{d R_g}{dt} \right| = 2 \left| \frac{d E_g}{dt} \right|
   = 2 \sigmap T_g^4 A_g \, .
\eqe
Hence, following a similar discussion given in previous paragraph with setting $T_h = 0$, we require $v < 1$ and obtain $R_g^2 > 1/(16 \pi^2 \beta) \simeq 10^{-2}$, where $\beta \simeq 1$ is used. This gives
\eqb
 R_g \gtrsim 0.1 \, .
\label{eq-sc.fast.2}
\eqe
This denotes that the fast evaporation $v = 1$ occurs at $R_g \simeq 0.1$, and that the highly nonequilibrium dynamical stage does not occur for a black hole evaporation in an empty space with ignoring grey body factor.

\subsubsection{Abrupt catastrophic evaporation}
\label{subsubsec-ace}

Let us point out again a numerical evidence shown in list \eqref{eq-evapo.ne.list} that the black hole radius $R_g^{\ast}$ at time $\tn$ is greater than unity $R_g^{\ast} > 1$. 
According to the discussion in last paragraph in Subsec.\ref{sec-evapo.trans}, we can expect a highly nonequilibrium dynamical stage of evaporation process will occur at semi-classical level $R_g^{\ast} > 1$ in the framework of NE model. 
After the time $\tn$, the black hole and radiation fields should be described as highly nonequilibrium dynamical ones.

In the following discussion, we make two steps: 
Firstly, to confirm the numerical evidence, we show $R_g^{\ast} \gtrsim 1$ analytically. 
Secondly, a physical implication of $R_g^{\ast} \gtrsim 1$ is discussed and a specific nonequilibrium phenomenon is suggested.

For the first, we analyze the energy transport equations \eqref{eq-evapo.trans.transport.2}. 
Due to definition \eqref{eq-evapo.ne.tne} of time $\tn$, we consider two cases, $\tn = t_1$ and $\tn = t_2$, where $t_1$ and $t_2$ are given in definition \eqref{eq-evapo.ne.tne}. 
But before proceeding to the analysis of these cases, we should point out the following: 
As explained at the end of Subsec.\ref{subsubsec-CXCYg}, equation $C_X(R_g) + C_Y^{(g)}(R_g) = 0$ has one, two or three solutions of $R_g$ for a sufficiently small $T_h$. 
However even if there are two or three solutions of $R_g$ for our choice of $T_h$, it is the lowest solution at which a highly nonequilibrium dynamical evaporation stage starts towards a quantum evaporation stage. 
Hence the time $t_1$ in definition \eqref{eq-evapo.ne.tne} is the lowest solution of equation $C_X(t) + C_Y^{(g)}(t) = 0$.

Here we estimate the order of $R_g^{\ast}$. 
Consider the case $\tn = t_1$, where $R_g^{\ast} = R_g(t_1)$ and $C_X(R_g^{\ast}) + C_Y^{(g)}(R_g^{\ast}) = 0$. 
Because of $C_Y^{(g)} > 0$ by definition, $C_X(R_g^{\ast}) = - C_Y^{(g)}(R_g^{\ast}) < 0$ holds. 
Consequently, according to a behavior of $C_X(R_g)$ explained in Subsec.\ref{subsubsec-CX.Rg} (left panel in Fig.\ref{fig-3}), we find $R_g^{\ast} > \tilde{R}_g \simeq 0.055 \times \left( N R_h \right)^{1/3}$. 
Hence together with Eq.\eqref{eq-intro.N}, we find $R_g^{\ast} \gtrsim 1$ for the situation $R_h \gtrsim 60$. 
And next consider the other case $\tn = t_2$, where $|\dot{R}_g(t_2)| = 1$. 
Because of $t_2 < t_1$, we find $C_X(t_2) + C_Y^{(g)}(t_2) < 0$. 
Then, because it is assumed that $T_h$ is small enough so that the validity condition \eqref{eq-evapo.trans.validity.1} holds, we find by Subsec.\ref{subsubsec-CXCYg} (Fig.\ref{fig-4}) that $R_g^{\ast} = R_g(t_2) > R_{g0}$ holds, where $R_{g0}$ is the lowest solution of $C_X(R_{g0}) + C_Y^{(g)}(R_{g0}) = 0$. 
Therefore, following the same discussion given for the case $\tn = t_1$, we obtain $R_g^{\ast} > R_{g0} > \tilde{R}_g \simeq 1$ for the situation $R_h \gtrsim 60$. 
In summary, the black hole radius $R_g^{\ast}$ at time $\tn$ is greater than unity $R_g^{\ast} \gtrsim 1$, when the black hole evaporates in the framework of NE model under the condition $R_h \gtrsim 60$. 
Here we have to note two points: 
First is that the condition $R_h \gtrsim 60$ is not a necessary condition but a sufficient condition for $R_g^{\ast} \gtrsim 1$, and there may remain a possibility that $R_g^{\ast} \gtrsim 1$ holds even if $R_h \lesssim 60$. 
Second point is that, if the nonequilibrium nature of black hole evaporation is not taken into account, the radius $R_g^{\ast}$ can not be greater than unity but it becomes less than Planck length as seen in Eqs.\eqref{eq-sc.fast.1} and~\eqref{eq-sc.fast.2}. 
The relation $R_g^{\ast} \gtrsim 1$ is a peculiar property of the NE model.

We proceed to the second part of this subsection, an implication of the above result, $R_g^{\ast} \gtrsim~1$. 
Recall a highly nonequilibrium dynamical stage of evaporation process occurs after the time $\tn$. 
Because of $R_g^{\ast} \gtrsim 1$, a semi-classical (but not quasi-equilibrium) discussion is available for the highly nonequilibrium dynamical stage while the black hole radius shrinks from $R_g^{\ast}$ to Planck length $l_{pl} \defeq 1$. 
Then it is appropriate to consider that the mass energy of black hole evolves from $E_g^{\ast}$ ($= R_g^{\ast}/2$) to $E_p \defeq l_{pl}/2$. 
Energy difference $\Delta E_g \defeq E_g^{\ast} - E_p$ is emitted during highly nonequilibrium dynamical stage. 
Furthermore, for example, Fig.\ref{fig-7} and fourth line in list~\eqref{eq-evapo.ne.list} of our numerical example imply a very strong luminosity $\jn$ of Hawking radiation in the NE model in comparison with the luminosity $\je$ in the evaporation in an empty space with ignoring grey body factor. 
On the other hand $\je$ is very strong as explained in \S1 of~\cite{ref-hr}. 
Hence, $\jn$ may be a huge luminosity. 
The energy emission by a black hole in NE model may be understand as a strong ``burst''.

In addition to the luminosity of Hawking radiation, we consider the duration $\delta t_{dyn}$ of the highly nonequilibrium dynamical stage. 
Since the shrinkage speed of black hole radius $v \defeq \left|dR_g/dt\right|$ is approximately unity $v \sim 1$ during highly nonequilibrium dynamical stage (see condition~\eqref{eq-evapo.trans.validity.2}), the duration is estimated as $\delta t_{dyn} \sim R_g^{\ast}/v \sim R_g^{\ast}$, and the following relation is obtained,
\eqb
 \delta\tau \defeq \frac{\delta t_{dyn}}{\te} \sim \frac{R_g^{\ast}}{\te}
 < \frac{R_g(0)}{\te} = \frac{N}{1280 \pi R_g(0)^2} \sim \frac{1}{40 R_g(0)^2} \ll 0.025 \, ,
\label{eq-evapo.ace.duration}
\eqe
where Eq.\eqref{eq-intro.N} is used, and, since the initial radius $R_g(0)$ should be large enough to consider a semi-classical evaporation stage, we introduced relations $R_g^{\ast} < R_g(0)$ and $1 \ll R_g(0)$. 
This denotes $\delta t_{dyn} \ll \te$. 
Furthermore, for example, we find the shortest $\tn = O(0.1)\times \te$ from list~\eqref{eq-evapo.ne.list}. 
This together with~\eqref{eq-evapo.ace.duration} imply $\delta t_{dyn} \ll \tn$. 
Hence it seems reasonable to consider that $\delta t_{dyn}$ is very shorter than $\tn$. 
(For example it seems that the tangent $dT_g/dt$ seen in each panel in Fig.\ref{fig-6} becomes very large quickly as $t \to \tn$, and $T_g$ will reach Planck temperature quickly just after $\tn$.) 
Hence it is suggested that the energy $\Delta E_g \defeq E_g^{\ast} - E_p$ bursts out of black hole with a very strong luminosity within $\delta t_{dyn}$ which is negligibly short in comparison with $\tn$.

From the above, we suggest the following: 
When a black hole evaporates in the framework of NE model under the condition $R_h \gtrsim 60$, a quasi-equilibrium evaporation stage continues until $\tn$. 
Then a highly nonequilibrium dynamical evaporation stage occurs at $\tn$. 
In that stage, a semi-classical black hole of radius $R_g^{\ast} \gtrsim 1$ evaporates abruptly (within a negligibly short time scale $\delta t_{dyn}$) to become a quantum one. 
This abrupt evaporation in the highly nonequilibrium dynamical stage is accompanied by a burst of energy $\Delta E_g$. 
We call this phenomenon {\it the abrupt catastrophic evaporation} at semi-classical level $R_g^{\ast} \gtrsim 1$, where ``catastrophic'' means the shrinkage speed of black hole radius is very high $v \sim 1$ and the energy $\Delta E_g$ bursts out of black hole with a huge luminosity $\jn \gg \je$ within a negligibly short time scale $\delta t_{dyn} \ll \tn$.

The above discussion is based on the NE model. 
As shown in previous subsection, for the equilibrium model used in~\cite{ref-hr} and the black hole evaporation in an empty space with ignoring grey body factor, the black hole radius becomes Planck size before the shrinkage speed of black hole radius reaches unity. 
The highly nonequilibrium dynamical stage and the abrupt catastrophic evaporation at semi-classical level do not occur in those models. 
Hence the abrupt catastrophic evaporation at semi-classical level seems to be a specific nonequilibrium phenomenon suggested by NE model.

Here we discuss about a black hole evaporation in an empty space in a full general relativistic framework. 
Note that, even if a black hole is in an empty space, there should exist an incoming energy flow onto the black hole due to the curvature scattering. 
When the curvature scattering is taken into account for the case of a black hole evaporation in an empty space, we can interpret the whole system as if a black hole is surrounded by some nonequilibrium matter fields which possess outgoing and incoming energy flows of Hawking radiation under the effects of curvature scattering. 
Furthermore, since the curvature scattering occurs whole over the spacetime, it is expected that the nonequilibrium region is so large that a condition corresponding to $R_h \gtrsim 60$ in NE model holds. 
Hence, if the NE model is extended to a full general relativistic model, we can expect that a black hole evaporation in an empty space can be treated in the framework of full general relativistic version of NE model (with removing the heat bath), and that the abrupt catastrophic evaporation at semi-classical level may occur as well since the nonequilibrium region is sufficiently large.

Finally we estimate a typical time scale of black hole evaporation with energy accretion. 
It is reasonable to consider the duration of quantum evaporation stage is about one Planck time. 
Then, the time scale of black hole evaporation $t_{ev}$ is estimated as
\eqb
 t_{ev} \simeq \tn + \delta t_{dyn} + 1 \sim \tn \, .
\eqe
The time $\tn$ gives a typical time scale of black hole evaporation with energy accretion.

\subsection{Beyond the NE model II: final fate of quantum black hole evaporation}
\label{sec-evapo.info}

So far we have considered semi-classical evaporation stages in the framework of NE model, and found it consists of two stages, quasi-equilibrium one and highly nonequilibrium dynamical one. 
This subsection discusses the quantum evaporation stage following the highly nonequilibrium dynamical stage.

Concerning quantum black hole evaporation, the so-called {\it information loss paradox} is an interesting and important issue~\cite{ref-info}: 
If the black hole mass is radiated out completely by the Hawking radiation, then the thermal spectrum of Hawking radiation implies that only a matter field which is in a thermal equilibrium state may be left after the evaporation. 
This implies that the initial condition of black hole formation (gravitational collapse) is completely smeared out. 
For example, even if the initial state of collapsing matter is a pure quantum state, the final state after a black hole evaporation must be transformed to a thermal state. 
This contradicts the unitary invariance of quantum theory. 
This is the information loss paradox.

The study on the final fate of black hole evaporation is usually carried out in the context of quantum gravity theory. 
However, since no complete theory of quantum gravity has yet been constructed, it is meaningful to some extent to study the final fate of black hole evaporation with an appropriate model of black hole evaporation without referring to present incomplete quantum gravity theories. 
Therefore, we use the NE model and try to extract what we can suggest about the final fate of black hole evaporation.

In this subsection, we consider whether some remnant remains after the quantum evaporation or not. 
If a remnant remains, then it is implied that the information loss paradox does not exit, because the complete evaporation may not be true and the remnant may preserve the information about initial condition of black hole formation to guarantee the unitary evolution of the system. 
For the time being, we assume that a black hole evaporates out completely and only equilibrium radiation fields remain at the end state of quantum evaporation stage. 
If a contradiction results from this assumption, we may conclude that a remnant will remain after a black hole evaporation. 
This subsection aims to suggest a necessity of some remnant by the reductive absurdity.

As discussed at the end of previous subsection, it is expected that the abrupt catastrophic evaporation at semi-classical level occurs in a full general relativistic framework. 
Therefore we consider the case $R_h \gtrsim 60$ in the NE model which denotes the occurrence of abrupt catastrophic evaporation. 
Then, under the assumption of complete evaporation of black hole, we can draw a scenario in the framework of NE model as follows:

\begin{figure}[t]
 \begin{center}
 \includegraphics[height=100mm]{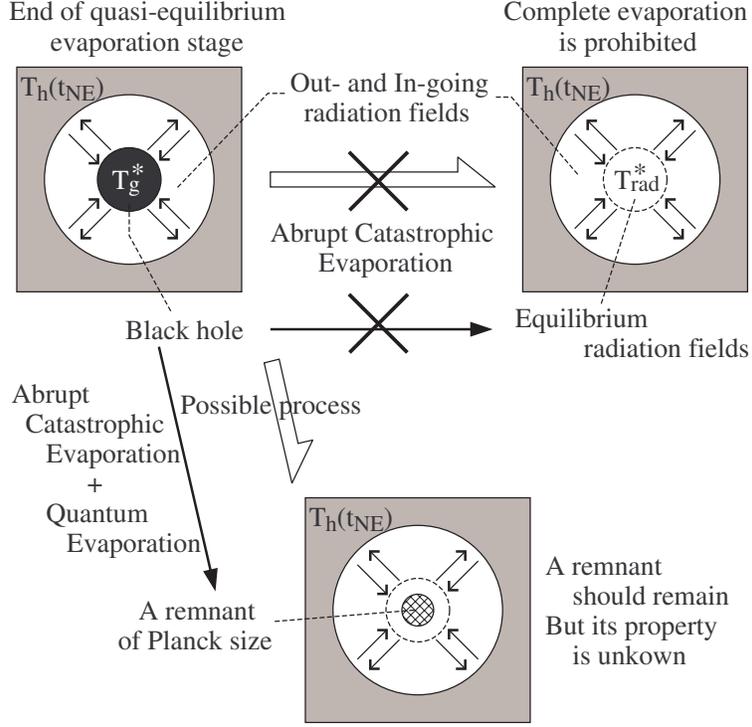}
 \end{center}
\caption{If a black hole evaporates out completely, when a black hole temperature $T_g$ reaches $T_g^{\ast}$, the black hole is suddenly replaced by radiation fields of temperature $T_{rad}^{\ast}$ which is determined by the energy conservation. However it is impossible from entropic viewpoint, and a remnant should remain. But its equations of states are unknown.}
\label{fig-8}
\end{figure}

As a quasi-equilibrium stage of black hole evaporation proceeds until $\tn$, the temperature $T_g$ approaches a critical value $T_g^{\ast} = 1/4 \pi R_g^{\ast}$, where $R_g^{\ast} \defeq R_g(\tn)$. 
Then a highly nonequilibrium dynamical stage and a quantum evaporation stage follow successively. 
Under the assumption of complete evaporation, these successive stages are described as one phenomenon; 
a black hole emits completely its mass energy $E_g^{\ast}$ ($= R_g^{\ast}/2$) within a negligibly short time scale $\delta t_{dyn} + 1$ ($\ll \tn$), and equilibrium radiation fields remain. 
Hereafter in this subsection until a contradiction will be derived, the abrupt catastrophic evaporation under the assumption of complete evaporation means those successive stages, highly nonequilibrium dynamical one and quantum evaporation one. 
Then the abrupt catastrophic evaporation is described by a replacement of a black hole of radius $R_g^{\ast}$ by equilibrium radiation fields of volume $V_g^{\ast} = (4 \pi/3) \, R_g^{\ast \, 3}$. 
In this replacement, thermodynamic states of heat bath and nonequilibrium radiation fields around the region of $V_g^{\ast}$ are not changed (see the upper part in Fig.\ref{fig-8}). 
Here we assume the energy conservation so that the energy of equilibrium radiation fields in volume $V_g^{\ast}$ equals the mass energy $E_g^{\ast}$ of black hole,
\eqb
 E_g^{\ast} = 4 \, \sigmap \, T_{rad}^{\ast \, 4} \, V_g^{\ast} \, ,
\eqe
where $4 \sigmap \, T^4 \, V$ is the equilibrium energy of radiation fields. 
This gives
\eqb
 T_{rad}^{\ast} = \left(\, \frac{3}{32 \, \sigmap \, \pi \, R_g^{\ast \, 2}} \,\right)^{1/4} \, .
\label{eq-evapo.info.temperature.rad}
\eqe
When a sufficiently long time has passed after the abrupt catastrophic evaporation (under the assumption of complete evaporation), the whole system reaches an equilibrium state in which a heat bath and radiation fields in the hollow region have the same equilibrium temperature. 
However, without considering such totally equilibrium end state of the whole system, but with considering the states just before and just after the abrupt catastrophic evaporation, we can discuss whether a remnant remains or not after a quantum black hole evaporation.

It is reasonable to require the increase of total entropy along black hole evaporation, because of the isolated condition of the whole system in NE model. 
Therefore $S_{tot}^{\ast} < S_{tot}^{\ast \, \prime}$ holds, where $S_{tot}^{\ast}$ is the total entropy of the whole system at time $\tn$ just before the abrupt catastrophic evaporation, and $S_{tot}^{\ast \, \prime}$ is the total entropy of the whole system just after the complete evaporation of black hole. 
Because the abrupt catastrophic evaporation is described by a simple replacement of a black hole by equilibrium radiation fields without changing thermodynamics states of heat bath and nonequilibrium radiation fields, the entropy difference is given by
\eqb
 \Delta S_{tot}^{\ast} \defeq S_{tot}^{\ast \, \prime} - S_{tot}^{\ast}
 = S_{rad}^{\ast} - S_g^{\ast} \, ,
\eqe
where $S_g^{\ast}$ is black hole entropy of temperature $T_g^{\ast}$ and $S_{rad}^{\ast}$ is equilibrium entropy of radiation fields of volume $V_g^{\ast}$ and temperature $T_{rad}^{\ast}$. 
If the complete evaporation of black hole is true of the case, a relation $\Delta S_{tot}^{\ast} > 0$ must hold. 
However it is impossible as follows:

\begin{figure}[t]
 \begin{center}
 \includegraphics[height=30mm]{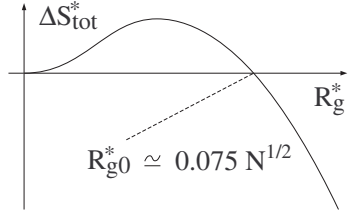}
 \end{center}
\caption{Schematic graph of $\Delta S_{tot}^{\ast}$. A prohibition against complete evaporation of black hole is suggested.}
\label{fig-9}
\end{figure}

Using equations of states \eqref{eq-model.eos} and the equilibrium entropy of radiation fields $(16 \sigmap/3)\, T_{rad}^{\ast\,3} \, V_g^{\ast}$ together with Eq.\eqref{eq-evapo.info.temperature.rad}, the entropy difference becomes
\eqb
 \Delta S_{tot}^{\ast}
 = \frac{2}{3} \left(\, \frac{32 \, \pi \, \sigmap}{3} \,\right)^{1/4} R_g^{\ast \, 3/2}
   - \pi \, R_g^{\ast \, 2} \, .
\label{eq-evapo.info.entropy.diff}
\eqe
A schematic graph of $\Delta S_{tot}^{\ast}$ is shown in Fig.\ref{fig-9}, where $R^{\ast}_{g 0} \defeq (8/27 \sqrt{5 \pi}) \sqrt{N} \simeq 0.075\sqrt{N}$ is given by $\Delta S_{tot}^{\ast} = 0$ and Eq.\eqref{eq-intro.sigma}. 
We find $\Delta S_{tot}^{\ast} > 0$ for $R_g^{\ast} < R_{g 0}^{\ast}$ and $\Delta S_{tot}^{\ast} < 0$ for $R_g^{\ast} > R_{g 0}^{\ast}$. 
Here recall that, because of $C_Y^{(g)} > 0$ by definition and a behavior of $C_X(R_g)$ explained in Subsec.\ref{subsubsec-CX.Rg} (left panel in Fig.\ref{fig-3}), an inequality $R_g^{\ast} > \tilde{R}_g$ holds due to the validity condition \eqref{eq-evapo.trans.validity.1}, where $\tilde{R}_g$ is given by $C_X(R_g) = 0$. 
Furthermore, using Eq.\eqref{eq-CX.tRg}, we find $\tilde{R}_g/R_{g 0}^{\ast} \simeq (27 \sqrt{5 \pi}/32 (30 \pi)^{1/3} ) \times (R_h^2/N)^{1/6} \simeq 0.734 \times (R_h^2/N)^{1/6}$. 
Then, by Eq.\eqref{eq-intro.N} and requirement $R_h \gtrsim 60$ of the occurrence of abrupt catastrophic evaporation, we obtain
\eqb
 \frac{R_g^{\ast}}{R_{g 0}^{\ast}} > \frac{\tilde{R}_g}{R_{g 0}^{\ast}} \simeq 1 \, .
\label{eq-evapo.info.entropy.absurd}
\eqe
This indicates $\Delta S_{tot}^{\ast} < 0$. 
Consequently, by the reductive absurdity as explained in fourth paragraph in this subsection, a complete evaporation of black hole is impossible (see Fig.\ref{fig-8}).

From the above, we give a suggestion as follows: {\it a complete evaporation of black hole is prohibited and a remnant should remain after a black hole evaporation. The entropy of remnant should guarantee the increase of total entropy}. This implies disappearance of the information loss paradox.

Here we comment about the mass of remnant. 
One may think that the mass energy (or internal energy) of remnant may be extracted by some energetics. 
One may obtain a minimum entropy $S_{min}^{(rem)}$ of remnant as $S_{min}^{(rem)} = S_g^{\ast}$, and a mass energy of remnant may be obtained from $S_{min}^{(rem)}$. 
However, because equations of states of remnant is unknown, it is impossible to obtain an energy of remnant from $S_{min}^{(rem)}$. 
On the other hand, as discussed in Subsec.\ref{sec-evapo.ace} (in Eq.\eqref{eq-evapo.ace.duration}), the mass of quantum black hole is expected to be about one Planck energy. 
Therefore, as long as the NE model is extrapolated over its validity, the mass of remnant seems to be of the order of one Planck energy.

In the rest of this subsection, we discuss more about the validity of discussion given in deriving inequality \eqref{eq-evapo.info.entropy.absurd}. 
There are three points we are going to discuss here. For the first, we discuss about the usage of NE model in this subsection. 
If we want to analyze details of the final stage of black hole evaporation process, then a quantum gravity is necessary, since a black hole becomes quantum, $R_g < 1$. 
On the other hand, because of the quasi-equilibrium assumption, the black hole radius has to be restricted as $( R_g(0) > ) R_g^{\ast} \gtrsim 1$ in the framework of NE model. 
However this restriction $R_g^{\ast} \gtrsim 1$ does not mean an impossibility of NE model for considering the final fate of black hole evaporation. 
Although the NE model is based on a classical Schwarzschild black hole, once one assumes a complete evaporation, then it is true that radiation fields remain and their entropy can be counted after a quantum evaporation process ended. 
This means, even though we never refer to any detail of present incomplete quantum gravity theories, we can compare an entropy of black hole before the start of quantum evaporation process with an entropy of radiation fields after the end of quantum evaporation process. 
Hence our analysis done above seems to be reasonable to approach the final fate of black hole evaporation, and does not include the ``uncertainty'' of present incomplete quantum gravity theories. 
Our result seems more universal than the other discussions based on present incomplete quantum gravity theories.

For the second point, we turn our discussion to the starting assumption that a black hole evaporates out completely and equilibrium radiation fields remain. 
As seen above, Eq.\eqref{eq-evapo.info.entropy.diff} has led a contradiction to deny the assumption and to result in necessity of a remnant. 
Here recall that Eq.\eqref{eq-evapo.info.entropy.diff} is calculated with using the equilibrium entropy of radiation fields. 
Then, one may think it is more general to modify the starting assumption so that radiation fields after complete evaporation are not necessarily in an equilibrium. 
However, as explained in detail in~\cite{ref-sst.rad}, the nonequilibrium entropy of radiation fields should be smaller than the equilibrium one $S_{rad}^{\ast}$, since the equilibrium state is the maximum entropy state. 
Hence it is reasonable to expect that an inequality $\Delta S_{tot}^{\ast} < 0$ holds stronger for a modified starting assumption which requires nonequilibrium radiation fields after complete evaporation of black hole.

For the third point, we comment about black hole evaporation in an empty space (a situation without heat bath). 
As discussed at the end of Subsec.\ref{sec-evapo.ace}, the abrupt catastrophic evaporation at semi-classical level is expected to occur for a black hole evaporation in an empty space in a full general relativistic framework. 
Recall that calculations of entropy difference $\Delta S_{tot}^{\ast}$ in Eq.\eqref{eq-evapo.info.entropy.diff} do not depend on the outside of black hole, but depend only on the black hole entropy $S_g^{\ast}$ and the equilibrium entropy $S_{rad}^{\ast}$ of radiation fields. 
This means, without respect to the degree of nonequilibrium nature of the environment around black hole, once the abrupt catastrophic evaporation of black hole occurs, the same discussion as given in this subsection seems applicable to a black hole evaporation in an empty space. 
Therefore, a full general relativistic treatment of black hole evaporation in an empty space may result in necessity of a remnant at the end state of black hole evaporation.

\subsection{Summary of this section}
\label{sec-evapo.sum}

In this section, using the NE model, we found followings:
\begin{itemize}
\item
The semi-classical evaporation stage consists of two stages, quasi-equilibrium one ($t < \tn$) and highly nonequilibrium dynamical one ($\tn < t$). 
The former is treated by the NE model, while the latter is not. 
If the steady state thermodynamics for radiation fields were not applied, the latter stage did not found. 
The existence of the latter stage is the very result of NE model. 
\item
In the quasi-equilibrium evaporation stage, if the thickness of the hollow region is thin enough, then $\jn/\je < 1$ holds during the black hole evaporation process and $\tn > \te$ is obtained. 
However if the hollow region is thick enough, then $\jn/\je > 1$ holds during the evaporation process and $\tn < \te$ is obtained. 
The nonequilibrium effect of energy exchange between black hole and its outside environment tends to accelerate the evaporation process. 
Because of the negative heat capacity of black hole, an inverse sense against a naive sense based on positive heat capacity is offered; the more amount of energy is extracted from black hole by heat bath, the more rapidly the black hole emits its mass energy. 
\item
Duration of the highly nonequilibrium dynamical stage $\delta t_{dyn}$ is negligibly shorter than $\tn$. 
Energy emission rate by black hole in NE model $\jn$ at $\tn$ is very stronger than that in an empty space $\je$. 
These imply a huge energy burst during highly nonequilibrium dynamical stage. 
Such huge burst is the very result of NE model, and we call it the abrupt catastrophic evaporation.
\item
Since the duration of quantum evaporation stage following the highly nonequilibrium dynamical stage will be about one Planck time, the time scale of black hole evaporation $t_{ev}$ is estimated as $t_{ev} \simeq \tn + \delta t_{dyn} + 1$. 
This gives $t_{ev} \sim \tn$.
\item
By extrapolating the NE model to the highly nonequilibrium dynamical evaporation stage and quantum evaporation stage, a complete evaporation of black hole after the quantum evaporation stage is prohibited. 
This denotes a remnant of Planck size may remain at the end of quantum evaporation stage in order to guarantee the increase of total entropy along the whole process of evaporation. 
This implies a disappearance of the information loss paradox due to the nonequilibrium effect of energy exchange between black hole and its outside environment. 
This suggestion does not depend on present incomplete theories of quantum gravity. 
\end{itemize}

\section{Physical essence of the generalized second law}
\label{sec-gsl}

Now we apply the steady state thermodynamics for radiation fields to the NE model, and modify it to reveal the physical essence which guarantees the validity of generalized second law of black hole thermodynamics. 
Contents of this section are based on~\cite{ref-gsl.sst}, and written without referring to previous Sec.\ref{sec-evapo}. 
Readers interested in the generalized second law may skip over previous section.

\subsection{GSL in the context of black hole evaporation}
\label{sec-gsl.evapo}

According to the black hole thermodynamics~\cite{ref-hr,ref-bht,ref-gsl}, a classical size black hole ($R_g > 1$) is regarded as an object in thermal equilibrium, whose equations of states are given in Eq.\eqref{eq-model.eos}. 
In a general relativistic framework, the statement of generalized second law in the context of black hole evaporation is as follows~\cite{ref-gsl}:
\begin{description}
\item[Generalized second law (GSL):] When a black hole evaporates, its mass energy $E_g$ decreases and consequently the black hole entropy $S_g$ decreases because of the equations of states \eqref{eq-model.eos}. 
However the total entropy of the whole system which consists of the evaporating black hole, the Hawking radiation and any other matters around black hole, must increase. 
\end{description}
In this statement we consider all matter fields on black hole spacetime plus the black hole itself. 
This denotes the whole system under consideration is isolated. 
Hence the GSL requires that the time evolution of that system is a relaxation process and the total entropy increases.

The original form of GSL was a conjecture that the horizon area (divided by $4$) is regarded as a true black hole entropy~\cite{ref-gsl}. 
However refer to the work~\cite{ref-entropy} which were written after the suggestion of original GSL and revealed the horizon area (divided by $4$) is the ``equilibrium'' entropy of whole gravitational field on black hole spacetime. 
Therefore, throughout this article, we consider the horizon area is a true equilibrium entropy of black hole and the GSL has already been proven~\cite{ref-entropy,ref-gsl.proof}. 
The equilibrium black hole entropy given by the horizon area is $S_g$ in Eq.\eqref{eq-model.eos}, which is called the Bekenstein-Hawking entropy. 
This section aims not to prove GSL, but to reveal a physical essence which guarantees the validity of GSL.

When we consider GSL in the context of black hole evaporation, the nonequilibrium nature in the outside environment around black hole should be taken into account. 
This means that, while a classical size evaporating black hole is described by an equilibrium solution of Einstein equation under the quasi-equilibrium assumption, the outside environment should be treated as a nonequilibrium matter. 
Then we have to define the total entropy of the whole system, which consists of the evaporating black hole and the nonequilibrium matter fields around black hole including Hawking radiation. 
It has already been shown in~\cite{ref-entropy.add} that, for the equilibrium state of a black hole with general ``self-interacting'' matter fields (a heat bath), the total ``equilibrium'' entropy is given by a simple sum of Bekenstein-Hawking entropy $S_g$ and equilibrium entropy of those matter fields. 
This simple sum of equilibrium entropies is obtained in a general relativistic framework using the Euclidean path-integral method for a black hole spacetime and equilibrium matter fields on it. 
Extending those equilibrium results, we assume that the total entropy $S_{tot}$ of the whole system which consists of evaporating black hole and nonequilibrium matter fields is also given by a simple sum, 
\eqb
 S_{tot} \defeq S_g + S_m \, ,
\label{eq-gsl.evapo.total.general}
\eqe
where $S_g$ is the black hole entropy given in Eq.\eqref{eq-model.eos} and $S_m$ is the nonequilibrium entropy of the general self-interacting matter fields including Hawking radiation. 
Here we consider a quasi-equilibrium regime of black hole evaporation process and use the equilibrium Bekenstein-Hawking entropy for black hole entropy $S_g$. 
The GSL requires the following inequality during the black hole evaporation process,
\eqb
 dS_{tot} > 0 \, .
\label{eq-gsl.evapo.gsl}
\eqe
More discussions on the validity of additivity \eqref{eq-gsl.evapo.total.general} and of the existence of a well-defined nonequilibrium entropy $S_m$ for an arbitrary matter are given later in Subsec.\ref{subsubsec-noneq.matter}.

The problem is how to deal with the nonequilibrium entropy $S_m$ of matter fields. 
However, because a general definition of nonequilibrium entropy is not formulated, the existing proofs of GSL consider an equilibrium between a black hole and a heat bath surrounding it, or assume the existence of a well-defined nonequilibrium entropy of arbitrary matters~\cite{ref-entropy,ref-gsl.proof}. 
The equilibrium settings succeeded in proving the GSL, but unfortunately the physical essence of GSL remains veiled. 
To understand the veil over GSL, it is important to know explicitly the situation considered in the equilibrium settings; 
the existing proofs of GSL consider self-interacting matter fields as Hawking radiation and any other matters around black hole. 
Then, shift from equilibrium to black hole evaporation, we can recognize there are three physical origins of GSL:
\begin{description}
\item[Origin of GSL (a):] Self-interactions of the matter fields of Hawking radiation and around black hole. 
These interactions are collision of composite particles and self-gravitation of matter fields. 
The self-interactions causes a self-relaxation of matter fields and produce their entropy. 
Self-interacting matter fields have a positive entropy self-production rate.
\item[Origin of GSL (b):] Gravitational interaction between black hole and matter fields. 
This interaction consist of curvature scattering, gravitational redshift and so on. 
The gravitational field around black hole works as if a virtual medium on which matter fields propagate, then the composite particles of matter fields interact with the virtual medium to result in a relaxation towards some equilibrium state. 
Therefore the gravitational interaction between black hole and matter fields produces matter entropy as well as the origin~(a). 
A positive entropy production rate of matters results.
\item[Origin of GSL (c):] Increase of black hole temperature $T_g$ along black hole evaporation due to the negative heat capacity $C_g < 0$ given in Eq.\eqref{eq-model.capacity}, becomes one of origins of GSL as follows: 
When $dE_g < 0$ due to the evaporation, $T_g$ increases since $dT_g = dE_g/C_g > 0$. 
Then, because the matter fields of Hawking radiation are ordinary matters of positive specific heat, we find that, the more evaporation process proceeds, the more matter entropy the evaporating black hole radiates out in Hawking radiation. 
This is not the entropy production inside the matter fields during propagating outside the evaporating black hole like origins~(a) and~(b), but the growth of the entropy emission rate by black hole along its evaporation. 
Because the negative heat capacity is a peculiar property of self-gravitating systems, this origin~(c) is a self-gravitational effect of black hole on its own thermodynamic state.
\end{description}
In the existing proofs of GSL, all of these origins are included and it remains unclear which of these dominates over the others. 
If the GSL would be proven by considering a situation keeping one of them and discarding the others, then we can conclude that the one kept is the physical essence which guarantees the validity of GSL. 
This section aims to reveal the origin~(c) is the physical essence of GSL.

To do so, we utilize the NE model. 
To pick up the origin~(c), we remove the heat bath from NE model, and put the black hole in an empty and infinitely large flat spacetime. 
The black hole is bared in this situation. 
We consider this situation throughout the present section, and call it {\it the bare NE model}.

The bare NE model includes the self-gravitational effect of black hole on its own thermodynamic state through equations of states~\eqref{eq-model.eos}, but ignores the self-interactions (including self-gravitational interaction) of radiation fields and the gravitational interaction between black hole and radiation fields.
This means that the bare NE model includes only the origin~(c) and ignores the so-called {\it grey body factor}.

Here it should be pointed out that, while the quasi-equilibrium assumption is also valid for the bare NE model for classical size black hole evaporation, but the fast propagation assumption breaks down since the radiation fields spread out into an infinitely large space. 
Hence, we have to take the retarded effect on radiation fields into account. 
However because of the quasi-equilibrium assumption, we ignore the special relativistic Doppler effects due to the shrinkage of black hole surface.

Here one may show an example as an objection: 
When an inter-stellar gas collapses to form a star, the self-gravitational effect of that gas decreases its entropy. 
Then the origin~(c), self-gravitational effects of black hole, may not be the essence of GSL. 
However as discussed later in Subsec.\ref{subsubsec-star.formation}, this objection is not true of the black hole evaporation.

Because the bare NE model does not include the origins~(a) and~(b), the radiation fields have zero entropy production rate during propagating in an empty and infinitely large space. 
Therefore we find
\eqb
 dS_{tot} > dS_{NE} \, ,
\label{eq-gsl.evapo.ineq}
\eqe
where $S_{tot}$ is the total entropy with general self-interacting matter fields given in Eq.\eqref{eq-gsl.evapo.total.general}, and $S_{NE}$ is the total entropy of the bare NE model. 
Under the assumption of simple sum for nonequilibrium entropies considered in Eq.\eqref{eq-gsl.evapo.total.general}, $S_{NE}$ is decomposed as
\eqb
 S_{NE} \defeq S_g + S_{rad} \, ,
\label{eq-gsl.evapo.total.NE}
\eqe
where $S_{rad}$ is the nonequilibrium entropy of radiation fields propagating in an empty and infinitely large flat spacetime. 
Therefore, if an inequality $dS_{NE} > 0$ holds, then the GSL $dS_{tot} > 0$ follows and we can conclude the origin~(c) is the physical essence of GSL. 
Next subsection shows $dS_{NE} > 0$ holds.

\subsection{Time evolution of total entropy $S_{NE}$}
\label{sec-gsl.evol}

In this subsection, we calculate a time evolution of $S_{NE}(t)$ to show the inequality $dS_{NE}~>~0$ along black hole evaporation. 
Here the time $t$ corresponds to a proper time of a rest observer at asymptotically flat region if we can extend the bare NE model to a full general relativistic model. 
In the followings, we obtain explicit forms of $S_g(t)$ and $S_{rad}(t)$ as functions of $t$, then calculate the time evolution of $S_{NE}(t)$ under the assumption of simple sum given in Eq.\eqref{eq-gsl.evapo.total.NE}.

\subsubsection{$S_g$ as a function of time}

Time evolution of black hole radius $R_g$ is given by the Stefan-Boltzmann law,
\eqb
 \frac{dE_g}{dt} = - \sigmap\, T_g^4\, A_g \, ,
\eqe
where $\sigmap$ is the generalized Stefan-Boltzmann constant and $A_g \defeq 4\,\pi\, R_g^2$ is the surface area of black hole. 
This together with Eq.\eqref{eq-model.eos} gives
\eqb
 R_g(t) = R_0 \, \left( 1 - \frac{N\,t}{1280 \, \pi \, R_0^3} \right)^{1/3} \, ,
\label{eq-gsl.evol.radius}
\eqe
where $R_0 \defeq R_g(0)$ is the initial radius, and it is assumed that the emission of Hawking radiation starts at $t=0$. 
This $R_g(t)$ leads the time evolution of thermodynamic states of black hole and radiation fields. 
Eq.\eqref{eq-gsl.evol.radius} gives the evaporation time (life time) of black hole in the framework of the bare NE model,
\eqb
 \te \defeq 1280\,\pi\,\frac{R_0^3}{N} \, .
\label{eq-gsl.evol.lifetime}
\eqe
Using Eqs.\eqref{eq-gsl.evol.radius} and~\eqref{eq-model.eos}, we obtain $S_g(t)$ as a function of time.

Here let us consider about the quasi-equilibrium assumption. 
To validate this assumption, a sufficiently slow evaporation is required. 
Then we discuss $t_{spread}$ and $v$, where $t_{spread}$ is a time scale of radiation fields to spread out into infinitely large space, and $v$ is a shrinkage speed of black hole radius. 
For the first consider the time scale. 
$t_{spread}$ is given by a particle of radiation fields traveling across the size of black hole, 
\eqb
 t_{spread} \defeq R_0 \, . 
\eqe
This $t_{spread}$ gives a typical time scale of radiation fields to spread out into the empty and infinitely large space. 
If $\te$ is longer than $t_{spread}$, we can consider the black hole evaporation proceeds slowly. 
Hence to validate the quasi-equilibrium assumption, we consider the case satisfying the following inequality,
\eqb
 \lambda \defeq \frac{\te}{t_{spread}} = 1280 \pi \frac{R_0^2}{N} > 1 \, .
\label{eq-gsl.evol.qe.1}
\eqe
For classical size initial condition $R_0 > 1$ together with Eq.\eqref{eq-intro.N}, this requirement~\eqref{eq-gsl.evol.qe.1} is relevant. 
For the second consider the shrinkage speed of black hole radius, 
\eqb
 v \defeq \left| \frac{d R_g(t)}{dt} \right| = \frac{R_0^3}{3\, \te} \frac{1}{R_g(t)^2} \, ,
\eqe
where Eq.\eqref{eq-gsl.evol.radius} is used. 
To validate the quasi-equilibrium assumption, $v$ should be slow enough, $v < 1$. 
This gives
\eqb
 t < \te - \frac{\sqrt{N}}{48 \sqrt{15\,\pi}} \, .
\label{eq-gsl.evol.qe.2}
\eqe
This requirement together with Eq.\eqref{eq-intro.N} means that, because of $\sqrt{N/15\pi}/48 < 1$, the quasi-equilibrium assumption is valid at least until one Planck time before evaporation time $\te$.

\subsubsection{$S_{rad}$ as a function of time}

We will obtain the nonequilibrium entropy $S_{rad}$ as a function of $t$ by applying the steady state thermodynamics for radiation fields. 
But before proceeding to that calculation, the retarded effect on radiation fields has to be introduced into the steady state thermodynamics, since the fast propagation assumption breaks down in the bare NE model. 
In this subsection, firstly we construct a modified distribution function for composite particles of radiation fields, then obtain $S_{rad}$ by substituting that distribution function into formulae~\eqref{eq-sst.min.entropy.ll} and~\eqref{eq-sst.min.entropy.ll.fermion}. 
Because of the quasi-equilibrium assumption, we ignore the special relativistic Doppler effects due to the shrinkage of black hole surface.

\begin{figure}[t]
\centerline{\includegraphics[height=35mm]{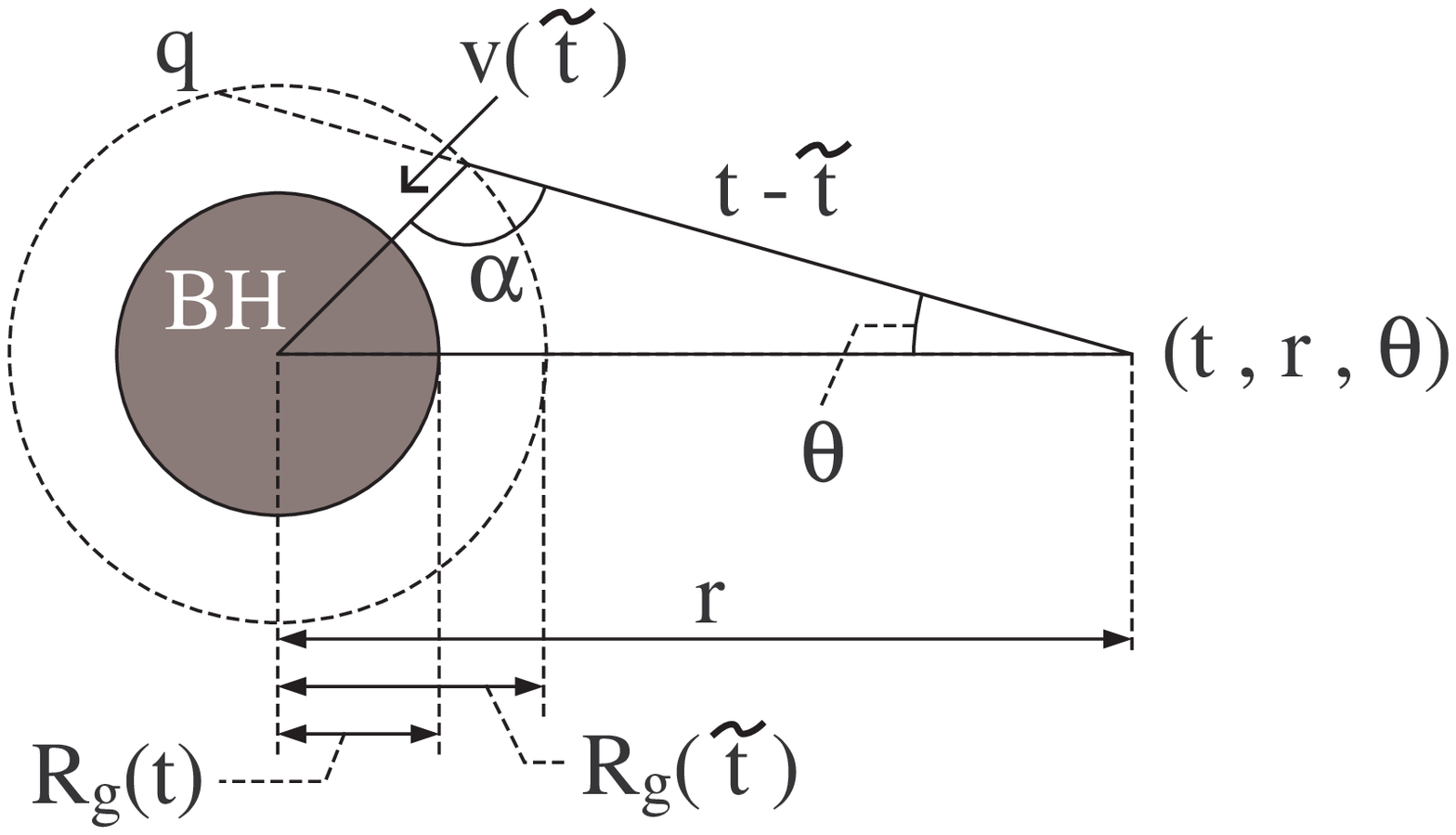} \qquad
            \includegraphics[height=35mm]{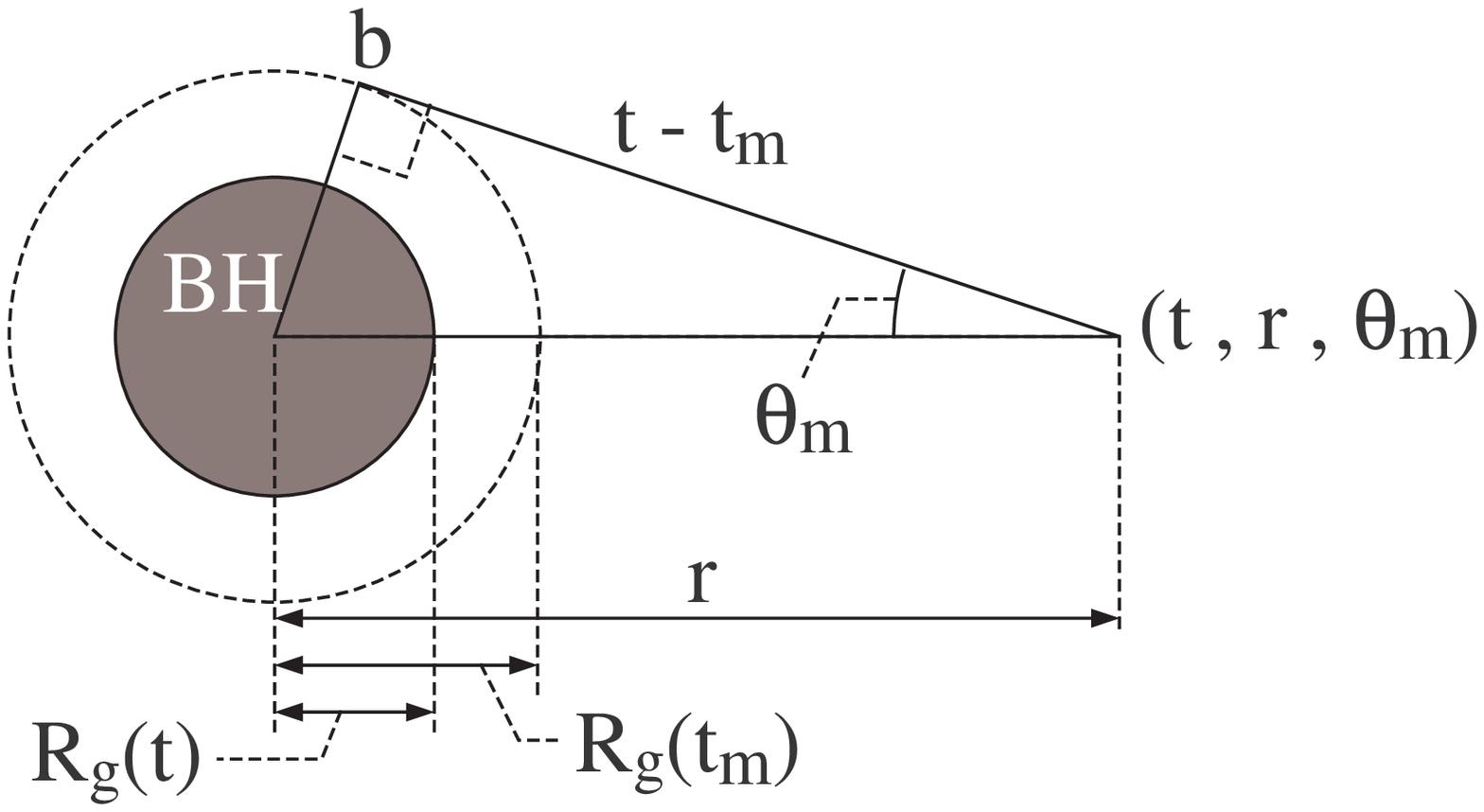}}
\caption{Left panel shows retarded effect on radiation fields. Right one shows the upper bound of $\theta$ for $r < R_0$, or $r < R_0$ and $t > \sqrt{r^2 - R_0^2}$.}
\label{fig-10}
\end{figure}

Hereafter $r$ denotes the areal radius from the center of black hole. 
To take the retarded effect into account, consider a particle of radiation fields emitted by black hole at time $\rt$ and reaches a spatial point of $r$ at time $t \, ( > \rt )$. 
The emission time $\rt$ depends on not only the coordinates $(t,r)$ but also an angle $\theta$ between radial direction of $r$ and momentum of the particle under consideration (see left panel in Fig.\ref{fig-10}). 
This emission time $\rt(t,r,\theta)$ is obtained as a root of the equation,
\eqb
 R_g(\, \rt \,)^2
  = \left(\, t - \rt \, \right)^2 + r^2 - 2 \left(\, t - \rt \, \right) r\, \cos\theta \, .
\label{eq-gsl.evol.emission}
\eqe
This is the equation of degree six about $\rt$, and an appropriate root as the emission time is the maximum root in range, $0 \le \rt \le t$. 
Although another root may exist in this range, however a non-maximum root corresponds to a particle emitted at point $q$ in left panel in Fig.\ref{fig-10} which is obviously unphysical. 
The other four roots of Eq.\eqref{eq-gsl.evol.emission} may be of complex valued.

Furthermore the angle $\theta$ has an upper bound, $0 \le \theta \le \anglm(t,r)$. 
To find an explicit expression for $\anglm$, look at a particle of radiation fields emitted by black hole at the initial time $t=0$, which we call {\it an initial particle}. 
Obviously, we have $\anglm = 0$ for $t < r - R_0$, since no particle has been reached a spatial point of radial distance $r$. 
The initial particles emitted in radial direction form the boundary of region filled with radiation fields. 
The initial particles emitted in off-radial directions propagate behind the initial particles emitted in radial directions. 
Therefore, the initial particles reach a point of radial distance $r  > R_0$ in a time interval, $r - R_0 \le t \le \sqrt{r^2 - R_0^2}$. 
Within this time interval, the upper bound $\anglm$ is given by initial particles, which is obtained with setting $\rt = 0$ in left panel in Fig.\ref{fig-10},
\eqb
 \cos\anglm(t,r)
 = \frac{t^2 + r^2 - R_0^2}{2\, t\, r} \quad , \quad
   \text{for $r > R_0$ and $r - R_0 \le t \le \sqrt{r^2 - R_0^2}$} \, .
\label{eq-gsl.evol.bound.1}
\eqe
For $t > \sqrt{r^2 - R_0^2}$, the upper bound $\anglm$ is not given by any initial particle but by a particle emitted at point $b$ at time $t_m$ shown in right panel in Fig.\ref{fig-10}. 
Then we find
\eqb
 \cos\anglm(t,r)
 = \frac{t - t_m(t,r)}{r} \quad , \quad
   \text{for $r > R_0$ and $\sqrt{r^2 - R_0^2} < t$} \, ,
\label{eq-gsl.evol.bound.2}
\eqe
where the time $t_m(t,r)$ is a real valued root of the equation, 
\eqb
 r^2 = R_g(\, t_m \,)^2 + \left(\, t - t_m \,\right)^2 \, .
\eqe
This is the equation of degree six about $t_m$. 
The appropriate root for $t_m$ should be in the range, $0 \le t_m \le t$. 
This root may be degenerated, since the trajectory of this particle is tangent to the sphere of radius $R_g( t_m )$, and the other four roots may be of complex valued. 
Finally turn to a point of radial distance $r \le R_0$. 
It is obvious for this point that the upper bound $\anglm$ is also given by formula \eqref{eq-gsl.evol.bound.2},
\eqb
 \cos\anglm(t,r)
 = \frac{t - t_m(t,r)}{r} \quad , \quad
   \text{for $r \le R_0$} \, .
\label{eq-upper.bound.3}
\eqe

From the above, the distribution function of radiation fields is given as
\eqb
 d(t , r ; \omega , \theta) =
  \begin{cases}
   \dfrac{1}{\exp\left[\, \omega / \tilde{T} \,\right] \pm 1}
                            &, \quad \text{for $\theta \le \anglm(t,r)$} \\
   0                        &, \quad \text{for $\theta > \anglm(t,r)$}
  \end{cases}
\label{eq-gsl.evol.distribution}
\eqe
where $\tilde{T} = T_g(\, \tilde{t}(t,r,\theta) \,)$, $\omega = \left| \vec{p} \right|$, and the signatures ``$-$'' and ``$+$'' are respectively for bosons and fermions. 
Here we note, while $(t,r)$-dependence in $d(t,r;\omega,\theta)$ expresses a spacetime dependence, $(\omega,\theta)$-dependence expresses a dependence on momentum $\vec{p}$ of particles of radiation fields. 

The nonequilibrium entropy of radiation fields $S_{rad}$ is obtained by substituting Eq.\eqref{eq-gsl.evol.distribution} into Eqs.\eqref{eq-sst.min.entropy.ll} and~\eqref{eq-sst.min.entropy.ll.fermion},
\eqb
 S_{rad}(t) = \int_{R_g(t)}^{t + R_0} dr\, 4 \, \pi\, r^2\, s_{rad}(t,r) \, ,
\label{eq-gsl.evol.entropy.rad} 
\eqe
where
\eqb
 s_{rad}(t,r) \defeq \frac{N}{2880 \, \pi} \int_{y_m(t,r)}^1 dy \,
                \frac{1}{R_g(\, \rt(t,r,y) \,)^3} \, ,
\eqe
where the integrals in Eq.\eqref{eq-sst.min.integrals} are used, and we set $y \defeq \cos\theta$, $y_m \defeq \cos\anglm$ and $g_{\vec{p} , \vec{x}} = n_b$ and $n_f$ for bosons and fermions respectively. 
Since the emission of radiation fields starts at $t=0$, the radiation fields fill the space in range, 
\eqb
 R_g(t) < r < t + R_0 \, .
\eqe

\subsubsection{Total entropy}

Now we obtain the total entropy $S_{NE}$ of the bare NE model under the assumption of simple sum as in Eq.\eqref{eq-gsl.evapo.total.NE},
\eqb
 S_{NE}(t) \defeq S_g(t) + S_{rad}(t) \, ,
\eqe
where $S_g(t) = \pi \, R_g(t)^2$ is given by Eq.\eqref{eq-gsl.evol.radius}, $S_{rad}(t)$ is by Eq.\eqref{eq-gsl.evol.entropy.rad} and time $t$ corresponds to a proper time of a rest observer at asymptotically flat region if we can extend the bare NE model to a full general relativistic model. 
Analytic proof for $dS_{NE}/dt > 0$ is difficult. 
So we try to show it numerically. 
To do so, normalize $S_{NE}$ as follows,
\eqab
&&
 \tau \defeq \frac{t}{t_{ev}} \quad , \quad 
 \rtau \defeq \frac{\rt}{t_{ev}} \quad , \quad
 x \defeq \frac{r}{R_0} \quad , \quad
 X_g(\tau) \defeq \frac{R_g(t)}{R_0} = \left(\, 1 - \tau \,\right)^{1/3} \\
&&
 \Sigma_{NE}(\tau) \defeq \frac{S_{NE}(t)}{S_{NE}(0)} \quad , \quad
 \sigma_{rad}(\tau,x) \defeq \frac{s_{rad}(t,r)}{S_{NE}(0)} \,\, .
\eqae
Then the normalized total entropy is
\eqb
 \Sigma_{NE}(\tau)
 =  X_g(\tau)^2
  + \frac{16}{9\, \lambda} \int_{X_g(\tau)}^{\lambda\,\tau + 1} dx 
    \int_{y_m(\tau,x)}^1 dy \, \frac{x^2}{X_g(\, \rtau(\tau,x,y) \,)^3} \, ,
\label{eq-gsl.evaol.entropy.total}
\eqe
where $\lambda$ is given in Eq.\eqref{eq-gsl.evol.qe.1}. 
Because $\Sigma_{NE}$ does not explicitly depend on $R_0$ (or $N$), then, once the value of $\lambda$ is fixed, the choice of $R_0$ (or $N$) is arbitrary with adjusting the value of $N$ (or $R_0$) appropriately to match with $\lambda$. 
We assume $\lambda > 1$ as mentioned in Eq.\eqref{eq-gsl.evol.qe.1}.

\begin{figure}[t]
\centerline{\includegraphics[height=50mm]{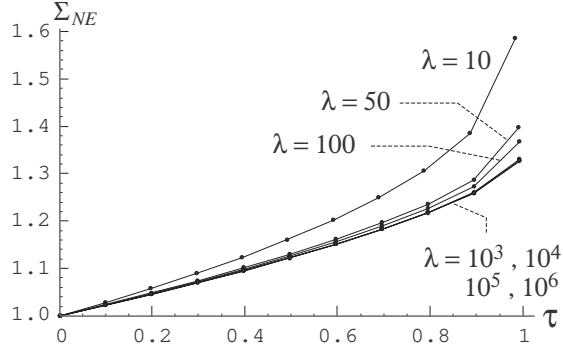}}
\caption{Time evolution of normalized total entropy $\Sigma_{NE}(\tau) \defeq S_{NE}(t)/S_{NE}(0)$ for $0 < \tau < 0.999$, where $\tau$ is a time normalized by evaporation time $\te$. The plotted curves are converging as $\lambda$ increases.}
\label{fig-11}
\end{figure}

\begin{figure}[t]
\centerline{\includegraphics[height=50mm]{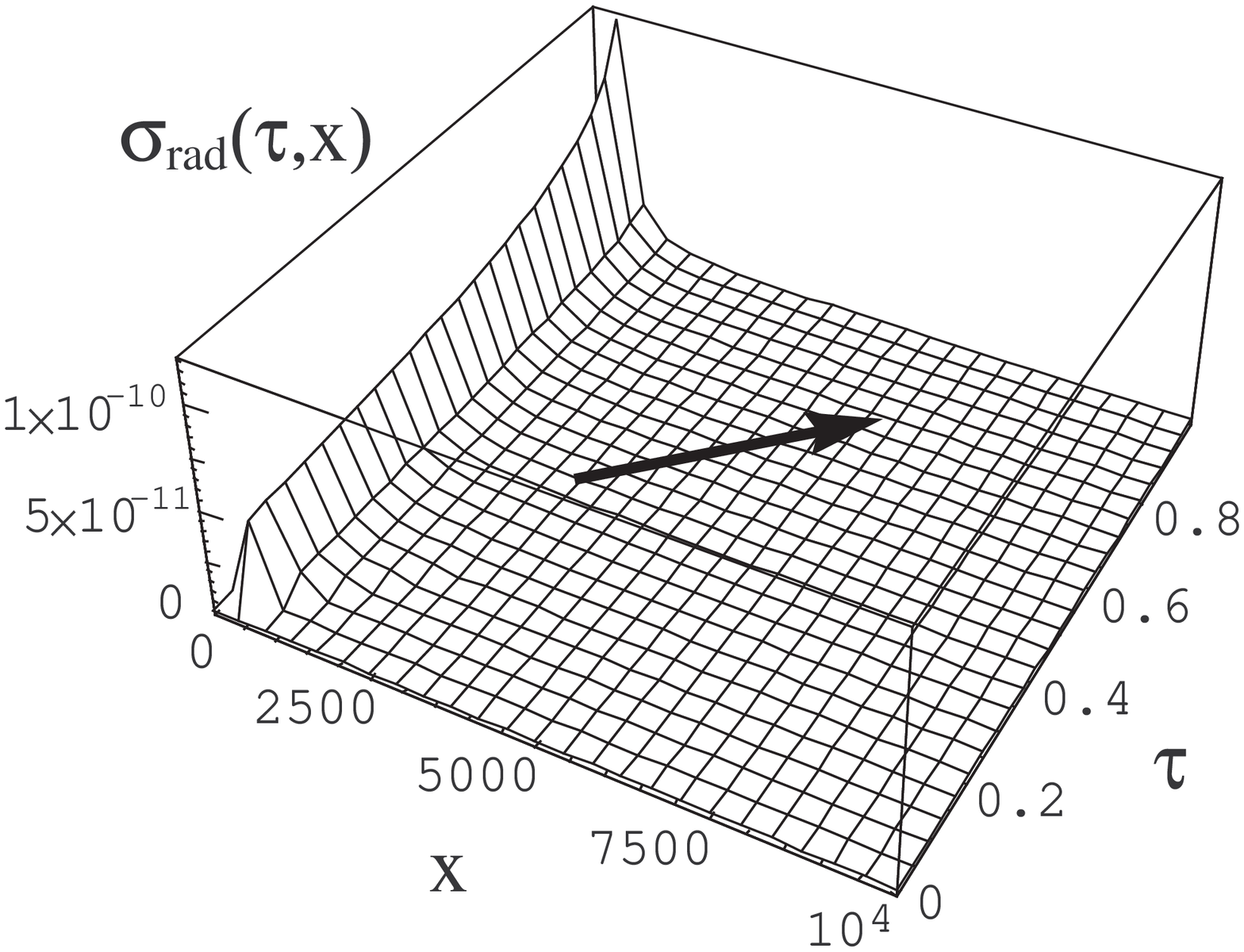} 
            \includegraphics[height=50mm]{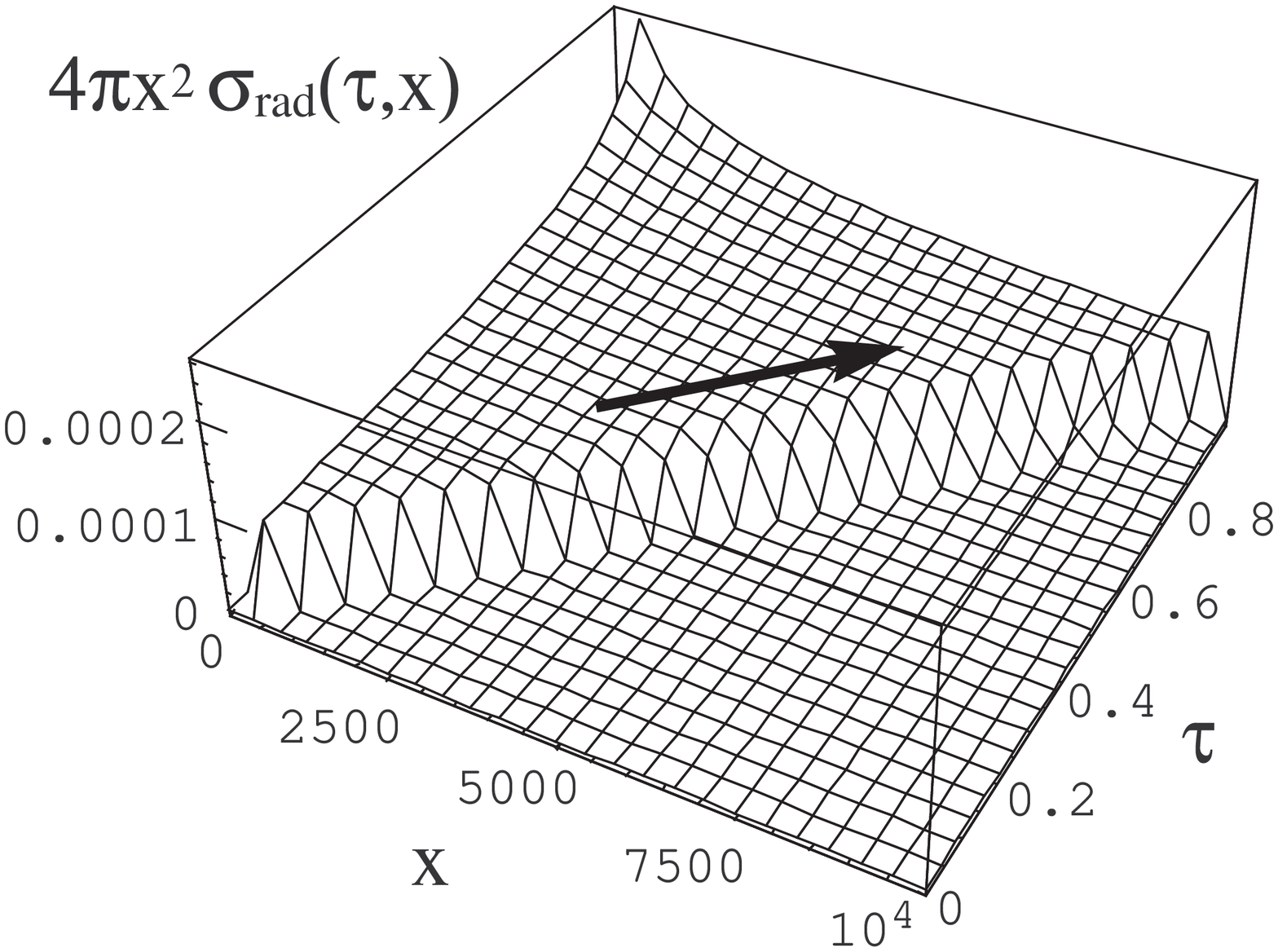}}
\caption{Spatial and temporal evolution of entropy density $\sigma_{rad}(\tau,x)$ and its spherical shell density $4 \pi x^2 \sigma_{rad}(\tau,x)$ for $\lambda  = 10^4$, where $\sigma_{rad}(\tau,x) \defeq s_{rad}(t,r)/S_{NE}(0)$ is a normalized spatial density of radiation fields, $x \defeq r/R_0$ is a normalized spatial distance and $\tau \defeq t/\te$ is a normalized time. The arrows denote direction of null world lines (geodesics).}
\label{fig-12}
\end{figure}

Numerical results are shown in figs.\ref{fig-11} and~\ref{fig-12}. 
Parameters $R_0$ and $N$ are chosen so that inequality~\eqref{eq-gsl.evol.qe.1} holds and time interval $0 \le \tau \le 0.999$ is included in range~\eqref{eq-gsl.evol.qe.2}. 
Those numerical calculations are carried out in that interval $0 \le \tau \le 0.999$ using {\it Mathematica} version 5.2.

Fig.\ref{fig-11} shows time evolution of total entropy $\Sigma_{NE}$ for $0 < \tau < 0.999$ and $\lambda = 10, 50, 100, 10^3, 10^4, 10^5$ and~$10^6$. 
The plotted curves in Fig.\ref{fig-11} are converging as $\lambda$ increases, and those for $\lambda \ge 10^3$ are almost coincident~\footnote{
A technical comment; since calculations by {\it Mathematica} takes a very long time for double integral, Fig.\ref{fig-11} plots $\Sigma_{NE}$ for $\tau = 0,\, 0.1,\, 0.2, \cdots, 0.9,$ and $0.999$ for each $\lambda$. 
However any singular behavior (oscillation, divergence and so on) of $\Sigma_{NE}$ is unexpected. 
Therefore, in order to find monotone increasing nature of $\Sigma_{NE}$, it is enough to show plots of some representative points like Fig.\ref{fig-11}.}. 
What we can find with this figure is as follows:
\begin{itemize}
\item 
Because the plotted curves are converging, it is concluded that the total entropy $\Sigma_{NE}$ is monotone increasing for $\lambda > 1$, and the GSL is well supported as explained in Eqs.\eqref{eq-gsl.evapo.ineq} and~\eqref{eq-gsl.evapo.total.NE}.
\item 
Since the normalized black hole entropy $X_g(\tau)^2$ is obviously monotone decreasing, the entropy of radiation fields increases monotonously faster than the decrease of $X_g^2$.
\item 
It is suggested rather well that $\Sigma_{NE}(\tau)$ gives almost the same curves in $\Sigma$-$\tau$ graph for all values of sufficiently large $\lambda \gtrsim 10^3$, and the final value is approximately given by $\Sigma_{NE}(0.999) \simeq 1.33$.
\end{itemize}

Left panel in Fig.\ref{fig-12} shows the normalized spatial entropy density for radiation fields $\sigma_{rad}(\tau,x)$, and right panel shows its normalized spherical shell density $4 \pi x^2 \sigma_{rad}(\tau,x)$. 
Fig.\ref{fig-12} is made with $\lambda = 10^4$, however the same behavior is obtained for all values of sufficiently large $\lambda \gtrsim 10^3$ as far as the author checked. 
The arrows in Fig.\ref{fig-12} denote a direction of null world lines of particles of radiation fields. 
What we can find with this figure is as follows:
\begin{itemize}
\item 
Tracking the graphs in Fig.\ref{fig-12} along a null word line, we recognize the following: 
Although the spatial entropy density of radiation fields decreases (left panel), the spherical shell density on each spherical shell remains constant while it spreads out into an empty and infinitely large flat spacetime (right panel). 
This is consistent with the absence of self-relaxation of radiation fields and gravitational interaction between black hole and radiation fields. 
\item 
The entropy of radiation fields at black hole surface (right panel), which equals the amount of matter entropy emitted by black hole at each moment, increases monotonously as the black hole evaporation proceeds. 
This means an accelerated entropy emission by black hole. 
\end{itemize}

Hence we find from figs.\ref{fig-11} and~\ref{fig-12} that the entropy emission rate by black hole increases faster than the decrease of black hole entropy. 
This indicates the physical essence of GSL is neither the origin~(a) nor~(b), but the origin~(c) which is the self-gravitational effect of black hole causing a negative heat capacity and an increase of black hole temperature along its evaporation.

\subsection{Summary and supplementary discussions of this section}
\label{sec-gsl.sum}

\subsubsection{Summary}

Considering the GSL in the context of black hole evaporation, we recognize there are three physical origins of entropy increase; (a)~self-interactions of matter fields around black hole, (b)~gravitational interaction between the matter and black hole, and (c)~self-gravitational effect of black hole.
The origins~(a) and~(b) give positive entropy production rates inside the matter fields. 
The origin~(c) appears as an increase of black hole temperature due to the negative heat capacity and gives an increasing entropy emission rate by black hole. 
Then we consider the bare NE model constructed by removing heat bath from the NE model. 
The bare NE model describes a black hole evaporation including only the origin~(c) and discarding the others~(a) and~(b). 
Applying the steady state thermodynamics to radiation fields in the bare NE model, we can calculate explicitly a time evolution of total entropy $S_{NE}$, and find $S_{NE}$ increases monotonously. 
This denotes the GSL holds as explained in Eqs.\eqref{eq-gsl.evapo.ineq} and~\eqref{eq-gsl.evapo.total.NE}. 
Hence we conclude as follows:
\begin{description}
\item[Physical essence of GSL:] The self-gravitational effect of black hole which appears as an increasing entropy emission rate by black hole guarantees a validity of GSL. 
The entropy emission rate by black hole increases faster than the decrease of black hole entropy. 
The entropy production inside the Hawking radiation and the other matter fields around black hole is not necessary for a validity of GSL. 
\end{description}
The increase of entropy emission rate is shown by an accelerated increase of entropy density $s_{rad}$ at black hole surface shown in Fig.\ref{fig-12} and causes the accelerated increase of total entropy $S_{NE}$ shown in Fig.\ref{fig-11}. 

Furthermore we find another interesting result from Fig.\ref{fig-11}: 
An inequality $\lambda > 10^3$ together with Eq.\eqref{eq-intro.N} corresponds to black holes of initial radius $R_0 \gtrsim 5$. 
Then Fig.\ref{fig-11} shows that the normalized total entropy $\Sigma_{NE}(\tau)$ for black holes of initial radius $R_0 \gtrsim 5$ evolves in almost the same increasing fashion and reaches almost the same final value $\simeq 1.33$. 
On the other hand inequality~\eqref{eq-gsl.evapo.ineq} denotes $S_{NE}$ is the lowest estimate for total entropy of a full general relativistic black hole evaporation including self-interaction of matter fields around black hole and gravitational interaction between black hole and the matters. 
Hence we find a result as follows:
\begin{description}
\item[Lower bound for growth of the  total entropy:] For black hole evaporation in a full general relativistic framework including self-interaction of matter fields and gravitational interaction between black hole and the matters, the final value of total entropy $S_{tot}$ (not its lowest estimate $S_{NE}$) should be larger than $1.33 \times S_g(0)$. 
\end{description}
Exactly speaking our analysis based on the bare NE model covers only a semi-classical evaporation stage until one Planck time before the time $\te$ as discussed in Eq.\eqref{eq-gsl.evol.qe.2}. 
However it is reasonable to expect that the total entropy after the end of quantum evaporation stage is greater than the total entropy just at the onset of quantum evaporation stage, since a well-defined entropy has to be of non-decreasing. 
Therefore the above result on the lower bound of total entropy should be true of the end state of quantum evaporation stage.

\subsubsection{Supplementary discussion 1}
\label{subsubsec-noneq.matter}

We discuss about two key assumptions remained to be proven, the existence of a well-defined nonequilibrium entropy $S_m$ of an arbitrary self-interacting matter fields and the additivity of equilibrium and nonequilibrium entropies~\eqref{eq-gsl.evapo.total.general}. 
These two assumptions seem to be reasonable as follows: 

On the existence of a well-defined nonequilibrium entropy $S_m$, we refer to the present status of study on laboratory systems which have self-interactions but not self- and external-gravitational interactions. 
For example, consider a laboratory system is in a nonequilibrium state which is far from an equilibrium but whose heat flux is not extremely strong. 
Then the extended irreversible thermodynamics \cite{ref-eit} gives a well-defined nonequilibrium entropy flux up to the second order in expansion by the heat flux. 
On the other hand, the evaporation time $\te$ of a black hole has a very long time scale as explained in Eq.\eqref{eq-gsl.evol.qe.1}. 
The black hole evaporation proceeds so slowly that the nonequilibrium state of self-interacting matter fields around black hole is not extremely far from an equilibrium. 
Therefore we can expect that the expansion of their thermodynamic quantities by the heat flux up to the second order is a good approximation. 
Hence it is reasonable to assume the existence of a well-defined nonequilibrium entropy of an arbitrary self-interacting matter fields according to extended irreversible thermodynamics.

Next consider about the additivity \eqref{eq-gsl.evapo.total.general}. 
As mentioned in the previous paragraph, the nonequilibrium entropy has already been defined well up to the second order in expansion by heat flux~\cite{ref-eit}. 
Furthermore that nonequilibrium entropy satisfies the additivity as in ordinary equilibrium thermodynamics. 
Therefore we may expect that, when an equilibrium entropy satisfies the additivity even under the effect of gravity, the nonequilibrium entropy in the framework of extended irreversible thermodynamics also satisfies the additivity. 
Hence, because of the additivity of equilibrium entropies of black hole and matter fields~\cite{ref-entropy.add}, it seems reasonable to assume the additivity \eqref{eq-gsl.evapo.total.general}.

\subsubsection{Supplementary discussion 2}
\label{subsubsec-star.formation}

Finally we try to answer an objection: 
When an inter-stellar gas collapses to form a star, the self-gravitational effect of that gas decreases its entropy. 
Then the black hole evaporation under the self-gravitational effect of black hole may not result in an increase of total entropy.

When an interstellar gas collapses to form a star, it is commonly believed that there arises the increase of net entropy of total system which consists of the collapsing gas and the radiated matters from the gas. 
However the self-gravitational effect of the collapsing gas causes the decrease of entropy of the collapsing gas. 
It is briefly explained as follows: 
Since the pressure of that gas at its surface is zero, the loss of energy of collapsing gas par a unit time $\Delta E$ due to energy emission is actually the loss of heat due to the first law of thermodynamics,
\eqb
 \Delta E \sim T \, \Delta S \, 
\label{eq-gsl.sum.1}
\eqe
where $T$ is the temperature of collapsing gas, and $\Delta S$ is the ``loss'' of entropy of collapsing gas par a unit time. 
The energy loss $\Delta E$ is the minus of the luminosity $L$ of collapsing gas, $L = - \Delta E$. 
Then, with the assumption of local mechanical and thermal equilibrium of collapsing gas at each moment of its collapse, relation \eqref{eq-gsl.sum.1} is rewritten more exactly to the following form~\cite{ref-sgs},
\eqb
 L = - \frac{d [ \, U + \Omega \,]}{dt} \sim - T \Delta S \, ,
\eqe
where $U$ is the total internal energy of collapsing gas, and $\Omega$ is the total self-gravitational potential given by
\eqb
 \Omega = \int_0^M \, dm \,\left( - \frac{G\, m}{r(m)} \, \right) \, ,
\eqe
where $G$ is Newton's constant, $M$ is mass of collapsing gas, and the radial distance from the center of gas $r(m)$ is expressed as a function of mass $m$ inside a sphere of radius $r$.

From the above, we recognize that the radius $r(m)$ becomes smaller as the gas collapses, then the self-gravitational potential $\Omega$ decreases to result in the radiation of energy $L >0$ and the loss of entropy $\Delta S < 0$. 
This is a similar phenomenon to the evaporation of black hole itself with decreasing its entropy $dS_g < 0$. 
On the other hand, if we consider not a collapsing but an expanding self-gravitating gas, the radius $r(m)$ increases to result in the increase of entropy $\Delta S >0$. 
Here we point out that the Hawking radiation in black hole evaporation corresponds not to a collapsing gas but to an expanding gas. 
If we consider the radiation fields of Hawking radiation including their self-gravitational effect, the entropy of radiation will increase during spreading out into an infinitely large space. 
This is just the origin~(b) of GSL. 
Hence it seems that the entropy of self-gravitating matter fields of Hawking radiation is larger than that of a non-self-gravitating Hawking field. 
This implies $S_m > S_{rad}$, and the objection mentioned above is not true of the black hole evaporation process.

Finally let us recall a statement given in the second paragraph; it is commonly ``believed'' that there arises the increase of net entropy of total system which consists of the collapsing gas and the radiated matters from the gas. 
One of the reasons why it is not proven but ``believed'' is that there has not been nonequilibrium thermodynamics to treat the net entropy. 
Although we considered the black hole evaporation in this section, a similar method based on the steady state thermodynamics will be applicable to a star formation process including radiations from collapsing gas. 
Then ``believed'' will become ``proven''.

\section{Concluding remarks}
\label{sec-conc}

Exactly speaking the black hole evaporation is a nonequilibrium process. 
We used the NE model as a simplified thermodynamic model of black hole evaporation, and applied the steady state thermodynamics for radiation fields~\cite{ref-sst.rad}. 
It is this steady state thermodynamics that enables us to treat the nonequilibrium nature of black hole evaporation. 
In the framework of NE model, we can find a detailed picture of evaporation process with energy accretion onto black hole~\cite{ref-evapo.sst}, which is summarized in Subsec.\ref{sec-evapo.sum}. 
Also, by modifying the NE model to describe the evaporation process in an empty space ignoring grey body factor for the Hawking radiation, we can find the essence of the generalized second law of black hole thermodynamics in the context of black hole evaporation~\cite{ref-gsl.sst}, which is summarized in Subsec.\ref{sec-gsl.sum}.

One of the results in Sec.\ref{sec-evapo} is that the larger the nonequilibrium region around black hole, the more accelerated the black hole evaporation under the influence of energy exchange between black hole and its outside environment (heat bath). 
On the other hand, Sec.\ref{sec-gsl} considered the bare NE model which is obtained by removing the heat bath from the NE model. 
The bare NE model describes the black hole evaporation in an empty space with ignoring grey body factor, which is considered in Subsec.\ref{sec-evapo.ne} to compare with the NE model. 
Then one may think that the bare NE model is a limiting case of NE model which has the infinitely large nonequilibrium region around black hole, and the evaporation time of the bare NE model would be zero. 
But here recall that the fast propagation assumption breaks down for such limiting situation, and the analysis given in Sec.\ref{sec-evapo} may not be applied to the limiting situation. 
Indeed, as discussed in Subsec.\ref{subsubsec-ne.empty}, the bare NE model seems not describe the limiting situation of NE model. 
The bare NE model is quite different from the NE model, because no energy accretion is in bare NE model while it is in NE model. 
The absence of energy accretion makes the bare NE model a qualitatively different situation from the limiting situation of NE model.

As mentioned Sec.\ref{sec-model}, the NE model is not a full general relativistic model, and ignoring gravitational redshift and curvature scattering on radiation fields propagating in hollow region. 
Towards a general relativistic NE model, we have to extend the steady state thermodynamics to its general relativistic version. 
If we will construct a general relativistic steady state thermodynamics for radiation fields, the gravitational effects on radiation fields will be included in the NE model. 
Even when the extended steady state thermodynamics for radiation fields are applied, the semi-classical evaporation stage is divided into two stages as shown in Sec.\ref{sec-evapo}, quasi-equilibrium one and highly nonequilibrium dynamical one. 
If we concentrate on the quasi-equilibrium evaporation stage, then the evaporating black hole can be expressed approximately by equilibrium solutions of Einstein equation. 
In that case, we can avoid the mathematical and conceptual difficulties for describing evaporating black holes (see the beginning of Sec.\ref{sec-intro}). 
However, in considering the highly nonequilibrium dynamical evaporation stage in a general relativistic setting, we will face those difficulties.

Next let us give a comment on the lower bound of total entropy of black hole and matter fields, summarized in Subsec.\ref{sec-gsl.sum}. 
Concerning the entropy in any spacetime, the so-called {\it covariant entropy bound conjecture} (holographic conjecture) is very interesting, which, inspired by some quantum gravitational evidences, conjectures a universal upper bound on the entropy of any gravitating system~\cite{ref-cebc}. 
As far as the present author considers, a weak point of the covariant entropy bound conjecture at present seems a lack of evidences related to some nonequilibrium thermodynamics and/or statistical mechanics, since the existing evidences seems based on the (local-)equilibrium hypothesis. 
When the quantum gravity will cover the very dynamical and nonequilibrium phenomena, the covariant entropy bound conjecture should also be related to some {\it nonequilibrium entropy}. 
Then, since our analysis in this article is based on the steady states of radiation fields, some insight into quantum gravity may be extracted by combining our result of the lower bound of total entropy with the covariant entropy bound conjecture. 
But at present, the research related with the conjecture is left as an open issue.

Finally focus our comments on ones given from the point of view of nonequilibrium physics. 
All of the results in this article are obtained by NE model and based on the steady state thermodynamics for radiation fields. 
Nonequilibrium thermodynamic approach may be a powerful tool for investigating the nonlinear and dynamical phenomena. 
While we considered the radiation fields which is of non-self-interacting, however nonequilibrium thermodynamics for ordinary dissipative systems has already been established in, for example, the extended irreversible thermodynamics~\cite{ref-eit}. 
It is applicable not to any highly nonequilibrium state but to state whose entropy flux is well approximated up to second order in the expansion by the heat flux of a nonequilibrium state under consideration. 
It is interesting to consider self-interacting matter field for the Hawking radiation and energy accretion, and apply the extended irreversible thermodynamics to those fields. 
Then we may find a variety of black hole evaporation phenomena. 
Furthermore apart from black hole evaporation, since accretion disks around black hole seem to consist of dissipative matters in realistic settings, the extended irreversible thermodynamics may give a unique new approach to investigate black hole astrophysics.

Apart from black hole physics, the steady state thermodynamics for radiation fields may be helpful to understand, for example, the free streaming in the universe like cosmic microwave background and/or the radiative energy transfer inside a star and among stellar objects. 
Also an example of possible application to a star formation process is explained at the end of Subsec.\ref{subsubsec-star.formation}. 
Keeping future expectation of such applications in mind, Subsec.\ref{sec-sst.detail} is devoted to exhibit a detail of the steady state thermodynamics for radiation fields.


\end{document}